%% file: 1ESpaper2.tex
\documentclass[twocolumn,apj]{aastex631}

\usepackage{graphicx}
\usepackage{subfigure}
\usepackage{multirow}
\usepackage{comment}
\usepackage{hyperref}
\usepackage{mathtools}
\usepackage{longtable}
\usepackage{mathrsfs}
\usepackage{fontenc}
\usepackage{xcolor}
\usepackage{url}
\usepackage{gensymb}
\usepackage{pifont}

\newcommand{\angs}{$\mathring{\mathrm{A}}$}

\newcommand{\oiii}{[O\,{\sc iii}] }

\begin{document}

%\slugcomment{Submitted to {\it The Astrophysical Journal}.}

\title{The Interplay between the Disk and Corona of the Changing-look Active Galactic Nucleus 1ES\,1927+654}

\shorttitle{The Interplay between the Disk and Corona of 1ES\,1927+654}

\shortauthors{Li et al.}

\author[0000-0001-8496-4162]{Ruancun Li}
\affil{Kavli Institute for Astronomy and Astrophysics, Peking University,
Beijing 100871, China}
\affil{Department of Astronomy, School of Physics, Peking University,
Beijing 100871, China}

\author[0000-0001-5231-2645]{Claudio Ricci}
\affil{Instituto de Estudios Astrof\'isicos, Facultad de Ingenier\'ia y Ciencias, Universidad Diego Portales, Av. Ej\'ercito Libertador 441, Santiago, Chile}
\affil{Kavli Institute for Astronomy and Astrophysics, Peking University, Beijing 100871, China}
\affil{George Mason University, Department of Physics \& Astronomy, MS 3F3, 4400 University Drive, Fairfax, VA 22030, USA}

\author[0000-0001-6947-5846]{Luis C. Ho}
\affil{Kavli Institute for Astronomy and Astrophysics, Peking University, Beijing 100871, China}
\affil{Department of Astronomy, School of Physics, Peking University, Beijing 100871, China}

\author[0000-0002-3683-7297]{Benny Trakhtenbrot}
\affil{School of Physics and Astronomy, Tel Aviv University, Tel Aviv 69978, Israel}

\author[0000-0003-0172-0854]{Erin Kara}
\affil{MIT Kavli Institute for Astrophysics and Space Research, 70 Vassar Street, Cambridge, MA 02139, USA}

\author[0000-0003-4127-0739]{Megan Masterson}
\affil{MIT Kavli Institute for Astrophysics and Space Research, 70 Vassar Street, Cambridge, MA 02139, USA}

\author[0000-0001-7090-4898]{Iair Arcavi}
\affil{School of Physics and Astronomy, Tel Aviv University, Tel Aviv 69978, Israel}
\affil{CIFAR Azrieli Global Scholars program, CIFAR, Toronto, Canada}

%XX I changed "four distinct models" to simply "different models", first because I want to avoid any possible confusion with the "four distinct stages" mentioned later, and also because "distinct" is superfluous. Of course, they are distinct models; it would be stupid if they were identical :)
%
\begin{abstract}
Time-domain studies of active galactic nuclei (AGNs) offer a powerful tool for understanding black hole accretion physics. Prior to the optical outburst on 23 December 2017, 1ES\,1927+654 was classified as a ``true'' type~2 AGN, an unobscured source intrinsically devoid of broad-line emission in polarized spectra. Through our three-year monitoring campaign spanning X-ray to ultraviolet/optical wavelengths, we analyze the post-outburst evolution of the spectral energy distribution (SED) of 1ES\,1927+654. Examination of the intrinsic SED and subsequent modeling using different models reveal that the post-outburst spectrum is best described by a combination of a disk, blackbody, and corona components. We detect systematic SED variability and identify four distinct stages in the evolution of these components. During the event the accretion rate is typically above the Eddington limit. The correlation between ultraviolet luminosity and optical to X-ray slope  ($\alpha_\mathrm{OX}$) resembles that seen in previous studies of type~1 AGNs, yet exhibits two distinct branches with opposite slopes. The optical bolometric correction factor ($\kappa_{5100}$) is $\sim 10$ times higher than typical AGNs, again displaying two distinct branches. Correlations among the corona optical depth, disk surface density, and $\alpha_\mathrm{OX}$ provide compelling evidence of a disk-corona connection. The X-ray corona showcases systematic variation in the compactness-temperature plot. Between 200 and 650 days, the corona is ``hotter-when-brighter,'' whereas after 650 days, it becomes ``cooler-when-brighter''. This bimodal behavior, in conjunction with the bifurcated branches of $\alpha_\mathrm{OX}$ and $\kappa_{5100}$, offers strong evidence of a transition from a slim disk to thin disk $\sim 650$ days after the outburst.
\end{abstract}


\section{Introduction}
\label{sec:sec1}

%XX check red text below, see if this is what you mean
%
Active galactic nuclei (AGNs) are powered by the accretion of mass onto supermassive black holes located at the center of galaxies \citep{Rees1984ARAA}. According to the unified model of AGNs \citep{Antonucci1985ApJ, Urry1995PASP}, differences between AGN types mainly stem from the orientation of the obscuring torus with respect to our line-of-sight. However, the discovery of changing-look AGNs (see \citealp{Ricci2023NatAs} for a review) has posed a substantial challenge to this model. These AGNs exhibit swift and dramatic changes in spectral properties, signifying inherent changes in the physical conditions of the AGN instead of mere orientation effects. In the X-ray band these objects are typically identified as chnaging-obscuration AGNs \citep{Mereghetti2021ExA}, which exhibit fluctuations in line-of-sight column densities, typically due to gas clouds surrounding the supermassive black hole (e.g., \citealp{Matt2003MNRAS,Risaliti2006ESASP,Rivers2015ApJ,Ricci2016ApJ}). Conversely, objects showing the emergence or disappearance of the blue continuum and broad ultraviolet (UV) or optical lines can be referred to as changing-state AGNs \citep{Graham2020MNRAS}. Interestingly, the emergence of broad emission lines are almost always accompanied by an increase in the continuum flux (e.g., \citealp{Yang2018ApJ,Temple2023MNRAS}). Different physical explanations have been proposed to account for the changing-state phenomenon (e.g., \citealp{Ricci2023NatAs}).  \cite{Noda2018MNRAS} and \cite{Ruan2019ApJ} surmise that geometric shifts due to disk state transitions between an advection-dominated accretion flow (ADAF; \citealp{Ichimaru1977ApJ,Narayan1994ApJ}) and a thin accretion disk (\citealp{Novikov1973blho,Shakura1973AA}), where the broad-line region (BLR) can be inactive during the faint phase \citep{Noda2023ApJ}. Alternatively, a BLR can be newly formed in response to external processes such as a tidal disruption event (TDE; see review in \citealp{Komossa2015JHEAp, Trakhtenbrot2019ApJ,Li2022paper1}).

To distinguish between the above scenarios, an effective approach is to directly measure the mass accretion rate ($\dot{M}$) before and after the changing-look event. In the former scenario, $\dot{M}$ is expected to fluctuate around $1\%$ of the Eddington accretion rate, $\dot{M}_\mathrm{E}$ \citep{Yuan2014ARAA}, while in the TDE scenario $\dot{M} \propto t^{-5/3}$ \citep{Rees1988Nature,Lodato2011MNRAS}. For a given black hole mass ($M_\mathrm{BH}$), $\dot{M}$ generally is linked directly to the bolometric luminosity ($L_\mathrm{bol} = \eta \dot{M}c^2$), where $\eta$ represents the radiation efficiency. However, estimating $L_\mathrm{bol}$ requires an accurate constraint of the peak of the spectral energy distribution (SED), which is located typically in the far-UV or ultra-soft X-ray band (10$-$2000 \angs) in nearby AGNs \citep{Elvis1994ApJS}. This ionizing flux is largely unobservable due to Galactic absorption, but can be interpolated from broadband (optical to X-ray) SED modeling \citep{Jin2012MNRAS,Collinson2015MNRAS} with the aid of either empirical models (e.g., a disk plus a broken power law; \citealp{Vasudevan2007MNRAS,Vasudevan2009MNRAS}) or recently developed physical models, such as that for a color temperature-corrected thermal disk \citep{Done2012MNRAS,Kubota2018MNRAS}, and simulated models with radiation post-processing \citep{Straub2011AA,Narayan2017MNRAS}.

The models mentioned above predominantly consist of thermal emission from the disk and Comptonized emission from the corona, the latter that contributes to the X-ray emission. The flux of these two components typically shows a rather tight correlation. The optical to X-ray slope ($\alpha_\mathrm{OX}$)\footnote{The relationship between X-ray and UV luminosity for AGNs can be defined using $\alpha_\mathrm{OX} \equiv -\log{(L_\mathrm{X-ray}/L_\mathrm{UV})}/\log{(\nu_\mathrm{X-ray}/\nu_\mathrm{UV})}$ \citep{Tananbaum1979ApJ}.} is a useful tool to understand these correlations. Past studies reveal that $\alpha_\mathrm{OX}$ tends to decrease as the luminosity at 2500~\angs\ ($L_\mathrm{2500}$) increases, indicating an enhancement in optical emission relative to the X-rays \citep{Steffen2006AJ,Just2007ApJ,Lusso2010AA,Lusso2017AA}. This suggests that changes take place in the accretion disk and X-ray corona as the accretion rate varies. Moreover, the strong correlation between $\alpha_\mathrm{OX}$ and the disk luminosity could indicate a coupling mechanism between the disk and the corona \citep{Haardt1993ApJ,Malzac2001MNRAS,Done2012MNRAS}. %Larger disk luminosities imply that more seed photons are available for Compton upscattering, generally leading to a surge in X-ray production.

Widely accepted models for the corona's geometry include a ``sandwich'' or a ``lamppost'' configuration. The sandwich model, also termed the slab or disk corona model, posits that the corona is a layer of hot plasma above and below the accretion disk \citep{Haardt1993ApJ,Dove1997ApJ}. Conversely, the lamppost model suggests that the corona is positioned on the black hole's rotation axis, above and/or below the accretion disk \citep{Miniutti2003MNRAS,Petrucci2013AA}. Another model, the spherical or ``blobby'' corona model, hypothesizes a hot corona composed of numerous small, localized, and potentially transient flares or blobs situated near the black hole, both within and above the plane of the disk \citep{Galeev1979ApJ, Beloborodov1999ApJ}. Each of these models carries different implications for the observed X-ray variability and spectral hardening. The exact configuration of the corona remains a topic of ongoing research and may differ across individual AGNs.

The nearby ($z=0.019422$) galaxy 1ES\,1927+654  recently underwent a changing-look event \citep{Trakhtenbrot2019ApJ,Ricci2020ApJL,Ricci2021ApJS},  providing a unique opportunity to test the nature of the accretion disk and its connection with the corona. The event was initially discovered by the All-Sky Automated Survey for Supernovae (ASAS-SN; \citealp{Shappee2014ApJ}). The source was observed to exhibit an increase in flux by at least 2 magnitudes in the $V$ band in March\,\,2018 \citep{Nicholls2018ATel}. Subsequent optical spectroscopy \citep{Trakhtenbrot2019ApJ} documented the emergence of broad Balmer emission lines, which were absent in earlier observations \citep{Boller2003AA,Tran2011ApJ}. These lines appeared several weeks after the initial outburst \citep{Trakhtenbrot2019ApJ}. The discovery of the event triggered a series of intensive UV and X-ray monitoring campaigns. Fourteen observations of 1ES\,1927+654 were conducted from 2018 to 2019 using the X-ray Telescope (XRT; \citealp{Burrows2005SSRv}) and the Ultraviolet Optical Telescope (UVOT; \citealp{Roming2005SSRv}) aboard the Neil Gehrels Swift Observatory \citep{Gehrels2004ApJ}. In addition, seven simultaneous observations were carried out using XMM-Newton \citep{Jansen2001AA} and the Nuclear Spectroscopic Telescope Array (NuSTAR; \citealp{Harrison2013ApJ}) from June 2018 to January 2021. This included observations with the Optical/UV Monitor Telescope (OM; \citealp{Mason2001AA}) on XMM-Newton, as well as over 800 high-cadence (2 days) observations with NICER \citep{Gendreau2012SPIE,Arzoumanian2014SPIE}. Given the broad wavelength coverage of these observations, comprehensive studies have been carried out to provide a detailed examination of the physical processes associated with the post-burst behavior of this source. A peculiar characteristic of the object during the changing-look event was that the optical and X-ray light curves were clearly decoupled  \citep{Trakhtenbrot2019ApJ,Ricci2021ApJS}. The disappearance of hard X-ray emission at around 200 days \citep{Ricci2020ApJL} has been attributed to the destruction of the corona due to shocks between debris from a tidally disrupted star and the accretion flow \citep{Ricci2020ApJL}. Alternatively, \cite{Scepi2021MNRAS} and \cite{Laha2022ApJ} argued that a magnetic flux inversion in the inner disk could explain the peculiar behaviour of the source in the X-ray band, particularly in the early phases of the event. Interestingly, the BLR appeared to be newly formed and underwent systematic dynamical evolution \citep{Li2022paper1}. Given the mass of the black hole ($1.38 \times 10^6 \,M_\odot$; \citealp{Li2022paper1}), the accretion rate after the outburst was well above the Eddington limit, a conclusion also supported by the detection of strong, relativistic reflection features that might originate from an optically thick outflow \citep{Masterson2022ApJ}.

In this paper, we present a comprehensive broadband analysis of the changing-look event in 1ES\,1927+654, focusing particularly on the evolution of the post-outburst SED. The paper is organized as follows. In Section~\ref{sec:sec2}, we report in detail our broadband observations and the accompanying analysis. Section~\ref{sec:sec3} illustrates the methodology adopted for our broadband SED modeling. The evolution of the intrinsic SED, broadband color, and hard X-ray properties are summarized in Section~\ref{sec:sec4}. Section~\ref{sec:sec5} discusses the potential correlation between the optical depth of the corona, the inner disk surface density, and the optical to X-ray slope. For the purposes of this study, we adopt the following parameters for a $\Lambda$CDM cosmology: $\Omega_m = 0.308$, $\Omega_\Lambda = 0.692$, and $H_0 = 67.8 \rm \; km \, s^{-1} \, Mpc^{-1}$ \citep{Planck2016AA}.

\section{Observations}
\label{sec:sec2}

We analyze 21 broadband SEDs obtained at different epochs, which include simultaneous optical and X-ray spectroscopy, as well as optical/UV photometry (Table~\ref{tab:obses}). The optical spectra are a subset of the 34 post-outburst spectra analyzed in \citeauthor{Li2022paper1} (\citeyear{Li2022paper1}; see also \citealp{Trakhtenbrot2019ApJ}), typically covering the $4000-9000\,$\angs\ wavelength range. A total of eight XMM-Newton observations are used in this study, one pre-outburst (May 2011; \citealp{Gallo2013MNRAS}) and seven post-outburst (see Section~\ref{sec:xmmobs} for a summary of the data analysis). The first NuSTAR observation from June 2018 did not detect the source \citep{Ricci2021ApJS}; hence, we utilized only the following six joint NuSTAR/XMM-Newton observations. The processing of the NuSTAR data is outlined in Section~\ref{sec:nuobs}. The XRT data from Swift, previously reduced in \citet{Ricci2021ApJS}, are supplemented here with UVOT photometry processed with two-dimensional image analysis (Section~\ref{sec:swiftobs}). Additionally, we studied a pre-outburst UV observation from GALEX, as detailed in Section~\ref{sec:GALEXobs}.

\subsection{XMM-Newton}
\label{sec:xmmobs}

We utilized eight XMM-Newton observations of 1ES\,1927+654 (Table\,\ref{tab:obses}) from June 2018 to January 2021, roughly 150 to 1100 days after the outburst. For each observation, we first extracted the X-ray spectra from the Original Data Files (ODF). The data reduction was performed using SAS v19.0.0. The process began with the {\tt emproc} task to extract the events. Subsequently, we employed the {\tt evselect} and {\tt tabgtigen} tasks to define good time intervals free from background flares and to compile clean event files. To generate an image from the clean events, we used the {\tt evselect} task. A 60\arcsec\ circular aperture was initially adopted to extract the source spectra. We used the {\tt epatplot} tool to identify any significant pile-up. For the observations in which pile-up was detected (see Table\,\ref{tab:obses}), we mitigated its effects by replacing the circular region with an annulus having inner and outer radii of 24\arcsec\ and 60\arcsec, respectively (see also \citealp{Ricci2021ApJS}). A source-free 60\arcsec\ circular aperture, located on the same CCD as the source, was utilized to extract the background spectra. Upon extracting the final source and background spectra, we used the {\tt rmfgen/arfgen} tools to create response matrices. The final spectra were grouped with a minimum of 25 counts per bin utilizing the \textsc{GRPPHA}\,V3.0.1 tool, in order to use $\chi^2$ statistics for the spectral fitting. We restricted our analysis to the EPIC/PN spectra because seven out of the eight MOS observations were conducted in small-window mode, making it difficult to select an uncontaminated background region on the same CCD as the source given the brightness of 1ES\,1927+654 across several epochs. Moreover, in certain epochs the effect of pile-up could not be eliminated entirely from the MOS spectrum (see \citealp{Ricci2021ApJS} for details).

We extracted the six-band UV (UVW2, 2120~\angs; UVM2, 2310~\angs; UVW1, 2910~\angs) and optical ($U$, 3440~\angs; $B$, 4500~\angs; $V$, 5430~\angs) images using the {\tt omchain} package. This enabled the generation of {\tt .SIMAGE} files, which were then used for our photometric analysis. The flux of 1ES\,1927+654 in each band was integrated via two-dimensional image decomposition using the \textsc{GALFIT} 3.0 package \citep{Peng2010AJ}, which allowed us to minimize the contamination from three nearby stars. To remove the effect of ghost images, a common issue with OM optical filters\footnote{XMM-Newton Users Handbook: {\url{https://xmm-tools.cosmos.esa.int/external/xmm_user_support/documentation/uhb/omlimits.html}}}, we created dedicated masks for each individual image to be used as input for \textsc{GALFIT}. The decomposition procedure is reported in detail in \citet{Li2022paper1}. In summary, besides the three contaminating sources, the model for 1ES\,1927+654 comprises a bulge and a disk for the host galaxy component, in addition to an AGN component represented by the point-spread function\footnote{The point-spread function model for the AGN component was not includd for the pre-outburst observation.}. Throughout different observation epochs, both the parameters and magnitude of the host galaxy component (Table~\ref{tab:hostflux}) were fixed to the values derived in \citet{Li2022paper1}, and only the normalization of the AGN component was allowed to vary. The final integrated magnitudes\footnote{We utilized the AB-magnitude system for both XMM-Newton OM and Swift UVOT images.}, derived from the sum of all three subcomponents, are listed in Table \ref{tab:photometry}. We utilize the X-ray spectrum from the XMM-Newton observation conducted on 12 December 2018 (observation number 14); however, as the OM only observed the UVW2 band, we also used the UVOT observation (observation ID 00010682013) from the same day for our photometric analysis.

\subsection{NuSTAR}
\label{sec:nuobs}

A total of seven NuSTAR observations were carried out simultaneously with XMM-Newton after the optical outburst. We used six of them in our analysis, since the source was not detected in the observation carried out in June 2018 (see \citealp{Ricci2021ApJS} for details). NuSTAR data were processed using the NuSTAR Data Analysis Software {\tt nustardas}\,v1.8.0 within \textsc{Heasoft}\,v6.28, using the latest calibration files. We used the {\tt nuproducts} task to extract the FPMA and FPMB spectra, using a 50\arcsec\ aperture for the source and a 60\arcsec\ aperture, in a source-free region, for the background. The spectra were binned to have at least 20 counts per bin, and $\chi^2$ statistic was used for the spectral fitting.

%XX There seems to be a problem in the formating transition between S2.2 and S2.3.  I had to insert a \bigskip, but that still doesn't work well

\bigskip
\subsection{Swift}
\label{sec:swiftobs}

Fourteen observations were conducted by Swift following the optical outburst, from May 2018 to March 2019 (\citealp{Ricci2021ApJS}; see Table \ref{tab:obses}). The XRT data from all these observations were reduced using {\tt xrtpipeline} v0.13.4, followig standard guidelines \citep{Evans2009MNRAS}. For our broadband SED analysis, we used data from 13 Swift observations, excluding the one performed on 12 December 2018, as an XMM-Newton observation was carried out on the same day. The UVOT data were processed using the {\tt UVOTimsum} and {\tt UVOTsource} routines. Each UVOT observation consisted of two exposures, which were co-added using {\tt ximage} to improve the signal-to-noise ratio (S/N). Unlike the OM images, there is no ghost image issue, and hence we generated a common mask image for all UVOT images as the \textsc{GALFIT} input. For each image, we then generated an empirical point-spread function following the same procedures extensively described in \citet{Li2022paper1}. To perform imaging decomposition and to extract the AGN emission, we applied the same model used for the OM images (see also Section~\ref{sec:xmmobs}), which includes a bulge, a disk, and a nucleus component. Compared to OM, UVOT has a very similar bandpass across the six bands but a slightly broader point-spread function \citep{Poole2008MNRAS}. Consequently, we fixed the host galaxy flux for the bulge and disk components to match precisely with the OM values (Table~\ref{tab:hostflux}). The final, integrated magnitudes for all UVOT images are listed in Table \ref{tab:photometry}.

\input{tab_opspec.tex}

\input{photometry.tex}

\subsection{GALEX}
\label{sec:GALEXobs}

\begin{figure*}
\centering
\includegraphics[width=\textwidth]{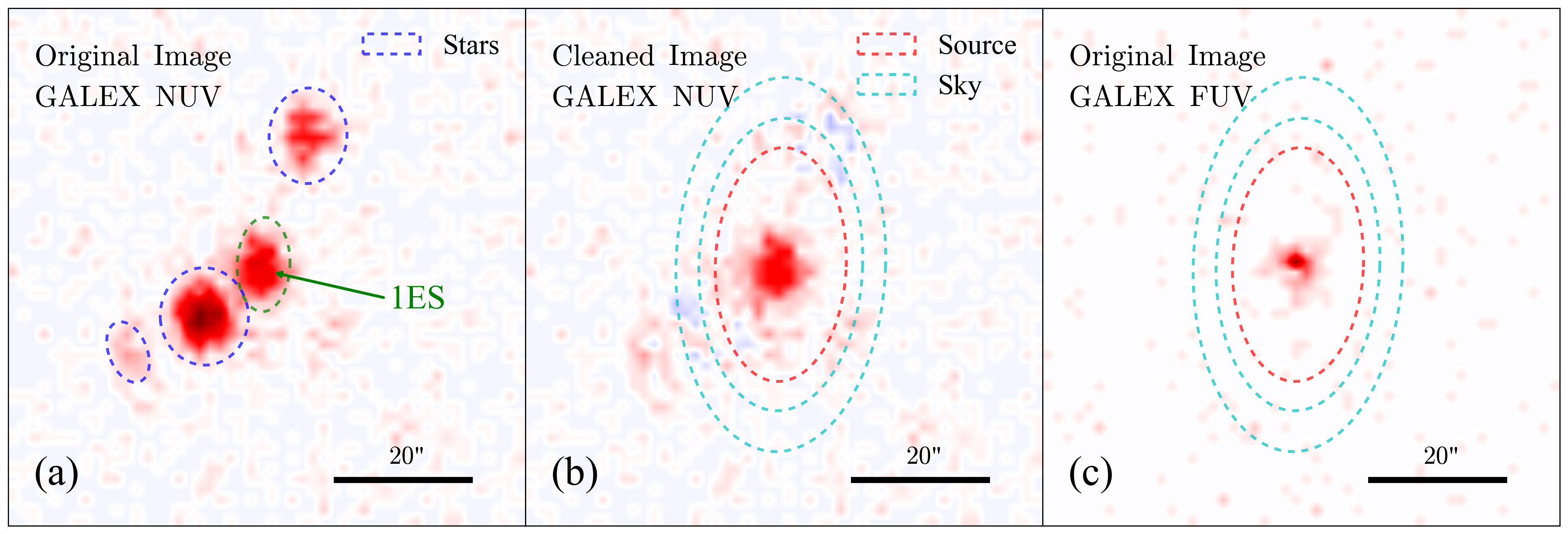}
\caption{GALEX images. The left panel displays the original NUV-band image, where the green ellipse shows the stellar disk derived from 2-D imaging decomposition \citep{Li2022paper1} and the blue ellipses show the mask generated through augmenting the {\tt SExtractor} segments. The central panel gives the cleaned NUV-band image after interpolating near-aperture contamination pixels by the source isophote. The red ellipse is the photometry aperture, while the cyan elliptical annulus with the same shape is used to estimate the nearby sky level. The same aperture and annulus are used for the FUV-band image in the right panel.}
\label{fig:GALEX}
\end{figure*}

Pre-outburst UV images, taken on 12 July 2007, are available from the Galaxy Evolution Explorer (GALEX; \citealt{Martin2005GALEX}). The stellar disk of 1ES\,1927+654, with a size of $r_e = 6\farcs83$ according to the imaging decomposition reported in \cite{Li2022paper1}, was marginally resolved in the GALEX images\footnote{Median point-spread function FWHM = $5\farcs3$ for the NUV and $4\farcs2$ for the FUV}. Besides their low resolution, the low-count nature of the UV-band images makes it difficult to perform the same image decomposition used for the other UV observations (see Section~\ref{sec:xmmobs}). The surface brightness profiles of GALEX images are quite asymmetric due to large Poisson noise.  Instead, we performed aperture photometry to extract the UV flux. We applied the same source detection procedure of \citet{Li2022paper1} to the sky-subtracted images, where the background image was obtained directly from the standard pipeline \citep{Bianchi2014AdSpR}. The contaminating stars are spectroscopically classified as G and K type, which display weaker emission in the UV. Therefore, the NUV-band image suffers much weaker blending, and the FUV-band image indicates that only 1ES\,1927+654 has enough counts to be detected (Figure~\ref{fig:GALEX}). For the NUV-band image (Figure~\ref{fig:GALEX}a), after masking the other sources, the one-dimensional surface brightness profile of the target was generated using the {\tt IRAF} task {\tt ellipse}. The pixel flux of the three contaminating stars were then interpolated by the flux of the nearest isophotes, generating a cleaned image (Figure~\ref{fig:GALEX}b). An elliptical aperture with the same shape as the stellar disk is adopted to integrate the NUV (on the cleaned image) and FUV flux (using the original image; Figure~\ref{fig:GALEX}c).  The size of the aperture is set to $2.5\,r_e$, which theoretically is expected to encompass $> 95\%$ of the total flux. To assess the local sky level, we designated an elliptical annulus with inner and outer semi-major axes set to 1.25 and 1.60 times the dimensions of the source region, respectively. We use the mean intensity of the sky pixels in the annulus as the local sky background per pixel. After integrating over the source region to obtain the total number of source counts ($C_{s}$), we calculated the photometric uncertainty taking into consideration both Poisson noise and sky subtraction noise,

\begin{equation}\label{equ:GALEXsig}
    \sigma_{s}^2=\frac{C_{s}}{T} + \frac{N_A^2}{N_B} \sigma_{B}^2,
\end{equation}

\noindent
where $T$ is the exposure time, $N_A$ and $N_B$ the number of pixels in the aperture and sky annulus, respectively, and $\sigma_{B}$ the variance in the sky-background annulus. We derive $\rm NUV = 18.17 \pm 0.06$ mag and $\rm FUV =18.79 \pm 0.11$ mag, which are $32\%$ and $55\%$ times larger, respectively, than the host fluxes derived by \cite{Li2022paper1} from modeling the pre-outburst SED ($\rm NUV = 18.59$ mag and $\rm FUV = 19.66$ mag; Table~\ref{tab:hostflux}).

%XX Can you format Table 3 better (wider), so that it doesn't look so weird (especially the captions)?
% LRC:  it's wider now

\input{host_phot}

\section{SED Analysis}
\label{sec:sec3}

To explore the evolution of the optical to X-ray SED during the post-outburst phase of 1ES\,1927+654, we must first amalgamate the broadband data and interpolate the extreme-UV emission through SED modeling. We outline how we derive the intrinsic AGN SED using our current dataset in Section~\ref{sec:intSED}, which includes host galaxy subtraction and adjustments to the dataset if the measurements were not taken simultaneously. Detailed modeling of the broadband SED is presented in Section~\ref{sec:sedmodel}.

\subsection{Intrinsic SED Data}
\label{sec:intSED}

The intrinsic AGN SED is the original emission emanating from the accretion disk, encompassing both thermal disk and Comptonized coronal emission. The emission from the BLR, dusty torus, and narrow-line region comes from reprocessing of the ionizing photons of this intrinsic emission. To construct the intrinsic AGN SED for each epoch, we primarily use X-ray spectra from XMM-Newton or Swift, which include simultaneous optical/UV photometry. For each X-ray spectrum, we only notice the energy band not dominated by background, as shown in Column (5) of Table~\ref{tab:obses}. For NuSTAR observations, we consistently choose the $3-10\,$keV band. We also use optical spectra from \citet{Li2022paper1}, by pairing them with the closest dates\footnote{We did not include any optical spectrum for the dataset number 0. This decision was based on the fact that the available pre-outburst spectrum was obtained in 2001 \citep{Boller2003AA}, a decade before the XMM-Newton observation conducted in 2011.}. We used a total of 21 epochs, comprising one observation before the optical outburst (No. 0) and 20 after the outburst (No. 1--20). The optical-to-X-ray data used for our SED analysis are summarized in Table\,\ref{tab:obses}. As discussed in Section\,\ref{sec:sec2}, several potential issues must be addressed before fitting the multiwavelength SED: (1) the flux contribution from the host galaxy needs to be subtracted; (2) we must account for the potential discrepancy between the non-simultaneous photometric data and the optical spectra on account of AGN variability; and (3) the data should be corrected for dust reddening and absorption from the interstellar medium.

Both the photometric data and optical spectra require the removal of the host galaxy flux. The AGN flux from the photometric data was derived from imaging decomposition in Sections~\ref{sec:xmmobs} and \ref{sec:swiftobs}. Although we did not list the values specifically in Table~\ref{tab:photometry}, they can be calculated by subtracting the integrated flux of the host galaxy from Table~\ref{tab:hostflux}, as we fixed the host galaxy flux during the imaging decomposition. The host galaxy contribution to the optical spectra was estimated through spectral decomposition. We initially determined the stellar population of the host by fitting the pre-outburst spectrum from \citet{Boller2003AA}, along with infrared photometry from the Two-Micron All Sky Survey (\citealp{Skrutskie2006AJ}) and Wide-field Infrared Survey Explorer (\citealp{Wright2010AJ}). Detailed information about the fitting procedure can be found in \citet{Li2022paper1}. The host galaxy emission can be modeled with a combination of a young ($t = 0.06$~Gyr) and an old ($t = 0.98$~Gyr) stellar population, which contribute to more than $\sim95\%$ of the emission in the optical band. We calculated the expected FUV and NUV flux of the stellar emission model (Table~\ref{tab:hostflux}). Because \cite{Li2022paper1} calibrated the optical spectra using the flux of the \oiii\ $\lambda 5007$ emission line, the host galaxy flux in each spectrum was removed through spectral decomposition, where the normalization is allowed to vary \citep{Li2022paper1}. The contribution from all prominent emission lines, including the optical Fe\,II complexes, was removed based on the spectral decomposition of \citet{Li2022paper1}.

Although we carefully selected the closest optical spectra to accompany the simultaneous X-ray and UV/optical photometric observations, there are still normalization discrepancies. These discrepancies arise from two main factors. First, 1ES\,1927+654 exhibited significant stochastic variability in the optical bands after the outburst \citep{Hinkle2023MNRAS}. Second, we calibrated the spectra using the \oiii\ flux, which will be affected by variations in the seeing. To reduce the potential impact of these effects, we freely adjusted the scale factor for the host emission subtraction from the different optical spectra. The median subtraction factor for host emission, calibrated to align with results obtained from the host-subtracted photometry, is 41.9\%, with a standard deviation of 26.3\%. Based on the best-fit \textsc{GALFIT} model of the host galaxy in the XMM-OM $V$-band image \citep{Li2022paper1}, we anticipated that $45.5\%$ of the host galaxy flux contributed to the spectra obtained through a 2\arcsec\ aperture from the FLOYDS spectrograph at Las Cumbres Observatory ($36.6\%$ for the 1\farcs5 aperture). Hence, our renormalization and host subtraction for the optical spectra generally agree with our expectations. However, considering the relatively large $26.3\%$ scatter of the scale factor, representing potential uncertainties introduced by the host galaxy subtraction, we incorporated $26.3\%$ of the host galaxy flux into the total error bars of our spectra. This adjustment not only balanced the weight distribution of the optical and X-ray portions of the SED during our fitting procedure but also accounted for potential uncertainties in the host galaxy subtraction. Subsequently, we rebinned all spectra into bins of 250 \angs\ to increase the S/N in each bin, and the flux uncertainty was calculated through summation in quadrature. To ensure consistent final data products, we utilized the {\tt flx2xsp} tool within Heasoft v6.28 to create .pha files and dummy responses.

Our last step in creating the intrinsic AGN SED was to correct for dust reddening and absorption by the interstellar medium. This correction was implemented in \textsc{XSPEC} by modifying the continuum model through the following steps: (1) including a multiplicative Galactic X-ray absorption component ({\tt TBabs}; \citealp{Wilms2000ApJ}) with fixed $N_\mathrm{H} = 6.58 \times 10^{20} \,\rm cm^{-2}$ as recorded by the Galactic H~I survey of \cite{Kalberla2005AA}; (2) incorporating a Galactic dust reddening component ({\tt zdust}), which uses the extinction curve from \citet{Cardelli1989ApJ} with $R_V = 3.1$ and $E(B-V)=0.0742\,\rm mag$, based on the Galactic dust map of \cite{Schlafly2011ApJ}; and (3) adding an intrinsic X-ray absorption element ({\tt zTBabs}; \citealp{Wilms2000ApJ}) as a free parameter during the fit (Section~\ref{sec:sedmodel}). We did not include intrinsic dust reddening, for the X-ray spectrum before the outburst revealed minimal nuclear absorption ($N_\mathrm{H} \approx 5-10 \times 10^{20}\, \rm cm^{-2}$; \citealp{Boller2003AA,Gallo2013MNRAS}). This is consistent with the detection of UV emission in excess of the host galaxy, which suggests minimal obscuration, as well as the Balmer decrement of the narrow lines typical of Case B$^\prime$ recombination in low-density photoionized gas \citep{Li2022paper1}.

\subsection{SED Modeling}
\label{sec:sedmodel}

We modeled the SED for the 21 epochs using \textsc{XSPEC}, considering the corrected broadband data ranging from $\sim 10^{-3}$ to $10\,$keV ($\sim 3\,$keV when NuSTAR data were not used). For the OM data, we used the standard response files\footnote{\url{https://www.cosmos.esa.int/web/xmm-newton/om-response-files}} to generate the .pha files that can be analyzed within \textsc{XSPEC}. For the UVOT data, we utilized the {\tt uvot2pha} tool within Heasoft v6.28 to produce a .pha file. Additionally, we downloaded the corresponding response matrices from the HEASARC Calibration Database\footnote{\url{https://swift.gsfc.nasa.gov/proposals/swift\_responses.html}}.

Including dust reddening and interstellar medium absorption, our SED model can be described as {\tt (continuum+zgauss)$\times$zTBabs$\times$TBabs$\times$zdust}.  The {\tt continuum} component represents the intrinsic AGN emission and has been incorporated using four different \textsc{XSPEC} models:

\begin{enumerate}
\item
A two-component phenomenological model consisting of a power law ({\tt zpower law}) dominating the flux in the hard ($1-10\,$keV) band, and a blackbody ({\tt zbbody}) dominating the soft ($0.3-1\,$keV) band (Section~\ref{sec:2cbbody}).

\item
A two-component phenomenological model consisting of a power law for the X-ray emission and a thin disk ({\tt diskbb}) for the optical/UV emission (Section~\ref{sec:2cdisk}).

\item
A three-component physical model ({\tt agnslim}), incorporating a color temperature-corrected disk  accounting for the optical/UV flux, a warm corona contributing to the extreme-UV and mainly soft X-ray emission, and a hot corona for the hard X-ray emission (Section~\ref{sec:wcoronamodel}).

\item
A three-component phenomenological model consisting of a thin disk for the optical/UV emission, a blackbody for the soft X-ray emission, and a Comptonized plasma ({\tt nthComp}) for the hard X-ray emission (Section~\ref{sec:finalmodel}).
\end{enumerate}

\noindent
Besides the {\tt continuum} component, an additional Gaussian emission feature ({\tt zgauss}) was considered to model the broad feature at $\sim 1\,$keV \citep{Ricci2021ApJS,Masterson2022ApJ}, which could be produced by reflected emission off an optically thick outflow \citep{Masterson2022ApJ}.

\begin{figure*}
\centering
\includegraphics[width=\textwidth]{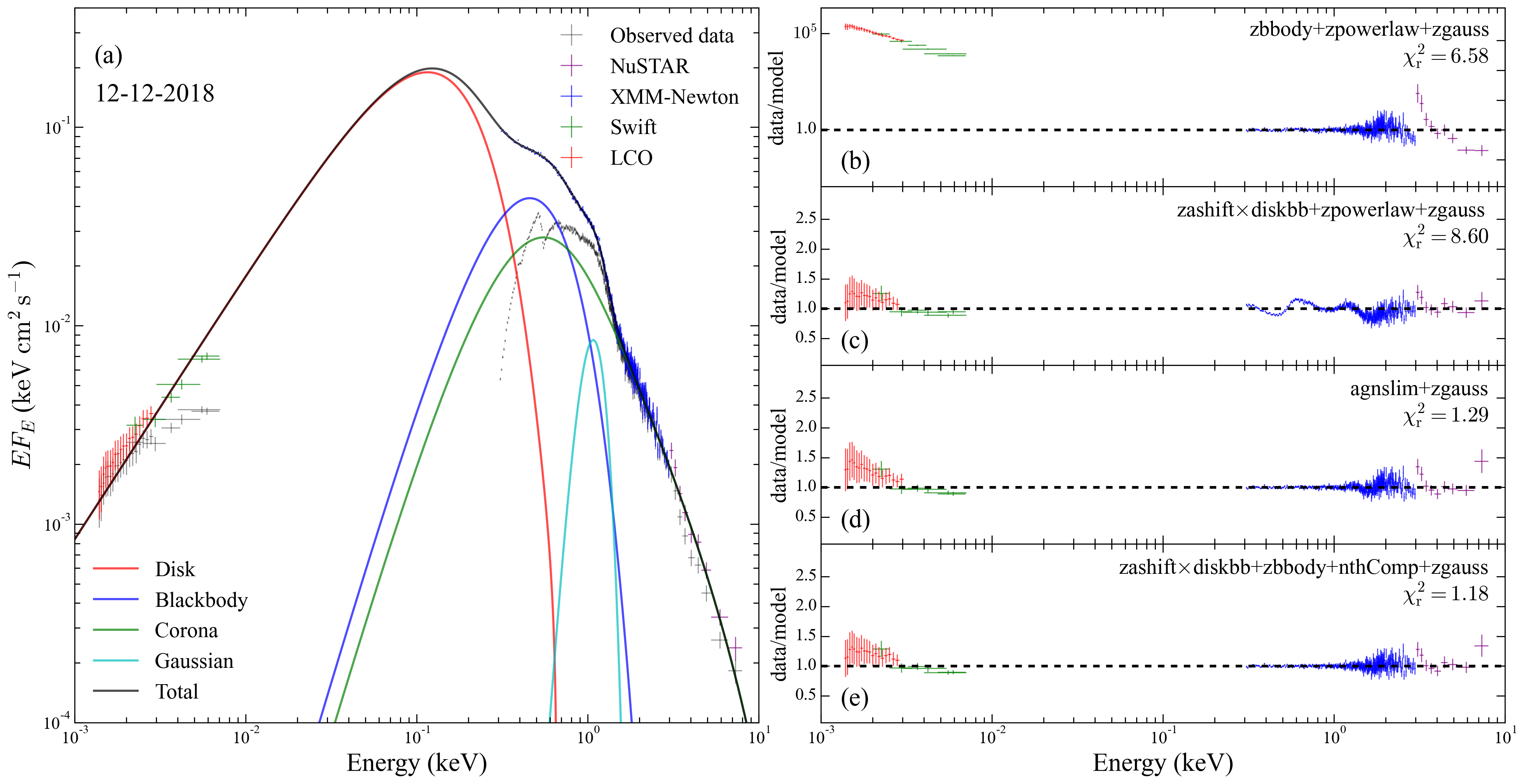}
\caption{Example of our broadband SED modeling for the 12 December 2018 observation. In the left panel (a), the gray points show the original broadband data, where all X-ray spectra were visually rebinned to have S/N = 10. To better illustrate the intrinsic SED, we show the data corrected for dust reddening and interstellar medium absorption based on our best-fit result. The data include the 250\,\angs-binned LCO spectrum (red), the photometric points from Swift UVOT images (green), the 0.3--3\,keV soft X-ray spectrum from the XMM-Newton EPIC\,PN camera (blue) and the 3--10\,keV hard X-ray spectrum from merging the NuSTAR FPMA and FPMB cameras (purple). The black curve shows our best-fit final model (see Section~\ref{sec:finalmodel}), which consists of four components: a geometrically thin accretion disk (red), a blackbody that peaks in the soft X-ray band (blue), a thermally Comptonized continuum (green), and a broad Gaussian feature at $\sim 1\,$keV (cyan). The right panel shows the data/model ratio for the four different models outlined in Section~\ref{sec:sedmodel} (from top to bottom): (b) a phenomenological model consisting of a single-temperature blackbody and a power law ({\tt zbbody+zpower law}); (c) a phenomenological model consisting of an accretion disk and a power law ({\tt zashift$\times$diskbb+zpower law}); (d) a physical model consisting of a thermalized disk, a warm Comptonized corona, and a hot Comptonized corona ({\tt agnslim}, \citealp{Kubota2019MNRAS}); (e) a physical model consisting of an accretion disk, a blackbody describing the inner hot accretion flow, and a thermally Comptonized continuum ({\tt zashift$\times$diskbb+zbbody+nthComp}). The reduced $\chi^2$ obtained by applying these models is listed in the four subpanels.
}
\label{fig:bbsed}
\end{figure*}

We use the {\tt chain} command in \textsc{XSPEC} to perform Markov chain Monte Carlo (MCMC) fitting, utilizing the \cite{Goodman2010CAMCS} algorithm. Although the NuSTAR and XMM-Newton observations were simultaneous, during our fitting process we introduced a multiplicative constant ({\tt cons} in \textsc{XSPEC}) to all models, to account for any potential cross-calibration offset between the two instruments. An example of the data used, along with the data/best-fit model ratios, is presented in the right panel of Figure\,\ref{fig:bbsed}. We employ 200 walkers and conduct a total of 25000 iterations, which include a burn-in phase of 10000 iterations. This burn-in duration is 3 times longer than the average auto-correlation time, ensuring adequate convergence. The results of our SED analysis are shown in Table \ref{tab:resubbf}. We calculated the $1\sigma$ uncertainties for each parameter based on the $16\%$ and $84\%$ values of the converged chains. We should note that when comparing the statistical performance of different models, we only included datasets with XMM-Newton, which has superior S/N compared to Swift.

\subsubsection{Two-component: Blackbody Model}
\label{sec:2cbbody}

Our initial approach assumed a two-component model to attempt to reproduce the broadband SED of 1ES\,1927+654. As reported in \citet{Ricci2021ApJS}, the $0.3-10\,$keV X-ray spectra of 1ES\,1927+654 can be modelled accurately by combining a blackbody, a cut-off power law, and multiple Gaussian components. \citet{Masterson2022ApJ} further explored over 800 NICER spectra of 1ES\,1927+654 using this model, with the spectroscopic follow-up spanning continuously from 22 May 2018 to 21 June 2021. The temperature of the blackbody component evolved over time, from approximately $0.1\,$keV to $\sim 0.2\,$keV during the first 600\,days, before declining back to the pre-outburst temperature at later stages ($\sim 0.15\,$keV; \citealp{Gallo2013MNRAS}). We initially employed the model {\tt zbbody+zpower law+zgauss} for our broadband data, with the {\tt zpower law} model applied solely to the X-ray data. This means that we fixed the normalization $N_\mathrm{pl}=0$ for the optical/UV data in \textsc{XSPEC}. The model adequately characterizes the X-ray data, as depicted in Figure\,\ref{fig:bbsed}b. However, a single-temperature blackbody cannot account for the optical/UV portion of the SED, which significantly exceeds the model prediction.

%XX The following refers to Table 5. Do you mean Table 4 or 5? Confusing. There is no L_bb given in Table 5. Check. You called the temperatures below kT_e. You mean k_T_bb, right?  I also changed last observations to latter observations.  Check that all Table numbers are correct...
%it's ocrrent
The best-fit parameters obtained by the blackbody model are shown in Table~\ref{tab:resublackbody}. The temperature of the blackbody component increased from $kT_{\rm bb}\approx 0.1$\,keV for the early observations to $kT_{\rm bb}\approx 0.15$\,keV for the latter observations. Meanwhile, its luminosity increased during the initial 650 days, from $L_\mathrm{bb} \approx 10^{43}\;\rm erg\,s^{-1}$ to $5\times10^{43}\;\rm erg\,s^{-1}$, before reaching a low state of $L_\mathrm{bb} = 2\times10^{42}\;\rm erg\,s^{-1}$. These results are in good agreement with those inferred from the NICER observations \citep{Masterson2022ApJ}. However, all the blackbody models resulted in a poor fit in the optical/UV bands, with median $\Bar{\chi_r} = 3.0_{-1.4}^{+40.8}$ for the XMM-Newton observations\footnote{We define the median reduced chi-square with $\Bar{\chi_{r}}\equiv M_{L}^{U}$, where $M$ represents the median value. Here, $M^U$ corresponds to the 84\% value, and $M_L$ refers to the 16\% value, crossing all the XMM-Newton observations.}.

\subsubsection{Two-component: Thin Disk Model}
\label{sec:2cdisk}

Next, we substituted the blackbody component with a thin accretion disk model ({\tt diskbb}; \citealp{Mitsuda1984PASJ}), which provided a much-improved representation of the optical/UV data, as demonstrated in the second row on the right panel of Figure\,\ref{fig:bbsed}. However, this model left clear residuals in the soft X-ray band (0.3--2\,keV), leading to a generally worse fit for all XMM-Newton observations, with $\Bar{\chi_r} = 8.8_{-5.4}^{+10.5}$. \citet{Ricci2020ApJL} discovered that after incorporating an ionized absorber component with a turbulent velocity of $v_\mathrm{turb} = 10^4 \rm\, km\,s^{-1}$, the chi-squared value can be reduced by $\Delta \chi^2 = 29$ for the 12 December 2018 observation. When the ionized absorber was applied to our model, the chi-squared value significantly decreased, with $\Delta \chi^2_r = 2.9$. The fit resulted in a redshift $z = + 0.166$ and a column density of $1.89\times10^{21}\,\rm cm^{-2}$. We then examined the RGS spectrum in the 0.35--1.3\,keV interval, where the best-fit ionized absorber yielded an improvement of $\Delta \chi_r^2 = 0.4$. However, by substituting the existing model with a three-component model (Section \ref{sec:finalmodel}), we can better fit the RGS spectrum without introducing any ionized absorber ($\Delta \chi_r^2 = 0.6$). Furthermore, \citet{Ricci2021ApJS} found that the best model for the observation carried out on 12 December 2018 indicated an insignificant amount of warm absorption, less than $1.1\times10^{20}\,\rm cm^{-2}$. Therefore, warm absorption is not included in our subsequent analysis using the three-component model.

%XX note that "g" in R_g should be in italics. The rule is one character = italics, unless it represents a proper noun, such as R_S, where S is for Schwarzschild, for N_H, for H is an element
%replaced
The best-fit paramters of the thin disk model is shown in Table~\ref{tab:resudbb}. The normalization of the {\tt diskbb} component can be expressed as $N_\mathrm{disk} \equiv (R_\mathrm{in}/D_{10})^2 \cos i$, where $R_\mathrm{in}$ represents an ``apparent'' inner disk radius in km, $D_{10}$ is the distance to the source in units of 10\,kpc, and $i$ denotes the inclination angle of the accretion disk (with $i=0^\circ$ for a face-on perspective). We assigned $i=60^\circ$, a value more appropriate based on our three-component physical modeling (Section~\ref{sec:wcoronamodel}). Subsequently, we found that $R_\mathrm{in}$ consistently decreased from approximately $8.1\times 10^{12}\rm \, cm$ to $7.4 \times 10^{11}\rm \, cm$ throughout our observations.  The black hole mass of $M_\mathrm{BH} = 1.9\times10^7 \, M_\odot$ for 1ES\,1927+654 estimated by \citet{Trakhtenbrot2019ApJ} using the single-epoch method (e.g., \citealp{Shen2013BASI}) yields a Schwarzschild radius $R_g \simeq 5.6\times10^{12}\rm \, cm$ that exceeds $R_\mathrm{in}$ in the later epochs. Alternatively, \citet{Li2022paper1} used the properties of the host galaxy to constrain $M_\mathrm{BH} = 1.38\times10^6 \, M_\odot$ from the mass of the stellar bulge \citep{Kormendy2013ARAA}. The resultant $R_g \simeq 4.1\times10^{11}\rm \, cm$ is now smaller than the minimum value of our best-fit $R_\mathrm{in}$. Since the equivalent photosphere of the accretion disk cannot be smaller than $R_g$, our thin disk SED model better accords with the smaller $M_\mathrm{BH}$ for 1ES\,1927+654 obtained by \citet{Li2022paper1}.

The soft X-ray emission is accounted for partially by the {\tt diskbb} component to achieve an effective overall fit. This suggests that the inner temperature of the thin disk is relatively high, with $kT_\mathrm{in}$ ranging from $\sim$50\,eV to 100\,eV. This value exceeds the estimate of $\sim$10\,eV reported by \citet{Ghosh2023ApJ}, which was based solely on optical/UV photometric data. We then computed the disk luminosity $L_\mathrm{disk} = 4\pi \sigma R_\mathrm{in}^2 T_\mathrm{in}^4$, as per the definition of the {\tt diskbb} model, and the corona luminosity $L_\mathrm{X}$ by integrating the best-fit {\tt zpower law} model from 0.3\,keV to 210\,keV, the median cut-off energy of nearby AGNs \citep{Ricci2018MNRAS}. The bolometric luminosity for this two-component model is then defined as $L_\mathrm{bol} \equiv L_\mathrm{disk} + L_\mathrm{X}$, which has a median value of $\sim 10^{45}\;\rm erg\,s^{-1}$ and a standard deviation of 0.2 dex. For $M_{\rm BH} = 1.38 \times 10^{6}\, M_\odot$, the observed $L_\mathrm{bol}$ is about 5 times larger than the Eddington luminosity, $L_\mathrm{E} \equiv 1.26 \times 10^{38} \,(M_\mathrm{BH}/M_\odot) \approx 1.8\times10^{44}\;\rm erg\,s^{-1}$. Consequently, even this simple two-component model already suggests that the changing appearance of 1ES\,1927+654 was associated with super-Eddington accretion.

\subsubsection{Three-component: Warm Corona Model}
\label{sec:wcoronamodel}

Besides allowing more freedom to minimize the fitting residuals, three-component models are also more practical for describing the broadband SED of AGNs (e.g., \citealp{Jin2012MNRAS,Ballantyne2022HEAD}). This is especially true for high-$\dot{M}$ AGNs, which ubiquitously exhibit a ``soft X-ray excess'' below $\sim 2$\,keV \citep{Zdziarski1996MNRAS,Laor1997ApJ}. This excess is generally too hot to be accounted for by thin disk emission (e.g., \citealp{Gierlinski2004MNRAS,Done2012MNRAS}). In Section~\ref{sec:2cdisk}, we have already shown that a thin disk provides a good description of the optical to UV emission but exhibits shortcomings in the soft X-ray band. Therefore, our three-component model includes a soft X-ray excess component. Although the physical origin of the excess remains unclear, two possible explanations have been widely discussed over the past decade. The observed soft X-ray spectral shape can be fit well by assuming Compton upscattering of disk photons, in which the excess would originate from a warm corona that has a lower temperature ($kT_e \approx 0.2$\,keV) and higher optical depth ($\tau \approx 10$) than the traditional hard X-ray corona (the warm corona model; e.g., \citealp{Czerny2003AA, Done2012MNRAS}). Alternatively, emission from relativistic reflection off optically thick material near the inner disk also can produce an apparent soft excess (the relativistic reflection model; e.g., \citealp{Fabian2002MNRAS}). Relativistic reflection models such as {\tt RelxilllP} or {\tt RelxilllPD} (\citealp{Garcia2013ApJ, Garcia2014aApJ}) applied to the X-ray data of 1ES\,1927+654 alone leave very strong residuals across the spectrum ($\chi_r \approx 2.7$; \citealt{Ricci2021ApJS}), a discrepancy that may arise because these models presuppose a typical AGN X-ray spectrum dominated by a power-law component, which is notably subdued in 1ES\,1927+654. Furthermore, the soft X-ray excess exhibits a light curve that differs markedly from that of the hard X-rays \citep{Ricci2021ApJS,Masterson2022ApJ}. This significant spectral variability does not support the reflection origin of the soft X-ray excess, for which a more consistent time evolution would be expected. Therefore, in this section we adopt the warm corona model, which also self-consistently calculates the disk spectrum.

The warm corona can be implemented in a three-component model that includes a color temperature-corrected disk, as well as a warm ($kT_{e} \approx 0.2\,$keV) and a hot ($kT_{e} \approx 100\,$keV) corona ({\tt optxagn}; \citealp{Done2012MNRAS}). This model was updated as {\tt agnsed} by \citet{Kubota2018MNRAS}, who identified three characteristic regions for the three components. A further version of the model ({\tt agnslim}; \citealp{Kubota2019MNRAS}) incorporates the slim disk emissivity from \citet{Abramowicz1988ApJ}, making it better suited for super-Eddington accretion. In light of the findings in Section \ref{sec:2cdisk}, which indicate that 1ES\,1927+654 underwent a period of super-Eddington accretion, we employ {\tt agnslim} as the intrinsic SED model of the AGN. Throughout our fitting process, we assume that during the campaign the variations in mass and spin ($a_{\star}$) of the central black hole, as well as the inclination angle ($i$) of the accretion disk, were negligible. The black hole mass is set to $M_\mathrm{BH}=1.38 \times 10^6\, M_\odot$ \citep{Li2022paper1}. To ascertain $a_{\star}$ and $i$, we perform the SED fits on all the XMM-NuSTAR observations for a range of pre-established values of spin and inclination angles.\footnote{We set $a_{\star}$ to 0, 0.4, 0.6, 0.7, 0.8, 0.9, and 0.998 and $i$ to $0^{\circ}$, $15^{\circ}$, $30^{\circ}$, $45^{\circ}$, $60^{\circ}$, $75^{\circ}$, and $85^{\circ}$.} The minimum of $\chi_r ^2$ is found for $a_{\star}=0.8$ and $i=60^{\circ}$. The inner radius of the disk was fixed to the value derived from the steady disk solution as given by Equation (1) in \citet{Kubota2019MNRAS}, which accounts for structural changes in the inner disk due to advection. Regarding the outer radius, since the outer edge of the accretion disk primarily contributes to the near-infrared emission, which is not covered in our SED, we fixed it to the self-gravity radius calculated according to \citet{Laor1989MNRAS}.

%XX R_o looks weird. I changed to R_out, here an in Table 7
%got it
For each epoch, we allowed seven of the parameters of the {\tt agnslim} model to vary: the dimensionless mass accretion rate, denoted as $\dot{m} \equiv \dot{M}/\dot{M}_\mathrm{E}$; the electron temperatures for the hot ($k T_{e}^\mathrm{hot}$) and warm ($k T_{e}^\mathrm{warm}$) Comptonizing components; the spectral indices of both the hot ($\Gamma^\mathrm{hot}$) and warm ($\Gamma^\mathrm{warm}$) Comptonizing components; and the outer radii of the hot ($R_\mathrm{out}^\mathrm{hot}$) and warm ($R_\mathrm{out}^\mathrm{warm}$) Comptonizing components. The parameter $R_\mathrm{out}^\mathrm{warm}$ was set free only for observations No.~14 and 16--20, which have better constraints from NuSTAR on the shape of the X-ray spectra. For the other observations, we linked $R_\mathrm{out}^\mathrm{warm} = 2 \, R_\mathrm{out}^\mathrm{hot}$ to mitigate parameter degeneracy, as suggested by \citet{Kubota2018MNRAS}. These adjustments led to significant improvements in the fits ($\Bar{\chi_r} = 1.7_{-0.4}^{+0.7}$) for all XMM-Newton observations compared to the earlier two-component models. The results of our fits are summarized in Table~\ref{tab:resuslim}. It is worth noting that in our datasets, observations No.~17 and 18 displayed a high ratio of hard X-ray to optical flux, along with a comparably weaker soft X-ray emission. During these two specific epochs, the {\tt agnslim} model did not fit the data well ($\chi_r \approx 2.1$ and 2.0, respectively).  The epochs are characterized by a constant soft X-ray energy, while both the hard X-ray and optical energies decrease, as shown in Figure~\ref{fig:lcall}. We attribute the poor fits to the model assumption that all energy originates from the disk, which is linked to the mass accretion rate. This assumption would lead the emission across different bands to evolve in a relatively coherent manner \citep{Done2012MNRAS,Kubota2019MNRAS}, which fails to describe accurately the data across all observed epochs for 1ES\,1927+654.

\subsubsection{Three-component: Final Model}
\label{sec:finalmodel}

The broadband SED properties of 1ES\,1927+654 can be summarized phenomenologically as follows:

\begin{itemize}

\item
The soft X-ray emission resembles a single-temperature blackbody.

\item
The optical/UV emission can be described as the thermal emission of a thin accretion disk.

\item
The hard X-ray emission, which is consistent with Comptonized coronal emission, exhibits a light curve that appears to be independent of the optical/UV light curves.

\end{itemize}

\noindent
We have therefore defined our final model as a phenomenological construction of a thin disk, a single-temperature blackbody, and a Comptonization component, which primarily contribute to the optical/UV, soft X-rays, and hard X-rays, respectively. A Gaussian component was introduced to account for the broad emission feature at $\sim$1\,keV, with the exception of epochs that exhibited a dim hard X-ray component (No.~4--7) or epochs that did not show any feature at this energy (No.~0, 19, and 20). Overall, we allowed the following eight parameters to vary for each epoch: (1) the effective temperature of the thin disk at the inner radius ($T_\mathrm{in}$); (2) the flux normalization of the thin disk ($N_\mathrm{disk}$); (3) the temperature of the blackbody component ($T_\mathrm{bb}$); (4) the flux normalization of the blackbody component ($N_\mathrm{bb}$); (5) the electron temperature of the hard X-ray corona ($T_\mathrm{cor}$); (6) the hard X-ray photon index ($\Gamma$); (7) the flux normalization of the corona ($N_\mathrm{cor}$); and (8) the column density of the intrinsic neutral absorber ($N_\mathrm{H}^{i}$). When a Gaussian component was included, we allowed three additional parameters to vary, the energy ($E_\mathrm{Gau}$), width ($\sigma_\mathrm{Gau}$), and normalization ($N_\mathrm{Gau}$) of the line. For all the XMM-Newton + NuSTAR observations, the cross-calibration constant applied to the NuSTAR data ($N_\mathrm{Nu}$) was also allowed to vary. Limited by the low count rates of some of our Swift observations, not all free parameters could be sampled adequately using our MCMC approach, particularly the photon index $\Gamma$ and electron temperature of the hard X-ray corona $T_\mathrm{cor}$. To address this, we performed several MCMC fits over a sparse grid of these parameters. The best fit for each parameter was then determined based on the most robust statistical measures. The grid size of the parameter space was used as a rough estimate of the associated uncertainties\footnote{For clarity, in Table~\ref{tab:resubbf}, the precise upper and lower $1\sigma$ uncertainties based on the $16\%$ and $84\%$ values of the converged MCMC chains are displayed as $\rm best^{+upper}_{-lower}$, while $\rm best\pm grid$ represents the rough estimate based on a sparse grid. }.
%
%XX N_H^I looks weird. changed "I" to "i". In the tables, it looks even weirder and inconsistent, where you use N_{H I}. I changed all to N_H^i.  Please check carefully that I didn't introduce problems.

In summary, we present our best-fitting results for the 21 epochs in Table~\ref{tab:resubbf}, where the statistics have improved significantly compared to the previous three models ($\Bar{\chi_r} = 1.1$ for all XMM-Newton observations). By considering all the 21 epochs studied here, the median reduced chi-squares are $\Bar{\chi_r} = 22.5$ for the blackbody model, $\Bar{\chi_r} = 2.78$ for the thin disk model, $\Bar{\chi_r} = 1.48$ for the warm corona model, and $\Bar{\chi_r} = 1.43$ for the final SED model.  It should be noted that for observations near 200\,days after the optical outburst, the poor fits resulted from the discrepancy between the observed optical continuum slope ($\alpha_\nu \approx -0.5$; \citealp{Li2022paper1}) and the model ($\alpha_\nu = 0.33$). The fits could not be improved by either including disk color corrections ({\tt optxagn}; $\chi_r \simeq 80$) or by incorporating the slim disk emissivity profile ({\tt agnslim}; $\chi_r \simeq 120$).

\input{tab_result}

\section{Results}
\label{sec:sec4}

\begin{figure*}
\centering
\includegraphics[width=0.95\textwidth]{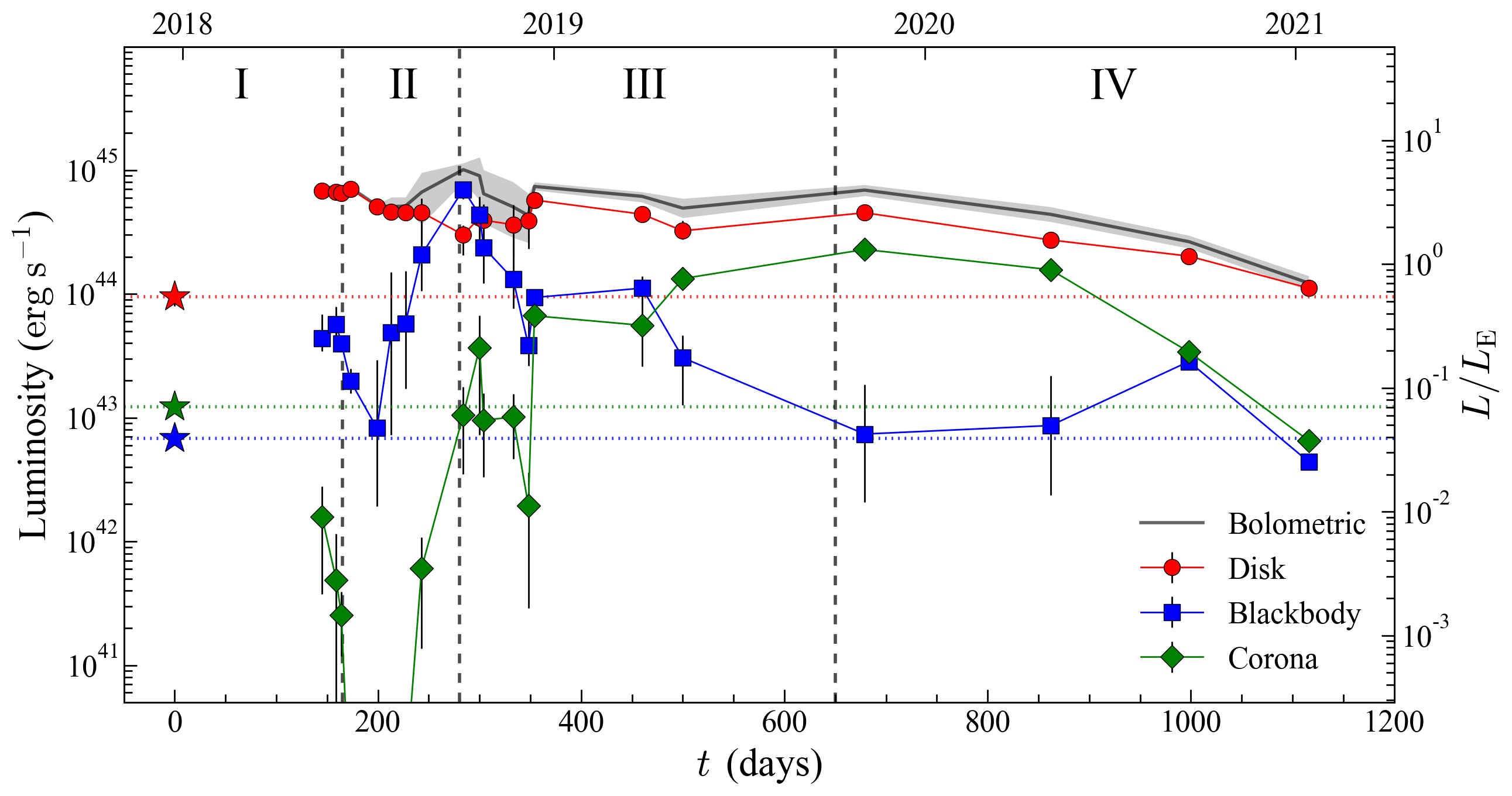}
\caption{Light curves of the disk (red circles), blackbody (blue squares), and corona (green diamonds) components. The x-axis shows days since the optical outburst on 23 December 2017. The bolometric light curve and its $1\sigma$ uncertainty are plotted as the black curve and the gray shaded region, respectively. The stars at $t=0$ and the horizontal dashed lines mark the luminosity of individual subcomponents before the optical outburst (No.~0). The vertical dashed lines separate the four evolution epochs discussed in Section~\ref{sec:SEDevo}. The luminosity is in $\rm erg\,s^{-1}$ (left y-axis) and in Eddington units (right y-axis; for $M_\mathrm{BH} = 1.36\times \, 10^6 M_\odot$; \citealp{Li2022paper1}). }
\label{fig:lcall}
\end{figure*}

\subsection{Summary of the SED Evolution}
\label{sec:SEDevo}

\begin{figure*}
\centering
\includegraphics[width=\textwidth]{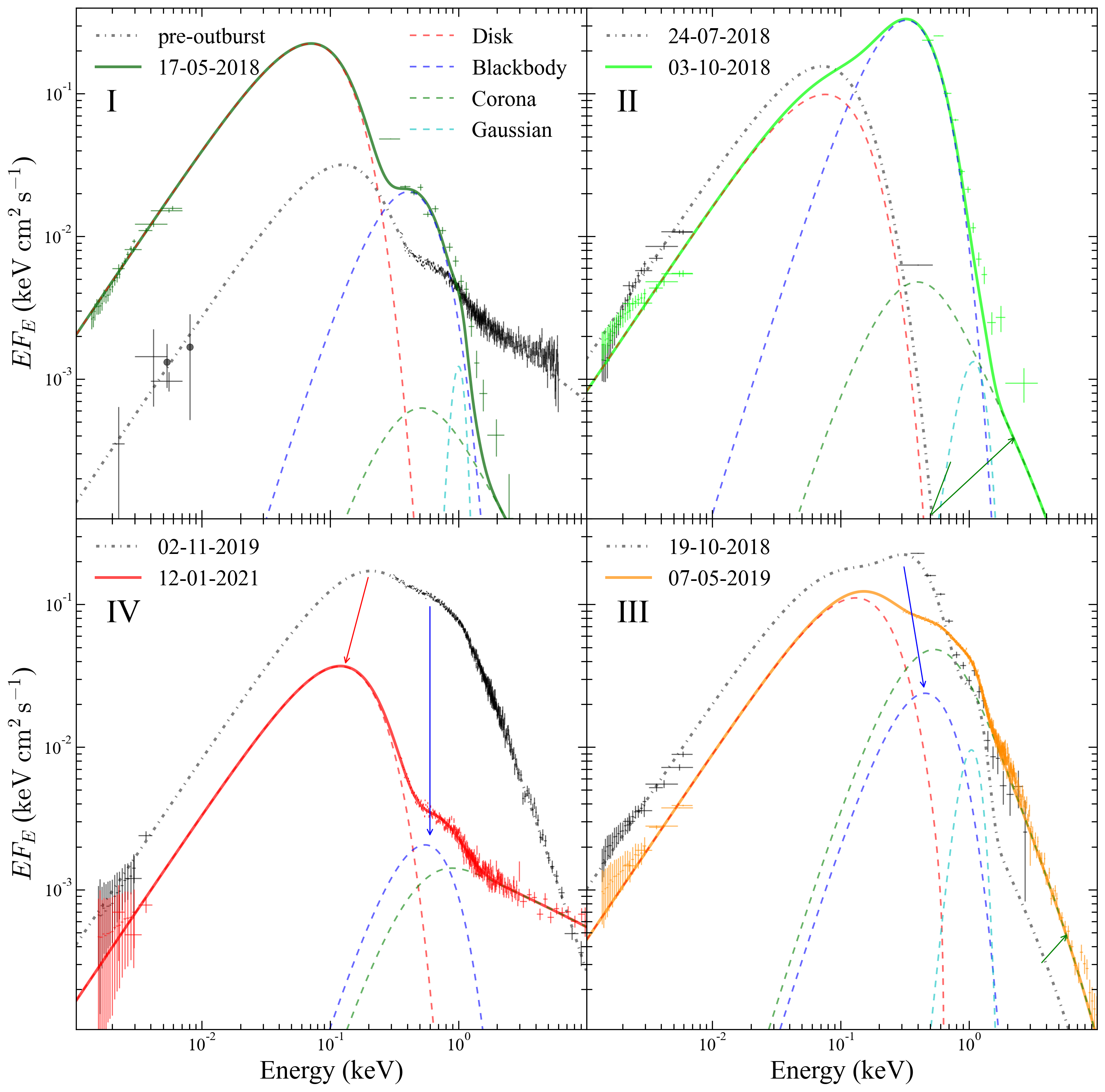}
\caption{Intrinsic SED models for the four evolution epochs discussed in Section~\ref{sec:SEDevo}. In each panel, the dash-dotted curve and black error bars show the best-fit intrinsic SED model and the broadband data used, at roughly the beginning of the epoch. The solid curve shows the best-fit intrinsic SED models at the end of each epoch: 17 May 2018 (I; dark green, upper left), 3 October 2018 (II; lime, upper right), 7 May 2019 (III; orange, lower right), and 12 January 2021 (IV; red, lower left). The dashed curves show the four subcomponents: thin disk (red), blackbody (blue), corona (green), and broad Gaussian profile (cyan). Arrows mark the evolution tracks for the different subcomponents, with red for thin disk, blue for blackbody, and green for corona. In panel I, the black circles with error bars show the host galaxy subtracted, pre-outburst UV flux from GALEX. }
\label{fig:SEDs}
\end{figure*}

After analyzing the four broadband models in Section~\ref{sec:sec3}, our SED analysis favors the three-component model outlined in Section~\ref{sec:finalmodel}, which offers a significantly better fit. This model consists of a multi-temperature blackbody component (hereafter referred to as the disk), a single-temperature blackbody (hereafter referred to as the blackbody), and a Comptonized component (hereafter referred to as the corona). These three components show markedly different light curves following the outburst (Figure~\ref{fig:lcall}). Based on their evolutionary behavior, we can identify four distinct stages, as depicted in Figure~\ref{fig:SEDs}.

\begin{figure}
\centering
\includegraphics[width=0.49\textwidth]{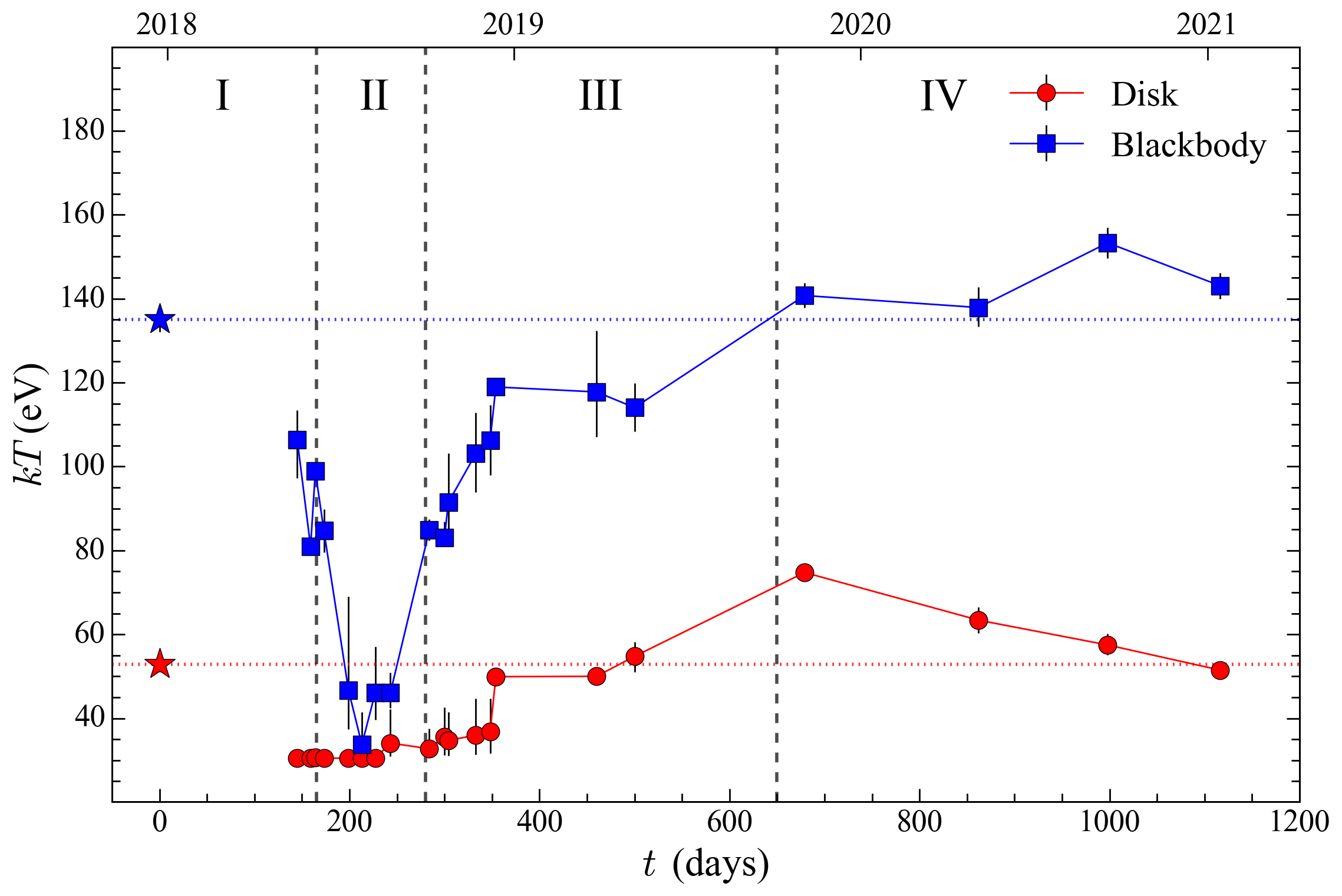}
\caption{Evolution of the maximum temperature of the disk (red circles) and blackbody (blue squares). The stars at $t=0$ and the horizontal dashed lines mark the luminosity of the individual subcomponents before the optical outburst (No.~0). The vertical dashed lines separate the four evolution epochs discussed in Section~\ref{sec:SEDevo}. }
\label{fig:tevo}
\end{figure}

\begin{enumerate}

\item{\bf Stage I: Outburst.} From the onset of the event ($t=0$; 23 December 2017) until $t = 165$\,days, the optical flux experienced a rapid ascent, reaching its maximum at approximately 80\,days, followed by a slower decline \citep{Trakhtenbrot2019ApJ}. Our multiwavelength observations started 145\,days after the outburst, coinciding with the moment when the broad emission lines attained their peak flux \citep{Li2022paper1}. We classify No.~1--3 into this stage, during which the hard X-ray emission ($> 1\,$keV) remained detectable. In this phase, the luminosity of the disk component surged to roughly 7 times the pre-outburst value. Intriguingly, the peak also shifted toward a lower temperature, dropping from $53\,$eV to roughly $\sim 30\,$eV (Figure~\ref{fig:tevo}). The blackbody component followed a comparable, albeit less pronounced pattern of luminosity amplification (around 6 times) and temperature reduction (from $135\,$eV to $93\,$eV). The corona component experienced a considerably steeper temperature drop. Based on No.~3, we could only make a rough estimation of $\sim 3\,$keV. This is more than 20 times lower than the pre-outburst value of approximately $72\,$keV.  Notably, the luminosity of the corona component diminished to merely $4.8$\% of its 2011 value, in conjunction with a significantly steeper photon index (from 2.6 to 3.4), exhibiting substantial variability \citep{Ricci2021ApJS,Masterson2022ApJ}.

\item{\bf Stage II: Fast X-ray evolution.} From $t=165$ to $t=280$ days, corresponding to No.~ 4--8, the three components exhibited disparate light curves. The luminosity of the disk component gradually declined from $7.0\times 10^{44}\rm \, erg\,s^{-1}$ to $4.6\times 10^{44}\rm \, erg\,s^{-1}$. By contrast, the blackbody and corona components fluctuated significantly. The blackbody luminosity rapidly diminished to nearly its pre-outburst intensity over a short timescale of 199 days (Figure~\ref{fig:lcall}). It subsequently began to ascend again, peaking in luminosity at 284\,days, at an intensity 10 times higher than at the onset of the outburst. The corona's light curve closely mirrored that of the blackbody. Around 200\,days into the event, its luminosity plunged substantially below $10^{40}\rm \, erg\,s^{-1}$. This vanishing of hard X-ray flux was also discernible in the NICER light curve (see Figure 1 of \citealp{Ricci2020ApJL}). Within 80\,days following the disappearance of hard X-ray emission, the X-ray flux soared by approximately 4 orders of magnitude, surpassing its initial intensity at the beginning of the outburst. This swift elevation of the X-ray flux coincided with a hardening of the X-ray emission \citep{Ricci2020ApJL}.

\item{\bf Stage III: X-ray bright.} From 280 to 650\,days, as evidenced by No.~ 9--16, the flux of the blackbody component underwent a significant decrease, returning to its pre-outburst flux. Meanwhile, the corona component rose to a higher level than before the event, such that the overall bolometric luminosity during this period reached $L_\mathrm{bol}/L_\mathrm{E} \approx 3$. The disk component continued its gradual decrease as in stage~II, exhibiting mild variations akin to those observed in the optical light curve \citep{Trakhtenbrot2019ApJ}. The coronal spectra manifested a ``softer-when-brighter'' characteristic (Figure~\ref{fig:gammatau}a), a behavior similar to AGNs of moderate to high accretion rates \citep{Markowitz2004ApJ,Sobolewska2009MNRAS}. This pattern also was seen distinctly in the analysis of NICER observations of 1ES\,1927+654 during this phase \citep{Masterson2022ApJ}.

\item{\bf Stage IV: Ebbing phase.} By 650 days after the event (No.~17--20), the flux of both the disk and corona components, being similar to the evolution of the blackbody component in stage~III, began to gradually drop toward the pre-outburst level over the span of $\sim$500\,days. Correspondingly, the photon index of the primary continuum, generated in the corona, hardened significantly from $\Gamma\simeq 4.5$ to $2.5$ (Figure~\ref{fig:gammatau}a). In addition to the decrease in the disk luminosity, the inner temperature of the disk also substantially diminished during this period (Figure~\ref{fig:tevo}), demonstrating a shift toward lower energies in the peak of the big blue bump (Figure~\ref{fig:SEDs}). By January 2021, both the blackbody and corona components had become $\sim$60\% dimmer compared to the 2011 observation. Concurrently, the temperature of the corona ($kT_\mathrm{cor}$) rose significantly during this stage, exhibiting a cooler-when-brighter behavior. According to NICER observations \citep{Masterson2022ApJ}, the decline in the X-ray flux was followed by a rebound characterized by a flux increase of approximately fourfold by mid-2021.
\end{enumerate}

For a black hole mass of $M_\mathrm{BH} = 1.36\times 10^6 \, M_\odot$ \citep{Li2022paper1}, the Eddington ratio $\lambda_\mathrm{E} \equiv L_\mathrm{bol}/L_\mathrm{E}$ stood at a value of 4.2 at the beginning of the observations, 145\,days post-outburst.  The source remained above the Eddington limit until the last observation studied here (January 2021) when $\lambda_\mathrm{E} \approx 0.7$. Following the event, $\lambda_\mathrm{E}$ witnessed a gradual decrease interspersed with fluctuations, exhibiting a median of $\lambda_\mathrm{E} = 3.3$ and a standard deviation of 1.1. Overall, the changing-look event subsequent to the outburst was predominantly super-Eddington, aligning with results inferred from other SED models (Sections \ref{sec:2cdisk} and \ref{sec:wcoronamodel}). Besides the luminosity, by the end of the observations studied here, the maximum temperature of the disk, the temperature of the blackbody, and the electron temperature of the corona all returned to their pre-outburst values (Figures~\ref{fig:tevo} and \ref{fig:thetal}). Furthermore, aside from the X-rays, the optical spectral shape also reverted to its pre-outburst state around 1200 days post-outburst, once again dominated by narrow emission lines \citep{Laha2022ApJ}. The rapid timescale of the changing-look phenomenon and its subsequent return, maintaining a type~1 phase for roughly 3\,years, is considerably shorter than the timescales observed for changing-look AGNs documented in the literature ($\sim 10-30$\,years; \citealp{Lawrence2016MNRAS,McElroy2016AA,Yang2018ApJ}; see \citealp{Ricci2023NatAs} for a review).

%XX This figure neesd a legend, similar to that of Fig. 3
%reploted

\subsection{Spectral Energy Partition}
\label{sec:energydis}

The systematic evolution of the SED highlights the peculiarities of 1ES\,1927+654 compared with other AGNs. We discuss here, in a more quantitative way, some of the differences between 1ES\,1927+654 and typical AGNs. Many studies have shown that $\alpha_\mathrm{OX}$ is correlated with the UV luminosity (e.g., \citealp{Avni1982ApJ,Strateva2005AJ}). For example, \citet{Lusso2010AA} studied a sample of X-ray-selected AGNs and found, choosing monochromatic luminosities at 2500~\angs\ ($L_\mathrm{2500}$) and 2\,keV, ($L_\mathrm{2\,keV}$),

\begin{equation}
    \alpha_\mathrm{OX} = (0.154 \pm 0.010)\log{L_\mathrm{2500}} - (3.176\pm 0.223).
\end{equation}
\noindent

\begin{figure}
\centering
\includegraphics[width=0.49\textwidth]{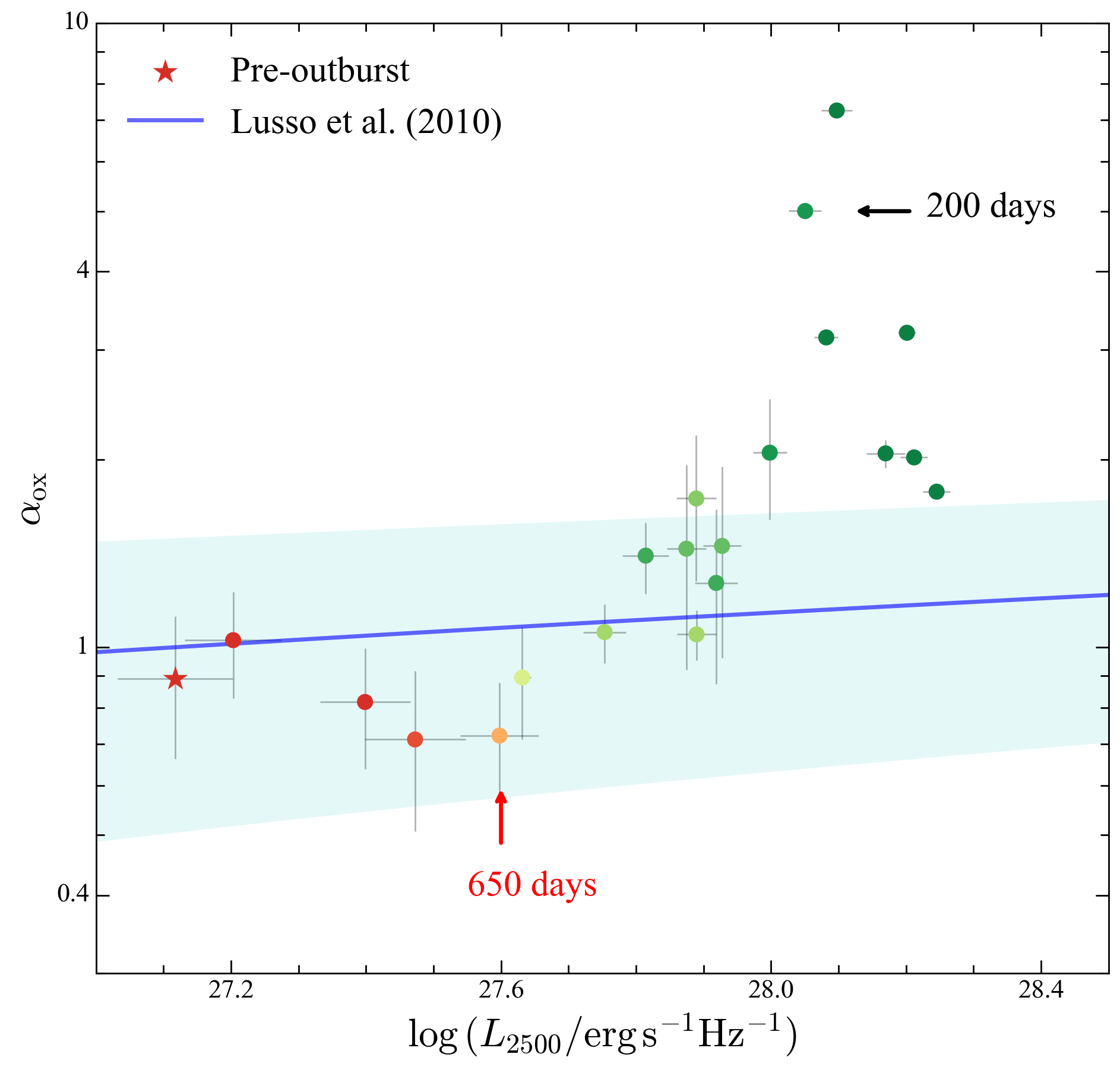}
\caption{The optical to X-ray slope ($\alpha_\mathrm{OX}$) as a function of the monochromatic luminosity at 2500~\angs\ ($L_\mathrm{2500}$), color-coded by the time since the optical outburst, from the beginning (green) to the end (red); the orange circle marks 650 days after the outburst. The red star shows the pre-outburst value. The blue line shows the $\alpha_\mathrm{OX}-L_\mathrm{2500}$ correlation for X-ray-selected type~1 AGNs \citep{Lusso2010AA}, with the cyan shaded region indicating the scatter. }
\label{fig:alphaox}
\end{figure}

For 1ES\,1927+654, we directly calculated $\alpha_\mathrm{OX} \equiv 0.3838\log{L_\mathrm{2500}/L_\mathrm{2\,keV}}$ using extinction/absorption-corrected data\footnote{For observations No.~17--20, which did not cover 2500\,\angs, we extrapolated $L_\mathrm{2500}$ from the best-fit SED model.}. The median value for all observations is $\alpha_\mathrm{OX} = 1.44$. The $\alpha_\mathrm{OX}-L_\mathrm{2500}$ correlation (Figure~\ref{fig:alphaox}) generally separates into two branches: (1) before $t\approx 650$ days ($\log{L_\mathrm{2500}} \gtrsim 27.5$), $\alpha_\mathrm{OX}$ is positively correlated with $L_\mathrm{2500}$, both of which were decreasing over time. An exception occurs at $\sim 200$ days ($\log{L_\mathrm{2500}} \simeq 28.1$), where $\alpha_\mathrm{OX}$ suddenly increases with the ``disappearance'' of the X-ray emission \citep{Ricci2020ApJL}, resulting in an extreme $\alpha_\mathrm{OX} \gtrsim 5$; (2) after $\sim 650$ days, while $L_\mathrm{2500}$ continues to decrease, $\alpha_\mathrm{OX}$ starts to increase, leading to a negative correlation. Overall, for epochs in which $\log{L_\mathrm{2500}} \lesssim 28.0$, the $\alpha_\mathrm{OX}$ of 1ES\,1927+654 is quantitatively consistent, within the scatter, with previous studies of type\,1 AGNs \citep{Lusso2010AA}. However, the slope of the correlation before 650\,days is significantly steeper than in \citet{Lusso2010AA}. More interestingly, after 650\,days 1ES\,1927+654 displays a negative slope in the $\alpha_\mathrm{OX}-L_\mathrm{2500}$ plot, which is opposite to that of normal type~1 AGNs.

%XX note that the "ox" here is in lower case.  In other places it is in upper case ("OX"). I changed all the main text to "OX". If so, the figure also needs to be consistent.  Same comment applies to Fig. 12.
%reploted

After removing the host galaxy and pseudo-continuum emission reported in \citet{Li2022paper1}, we can calculate the optical bolometric correction at 5100~\angs, $\kappa_{5100} \equiv L_\mathrm{bol}/L_\mathrm{5100}$. The rest-frame 5100~\angs\ monochromatic luminosity was derived from the photometry-corrected optical spectra (Section~\ref{sec:intSED}), ensuring consistency with $L_\mathrm{bol}$, which was obtained through SED fitting. The median value of $\kappa_{5100} \approx 120$ is significantly larger than that of typical luminous type~1 AGNs and quasars ($\kappa_{5100} = 10.33 \pm 2.08$; \citealp{Richards2006ApJS}). Interestingly, $\kappa_{5100} \approx 120$ is closely agrees with expectations for a thin accretion disk (e.g., \citealp{Shakura1973AA}). For a relativistic thin disk, the emitted spectrum can be integrated from the multi-color blackbody model of \citet{Novikov1973blho}, which has typically three parameters: $M_\mathrm{BH}$, the dimensionless mass accretion rate ($\dot{m}$), and the radiation efficiency ($\eta$). Since these parameters determine the spectral peak, $\kappa_{5100}$ is closely related to them in the case of thin disks. We integrated the luminosity $L^\mathrm{NT}$ from the \citet{Novikov1973blho} model, using a large grid of these parameters, and calculated the disk bolometric correction simply as $\kappa_{5100}^\mathrm{NT} \equiv L^\mathrm{NT}/L_\mathrm{5100}$. The calculation of $\kappa_{5100}^\mathrm{NT}$ did not take into account X-ray emission from the corona model, as for the case of 1ES\,1927+654 the median $L_\mathrm{cor}/L_\mathrm{bol} \simeq 1.5\%$. Hence, for the sake of comparison, we excluded corona emission and obtained

\begin{equation}
    \log{\kappa_{5100}^\mathrm{NT}} = 3.5 - 0.3\log{M_\mathrm{BH}/M_\odot} + 0.3\log{\dot{m}}+2.0\eta.
\label{equ:k5100}
\end{equation}

\begin{figure}
\centering
\includegraphics[width=0.49\textwidth]{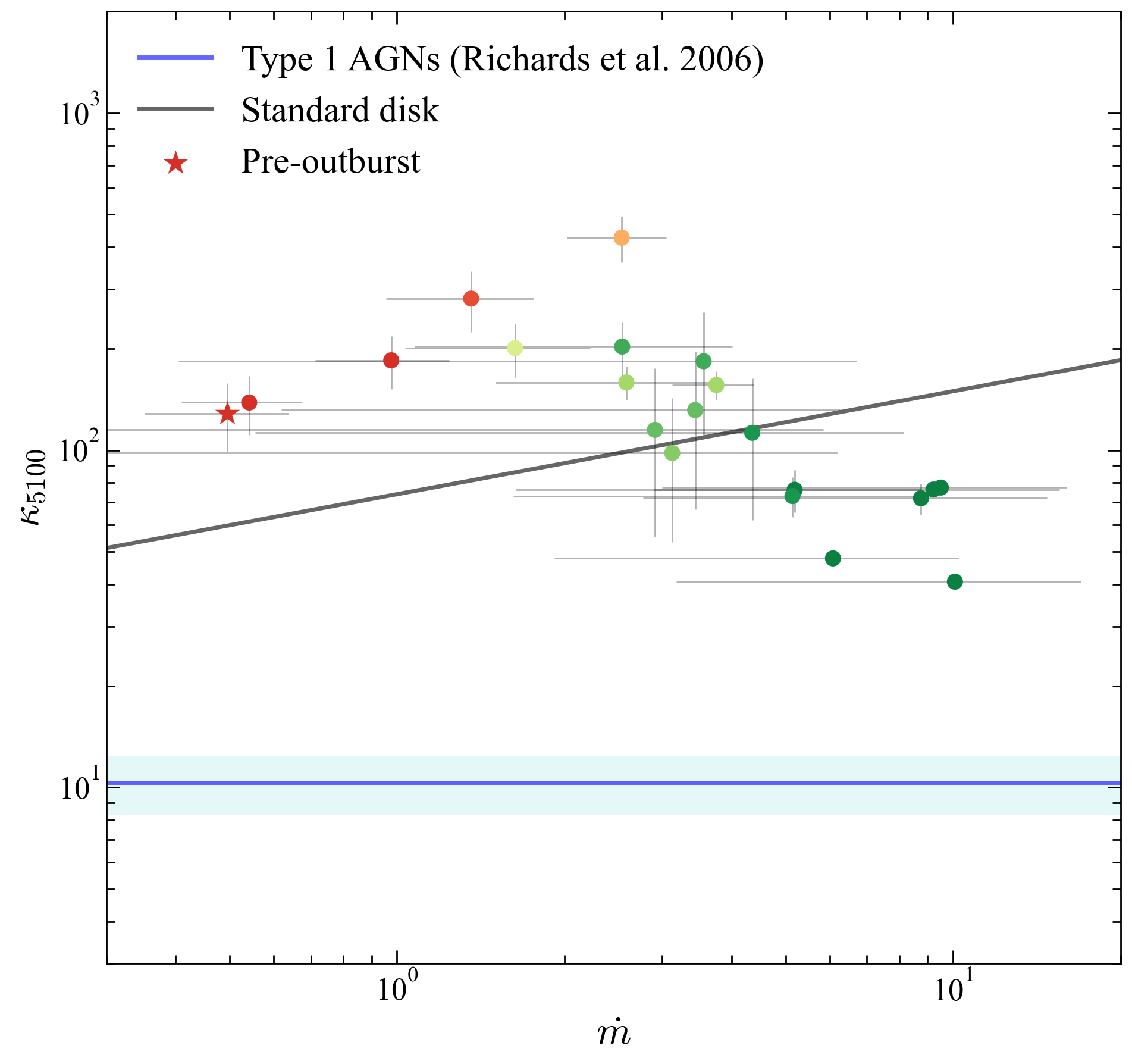}
\caption{The optical bolometric correction ($\kappa_\mathrm{5100}$) as a function of the dimensionless mass accretion rate ($\dot{m}$), from the beginning (green) to the end (red); the orange circle marks 650 days after the outburst. The red star shows the pre-outburst value. The black line shows the expected $\kappa_{5100}-\dot{m}$ correlation for a relativistic thin accretion disk (Equation~\ref{equ:k5100}). The blue line shows the observed $\kappa_\mathrm{5100}$ for type~1 AGNs \citep{Richards2006ApJS}, with the cyan shaded region indicating the scatter.
}
\label{fig:k5100}
\end{figure}

\noindent
To compare this with 1ES\,1927+654, we adopt $M_\mathrm{BH} = 1.36\times 10^6 \, M_\odot$ \citep{Li2022paper1} and $\eta = 0.1$, which is expected for a black hole with $a _\ast = 0.8$, as estimated from the {\tt agnslim} physical modeling in Section ~\ref{sec:wcoronamodel}. Using these parameters, we find that the median $\kappa_{5100}$ for 1ES\,1927+654 corresponds to a thin disk with $\dot{m} \approx 4$. This consistency can be attributed to the fact that the X-ray luminosity ($L_{\rm X}$) alone is already close to the Eddington limit during the period of $t=550-800$ days (also observed in NICER light curves; \citealp{Ricci2020ApJL,Masterson2022ApJ}), indicating that 1ES\,1927+654 is a super-Eddington AGN during its bright state. To further understand the evolution of $\kappa_{5100}$ and compare it with the theoretical expectation from Equation~\ref{equ:k5100}, we adopt the values of $\dot{m}$ calculated by \citet{Li2024paper3}, which are obtained by solving the disk equation independently at each epoch. We illustrate the $\kappa_{5100}-\dot{m}$ correlation in Figure~\ref{fig:k5100}. The evolution of $\kappa_{5100}$ (Figure~\ref{equ:k5100}) can also be separated into two branches, with a similar transition occurring at around $\sim 600$ days: (1) before 650 days, the SED is redder than would be expected from a standard disk, and becomes bluer as $\dot{m}$ decreases, leading to an increase in $\kappa_{5100}$; (2) after 650 days, the SED becomes even bluer than the standard disk, and $\kappa_{5100}$ is positively correlated with $\dot{m}$. Notably, the slope of this correlation is quite close to Equation~\ref{equ:k5100}.

\subsection{X-ray Corona Evolution}
\label{sec:corona}

\begin{figure}
\centering
\includegraphics[width=0.49\textwidth]{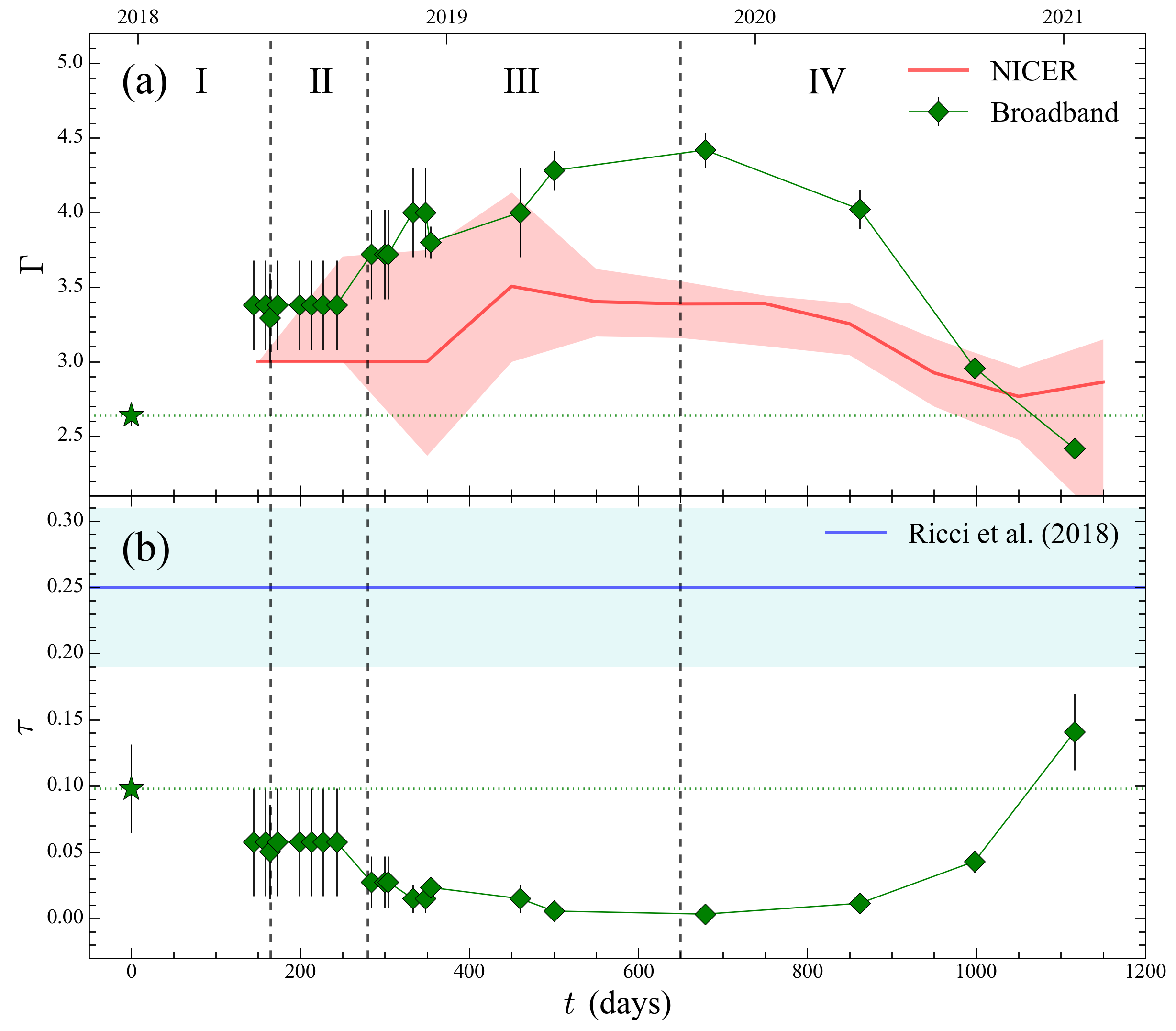}
\caption{Evolution of (a) the hard X-ray photon index ($\Gamma$) and (b) the plasma optical depth ($\tau$) after the outburst. The x-axis shows the time since the optical outburst on 23 December 2017. In panel (a), the green error bars are derived from our broadband SED fitting (Section~\ref{sec:intSED}). The red curve shows $\Gamma$ derived from the NICER observations \citep{Masterson2022ApJ}, binned to 100\,days; the $2\sigma$ range for each bin is shown as a red shaded region. The green star and the horizontal dashed line mark the pre-outburst photon index. In panel (b), the blue horizontal line shows the median $\tau$ for BASS AGNs \citep{Ricci2018MNRAS}. The green star and the horizontal dashed line mark the pre-outburst plasma optical depth. The vertical dashed lines separate the four evolution epochs discussed in Section~\ref{sec:SEDevo}. }
\label{fig:gammatau}
\end{figure}

After the outburst, 1ES\,1927+654 went through a fast X-ray evolution in stage~II. \citet{Ricci2021ApJS} studied the X-ray spectra with a phenomenological model consisting of a blackbody, a power law, and the Gaussian broad feature at $\sim 1\,$keV. During stage~II, the power-law component varied up to $\sim 4$ dex in $\sim 100$ days, and the ratio of the power law to the blackbody gradually increased. \citet{Ricci2020ApJL} argued that the disappearance of the hard X-rays ($1-10\,$keV) could be explained by the interaction between the accretion flow and debris from a tidally disrupted star. According to hydrodynamic simulations \citep{Chan2019ApJ}, the shocks excited in the disk by the debris empty the inner disk and lead to the destruction of the magnetic field, cutting off the energy source of the X-ray corona. Studying the three-year, high-cadence NICER observations, \citet{Masterson2022ApJ} found that after the corona was recreated the photon index was $\Gamma \geq 3$, which is much higher than seen in most AGNs ($\Gamma \approx 1.8-2.0$; e.g., \citealp{Nandra1994MNRAS,Ricci2017ApJS}). However, a clear softer-when-brighter-behavior, typical of moderately accreting AGNs, was observed after $\sim 300$\,days \citep{Masterson2022ApJ}. Using a Comptonization model ({\tt nthComp}) for the primary X-ray continuum, we derived an even higher photon index ($\Gamma \approx 4$), whose evolution is qualitatively consistent with that found from the NICER observations (Figure~\ref{fig:gammatau}a). The coronal temperature ($kT_\mathrm{cor}$) derived from our model allows us to estimate the optical depth ($\tau$) of the coronal plasma \citep{Poutanen1996ApJ,Ricci2018MNRAS}:

\begin{equation}
    \tau = 10^\frac{\Gamma-d}{f} \times (kT_\mathrm{cor})^{-0.3},
\label{equ:ctau}
\end{equation}

\noindent
where the constants are $d = 2.160$ and $f= -1.062$. The value of $\tau$ in 1ES\,1927+654 generally follows an opposite trend with respect to $\Gamma$ (Figure~\ref{fig:gammatau}b). Overall, we derived $\tau \simeq 0.04\pm 0.02$ after the outburst, which is significantly lower than in typical AGNs ($\tau \approx 0.25\pm 0.06$; \citealp{Ricci2018MNRAS}).

\begin{figure*}
\centering
\includegraphics[width=0.98\textwidth]{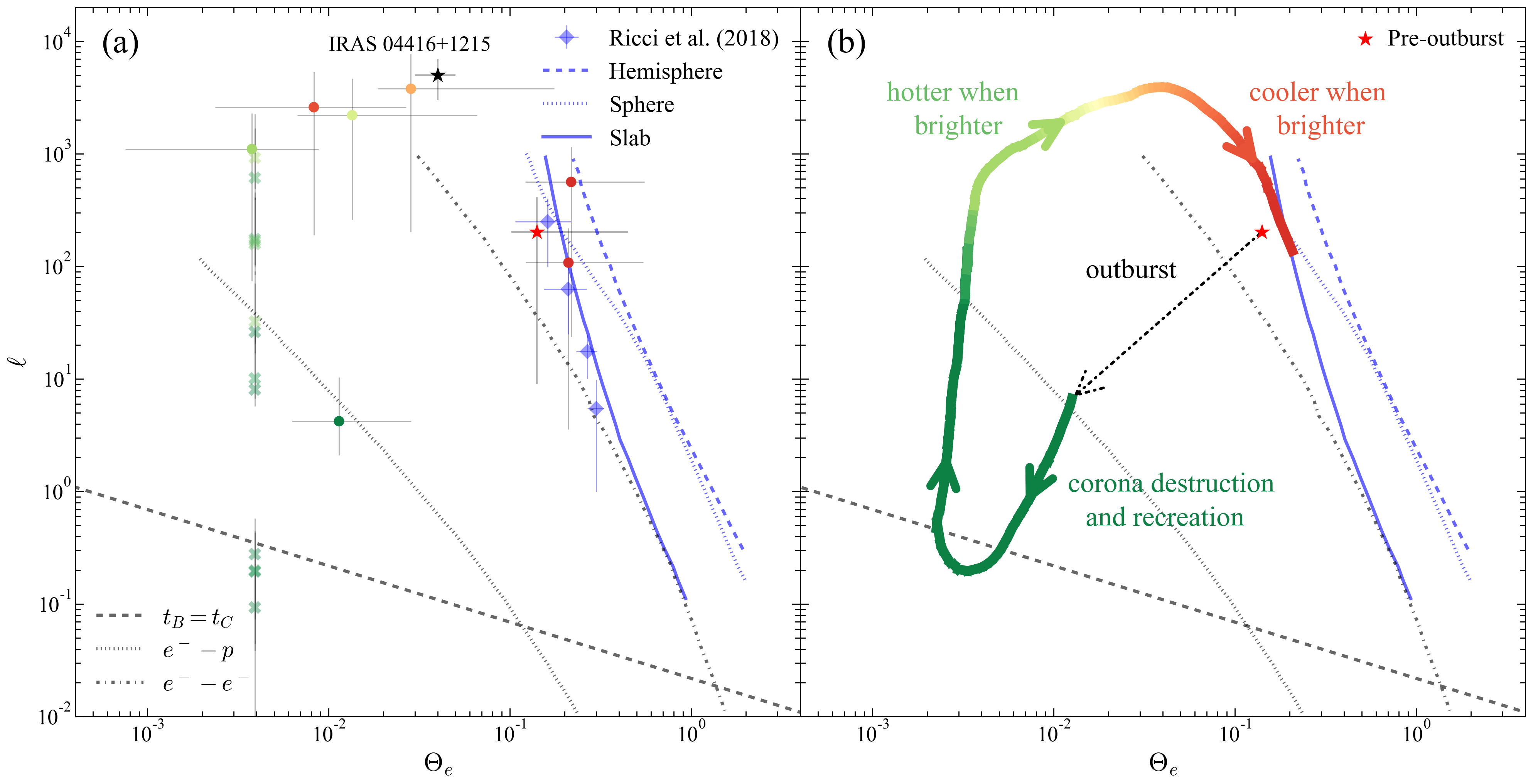}
\caption{Panel (a): compactness parameter ($\ell$) as a function of the dimensionless coronal temperature ($\Theta_e$), from the beginning (green) to the end of our observation (red) after the outburst; the orange circle marks 650 days after the outburst. The Swift observations are labeled with crosses, while the XMM-Newton observations are labeled with circles. The low S/N of the Swift/XRT observations results in loose constraints on the coronal temperature; therefore, we did not show the error bars for the Swift observations. The black dashed line shows the threshold between the dominance of Compton cooling and bremsstrahlung cooling (e.g., \citealp{Fabian2015MNRAS}). We also mark the threshold for thermal $e^{-}-p$ (black dotted line) and $e^{-}-e^{-}$ (black dash-dotted line; \citealp{Fabian1994ApJS}) coupling. The pair production lines \citep{Stern1995AA} are shown with different assumptions for corona geometry: slab (blue solid line), sphere (blue dotted line, and hemisphere (blue dashed line). The black star shows the hyper-Eddington AGN IRAS 04416+1215 \citep{Tortosa2022MNRAS}, while the blue error bars denote the median value of the X-ray-selected AGN sample of \citet{Ricci2018MNRAS}. Panel (b):  evolution track of 1ES\,1927+654 in the $\Theta_{e}-\ell$ plane, color-coded by the time after the outburst, from green to red. The color code of the points, as well as the lines, are the same as for panel (a). The red star shows the $\Theta_{e}$ and $\ell$ values in May 2011, before the outburst. }
\label{fig:thetal}
\end{figure*}

\citet{Masterson2022ApJ} studied the same XMM-Newton observations of 1ES\,1927+654 using a cutoff power law. They identified $E_\mathrm{cut} = 5.8^{+3.0}_{-1.8}\,$keV around $\sim 650$ days and $E_\mathrm{cut} = 47^{+73}_{-19}\,$keV at $\sim 1100$ days. For comparison, we used $kT_\mathrm{cor} = E_\mathrm{cut}/2$, which adheres to the approximation suitable for an optically thin plasma in a corona exhibiting slab geometry\footnote{Taking into account the $\tau \propto \Sigma$ results discussed in Section~\ref{sec:diskcorona}, we infer a slab geometry for the corona in 1ES\,1927+654.} \citep{Petrucci2000ApJ,Petrucci2001ApJ}. Consequently, the coronal temperature derived from our work aligns with that reported by \citet{Masterson2022ApJ}, when factoring in the uncertainties. In the last two epochs of our observations, we found the coronal temperature to be notably lower than the majority of AGNs ($kT_\mathrm{cor} \simeq 105 \pm 18\,$keV; \citealp{Ricci2018MNRAS}). To comprehend the origin of this cool corona, we traced the coronal evolution track in the $\Theta_e - \ell$ plane \citep[e.g.,][]{Guilbert1983MNRAS,Fabian2015MNRAS}, where $\Theta_e \equiv kT_\mathrm{cor}/m_{e}c^2$ signifies the dimensionless electron temperature and the compactness parameter

\begin{equation}
    \ell = 4\pi \frac{m_{p}}{m_{e}} \frac{R_g}{R} \frac{L_\mathrm{cor}}{L_\mathrm{E}},
\label{equ:compactness}
\end{equation}

\noindent
where the radius of the corona is assumed to be 8 times the gravitational radius ($R \simeq 8 R_g$), as found from reverberation studies of the highly accreting AGN I~Zwicky~1 \citep{Wilkins2017MNRAS}. Figure~\ref{fig:thetal} shows the evolution of the X-ray corona of 1ES\,1927+654 in the $\Theta_e - \ell$ plot. At the beginning of our X-ray monitoring (before $\sim 200$\,days), both the compactness and corona electron temperature were quite small, which locate them close to the green dotted $e^{-}-p$ equilibrium line. Shortly after, when the corona disappeared, the source moved close to the Compton cooling threshold (black dashed line), indicating that bremsstrahlung dominated the corona cooling. In stage~II, as detailed in Section~\ref{sec:SEDevo}, the corona underwent a recreation at about $200-300$ days, which is evidenced by a rapid increase depicted in Figure~\ref{fig:thetal}b. Assuming a constant size of the corona, the compactness parameter $\ell$ quickly escalated to $\sim 1000$. During stage~III, about $300-650$ days after the beginning of the event, the corona became hotter as it brightened (see also \citealp{Masterson2022ApJ}). The coronal temperature in this stage was still much lower ($kT_\mathrm{cor} \approx 10\,$\,keV) than that of typical AGNs ($105\,$keV; \citealp{Ricci2018MNRAS}). Interestingly, considering the super-Eddington nature of the disk \citep{Li2024paper3}, the value was close to that of the nearby hyper-Eddington AGN IRAS 04416+1215 ($3-22\,$keV; \citealp{Tortosa2022MNRAS}). In stage~IV, at $t\approx 650$ days, the X-ray luminosity decreased while $kT_\mathrm{cor}$ increased, indicating that the corona was then cooler when brighter. The coronal temperature finally reached $kT_\mathrm{cor} \approx 110$\,keV, where most AGNs are found (Figure~\ref{fig:thetal}a; \citealp{Ricci2018MNRAS}), a value also close to what we inferred for 1ES\,1927+654 before the outburst (red star).

%XX Take a good look at this and other similar figures. When shrunk to one column textwidth, the axis labels look too small. In these situations, you need to adjust the fonts sizes accordingly.
%
%  Legend: Broad-band --> Broadband
% Reploted

%XX Should \Theta be \Theta_e on the X-axis?
%Reploted

\section{Discussion}
\label{sec:sec5}

\subsection{The Disk-corona Correlation}
\label{sec:diskcorona}

\begin{figure}
\centering
\includegraphics[width=0.49\textwidth]{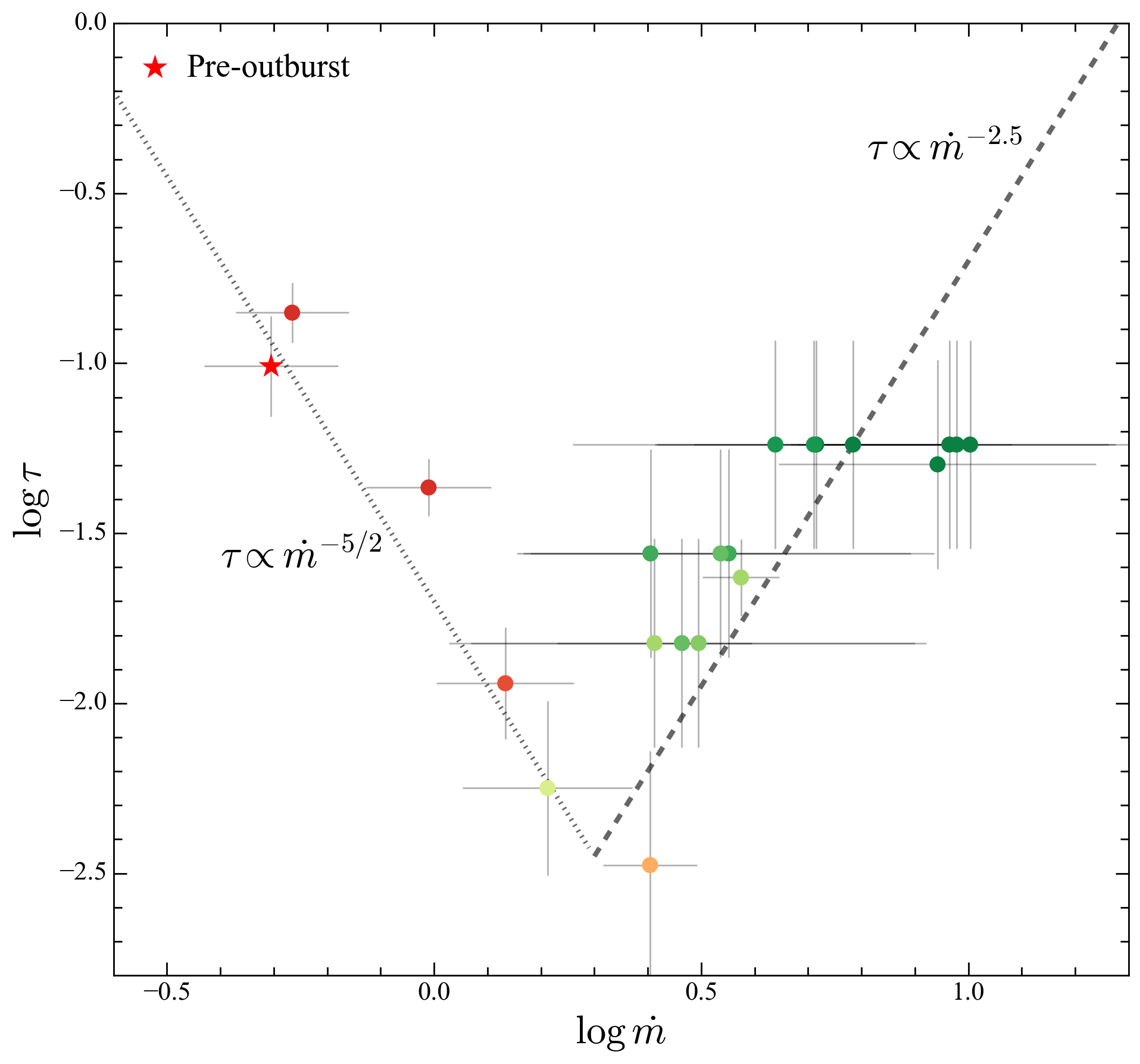}
\caption{The coronal plasma optical depth ($\tau$) as a function of dimensionless mass accretion rate ($\dot{m}$), color-coded by the time since the optical outburst (23 December 2017), from the beginning (green) to the end (red); the orange circle marks 650 days after the outburst. The red star shows  the value in May 2011, before the outburst. The black dotted line shows a power-law fit to the observations for the data on the left branch ($t > 650$\,days), with $\log{\tau} = -2.5\log{\dot{m}}-1.7$. The black dashed line gives a fit with an opposite slope for the data on the right branch ($t<650$ days).}
\label{fig:mdottau}
\end{figure}

As discussed in Section~\ref{sec:corona}, the plasma optical depth ($\tau$) for 1ES\,1927+654 is primarily deduced from the hard-band X-ray spectra. The progression of $\tau$ demonstrates a systematic pattern:  it first decreased before the 650-day mark and subsequently increased (Figure~\ref{fig:gammatau}). Upon substituting the x-axis in Figure~\ref{fig:gammatau} with the dimensionless mass accretion rate ($\dot{m}$) derived from the disk equations \citep{Li2024paper3}\footnote{This $\dot{m}$ is consistent with the values obtained from fitting the warm corona SED models, as shown in Figure~\ref{fig:mdotcompare}.}, we observe that the two branches in the $\tau-\dot{m}$ relationship are symmetrical. Given that $\dot{m}$ significantly influences the physical properties of the inner disk, close examination of this $\tau-\dot{m}$ correlation might provide insights into the relationship between the corona and the inner accretion flow during its evolution.

Assuming the $\alpha-$prescription, the advection cooling at a fixed radius can be expressed as $Q_\mathrm{adv}^- \propto \dot{m}^2\Sigma^{-1}$, where $\Sigma$ represents the disk surface density \citep{Kato2008book}, in units of $\rm g\,cm^{-2}$. Since gravitational energy is discharged through the viscous heating process, the heating rate $Q_\mathrm{vis}^+ \propto \dot{m}$. In the case of a slim accretion disk (e.g., \citealp{Abramowicz1988ApJ,Kato2008book}), where $Q_\mathrm{adv}^- \simeq Q_\mathrm{vis}^+$, we expect

\begin{equation}
    \Sigma \propto \dot{m}.
\label{equ:sigmaslim}
\end{equation}

\noindent
Taking into account $M_\mathrm{BH} = 1.38\times 10^6 \, M_\odot$ \citep{Li2022paper1}, in the inner region of the accretion disk the radiation pressure can be approximately $10^4$ times larger than the gas pressure for a $\dot{m} \simeq 1$ black hole \citep{Kato2008book,Jiang2019bApJ}. In this regime, the radiation cooling rate $Q_\mathrm{rad}^- \propto \dot{m}^{1/2}\Sigma^{-1/2}$. During the late epochs of our campaign, the disk falls into the regime of a thin disk, where $Q_\mathrm{rad}^- \simeq Q_\mathrm{vis}^-$, for which we anticipate

\begin{equation}
    \Sigma \propto \dot{m}^{-1}.
\label{equ:sigmathin}
\end{equation}

\noindent
Alternatively, in the case of a thin disk where gas pressure prevails over radiation pressure—commonly observed in AGNs with moderate accretion rates ($\dot{m} \lesssim 0.5$) or in the outer regions of the disk ($R \gtrsim 10 R_g$)—the radiation cooling rate can be described as $Q_\mathrm{rad}^- \propto \dot{m}^{4}\Sigma^{-5}$, which leads to

\begin{equation}
    \Sigma \propto \dot{m}^{3/5}.
\label{equ:sigmathinlow}
\end{equation}

The branches of $\Sigma \propto \dot{m}$ and $\Sigma \propto \dot{m}^{-1}$ typically correspond to the upper and middle part of the accretion disk state $S$-curve, while the $\Sigma \propto \dot{m}^{3/5}$ branch corresponds to the lower branch (see review in \citealp{Yuan2014ARAA}). \citet{Li2024paper3} identified the $\Sigma \propto \dot{m}$ and $\Sigma \propto \dot{m}^{-1}$ branches in the post-outburst evolution of 1ES\,1927+654, exhibiting a break at approximately 650 days. This timing coincides with the $\alpha_\mathrm{OX}$ and $\kappa_{5100}$ breaks discussed in Section~\ref{sec:energydis}, as well as with the $\tau-\dot{m}$ relation depicted in Figure~\ref{fig:mdottau}. We interpret this transition between the two branches as an indication that the inner accretion flow was initially a slim disk following the event, gradually evolving into a thin disk as the accreted material was consumed (Section~\ref{sec:phypic} and \citealp{Li2024paper3}). Consequently, the dashed and dash-dotted lines in Figure~\ref{fig:mdottau} physically delineate the evolution of the inner disk's $\Sigma$, as indicated by Equations~\ref{equ:sigmaslim} and \ref{equ:sigmathin}, respectively. This suggests a proportionality between the surface density of the inner disk and the optical depth of the coronal plasma of the form

\begin{equation}
    \Sigma \propto \tau^{2/5}.
\label{equ:sigmatau}
\end{equation}

\noindent
Given that in Figure~\ref{fig:mdottau} both the dashed and dotted lines closely align with our observations before and after 650 days, this suggests that the $2/5$ index of the $\Sigma-\tau$ correlation in Equation~\ref{equ:sigmatau} might be consistent across different states of the accretion disk.

According to the magnetohydrodynamic simulations of optically thick, geometrically thin AGN accretion flows of \citet{Jiang2019aApJ}, the fraction of dissipation in the optically thin corona drops as the disk surface density increases. In other words, $\alpha_\mathrm{OX}$ increases as $\Sigma$ decreases. Considering the two-branch evolution of $\alpha_\mathrm{OX}$ shown in Figure~\ref{fig:alphaox} and the $\tau-\Sigma$ relation in Equation~\ref{equ:sigmatau}, the properties of the corona of 1ES\,1927+654 are similar to those expected in the optically thin regions outside the disk photosphere, supporting the corona picture of \citet{Jiang2019aApJ}. Our results not only support what is expected from the simulations of \cite{Jiang2019aApJ}, but also generalize the results for the slim disk phase (i.e., $t \lesssim 650$ days of 1ES\,1927+654).

As outlined in Equations~\ref{equ:sigmaslim} and \ref{equ:sigmathin}, there is a strong correlation between the disk surface density and the accretion rate. This connection subsequently affects the energy distribution between the corona and the disk \citep{Jiang2019aApJ}. Our results offer a physical rationale for the $\alpha_\mathrm{OX}-\lambda_\mathrm{E}$ correlation observed in the study by \citet{Lusso2010AA}. For typical type~1 AGNs, where $\dot{m} \approx 0.1$, the disk is mostly  gas pressure-dominated \citep{Kato2008book} or magnetic pressure-dominated \citep{Jiang2019aApJ}. In this regime, we would expect $\Sigma \propto \dot{m}^{3/5} $ (Equation~\ref{equ:sigmathinlow}). Following \citet{Lusso2010AA}, $\alpha_\mathrm{OX} \propto \lambda_\mathrm{E}^{0.4}$ for type~1 AGNs. Assuming $\dot{m} \sim \lambda_\mathrm{E}$, we have $\Sigma \propto  \alpha_\mathrm{OX}^{1.5}$. Combined with the $\tau-\Sigma$ relation in Equation~\ref{equ:sigmatau}, we expect the correlation

\begin{equation}
    \alpha_\mathrm{OX} \propto \tau^{0.27}.
\label{equ:tau_alphaox}
\end{equation}

\begin{figure}
\centering
\includegraphics[width=0.49\textwidth]{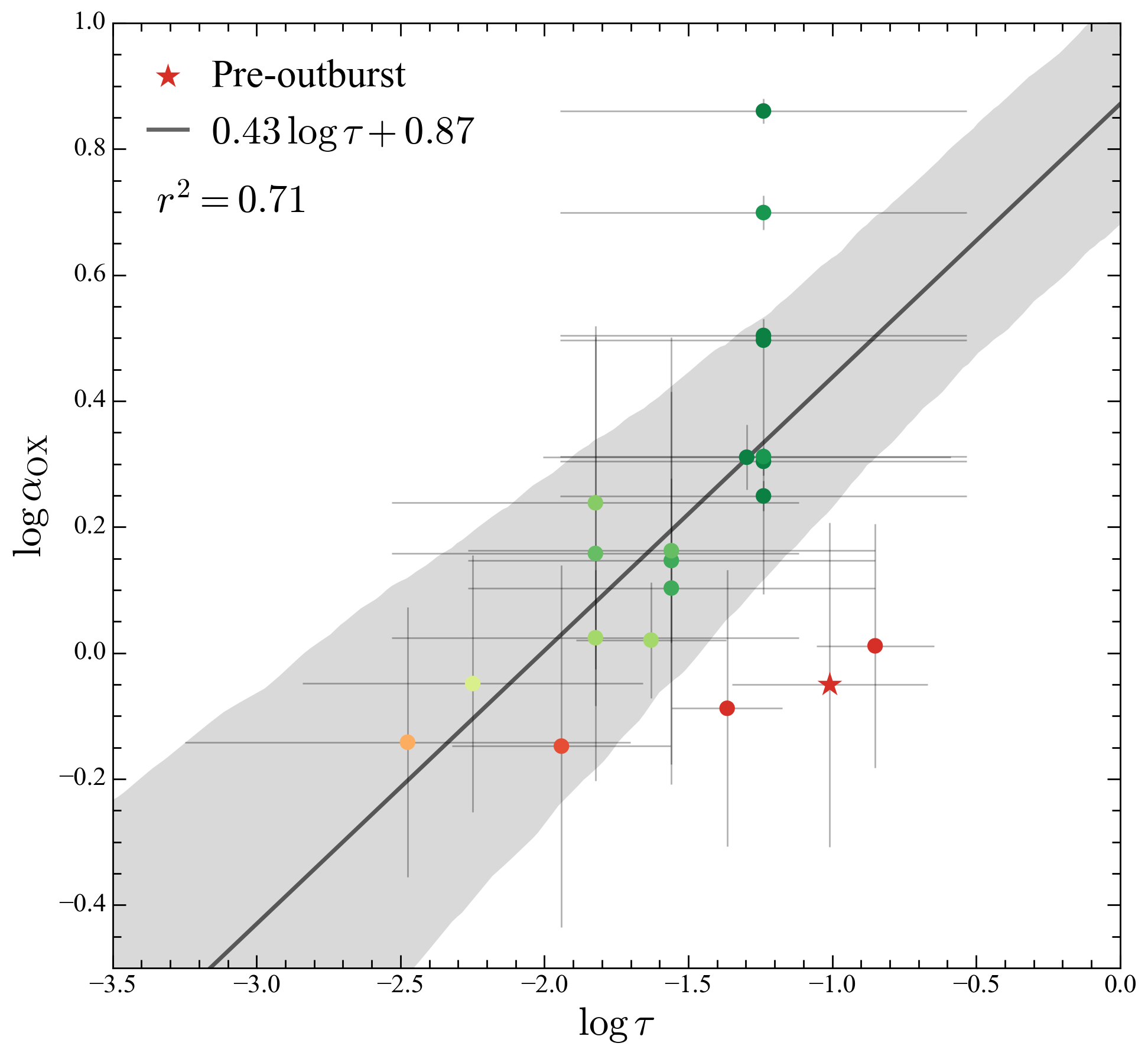}
\caption{The optical to X-ray slope ($\alpha_\mathrm{OX}$) versus the coronal plasma optical depth ($\tau$), color-coded by the time since the optical outburst (23 December 2017), from the beginning (green) to the end (red); the orange circle marks 650 days after the outburst. The red star shows the value in May 2011, before the outburst. The black line shows the best-fit relation $\log{\alpha_\mathrm{OX}} = 0.43\log{\tau} + 0.87$, with the gray shaded region indicating the $1\sigma$ uncertainty. }
\label{fig:aoxtau}
\end{figure}

\noindent
It is worth remarking that the intrinsic scatter of the $\alpha_\mathrm{OX}-\lambda_\mathrm{E}$ correlation is typically rather large (e.g., \citealp{Lusso2010AA}). Therefore, Equation~\ref{equ:tau_alphaox} in any AGN sample would have an even larger scatter due to different values of radiation efficiency, black hole mass, and viscosity in different sources. Therefore, detailed studies of individual AGNs are better suited to examine the $\Sigma-\alpha_\mathrm{OX}$ relation quantitatively. In the case of 1ES\,1927+654, a linear regression for $\log{\alpha_\mathrm{OX}}$ and $ \log{\tau}$ (Figure~\ref{fig:aoxtau}) yields the following best-fit correlation with 0.21\,dex intrinsic scatter,

\begin{equation}
\log{\alpha_\mathrm{OX}} = 0.43\pm 0.07 \,\log{\tau} + 0.87 \pm 0.12,
    \label{equ:tau_alphaox_fit}
\end{equation}

\noindent
which has a coefficient of determination $r^2 \simeq 0.71$. It is interesting to notice the four outliers in Figure~\ref{fig:aoxtau}: two points with high values of $\alpha_\mathrm{OX}$, corresponding to times when the corona had disappeared, and two values located at the radiation pressure-dominated thin branch in Figure~\ref{fig:mdottau}. Excluding the four outliers, the intrinsic scatter is 0.10\,dex, and the slope of $0.43\pm 0.07$ is also close to the expectation of Equation~\ref{equ:tau_alphaox}.

The correlation between disk surface density and coronal optical depth may indicate a significant interplay between the disk and the corona. Scenarios where alterations in one component directly influence the other are most congruent with the ``sandwich'' or slab corona model. This is particularly pertinent in the case of 1ES\,1927+654, where shifts in the disk's surface density could impact the corona's optical depth due to the interchange of mass and energy between these two regions \citep{Haardt1993ApJ,Dove1997ApJ}. The ``lamppost'' and ``blobby'' coronal models may not exhibit such a direct correlation. In the lamppost model, the corona is located away from the disk along the black hole's rotation axis, potentially limiting the direct interaction between the disk and corona \citep{Miniutti2003MNRAS,Petrucci2013AA}. The blobby model consists of the corona being formed by numerous localized, possibly transient flares, leading to a complex and not necessarily linear relationship between the disk and corona \citep{Galeev1979ApJ,Beloborodov1999ApJ}. Our observations offer a novel avenue to assess the geometry of the corona, utilizing time-domain X-ray studies in conjunction with investigations of X-ray polarization \citep{Schnittman2010ApJ,Marinucci2022MNRAS} or accretion disk emissivity profiles \citep{Gonzalez2017MNRAS}.

\subsection{The Disappearence of the Corona at $\sim 200$ Days}
\label{sec:200days}

From stage~I to stage~II, the X-ray properties of 1ES\,1927+654 went through dramatical changes in both flux and color \citep{Ricci2020ApJL,Ricci2021ApJS}, as shown by the black and cyan arrow on the $\Theta_e-\ell$ plane (Figure~\ref{fig:thetal}b). One of the most peculiar aspects of this event is the fact that the X-ray flux did not trace the 2-dex flux rise in the optical bands; instead, it decreased dramatically. Studying the UV/optical light curves, \citet{Trakhtenbrot2019ApJ} suggested that the changing-look event in 1ES\,1927+65 might have been triggered by a TDE. Based on this scenario, \citet{Ricci2020ApJL} suggested that shocks between the disk and the tidal debris of the disrupted star may have depleted the inner disk, leading to the disappearance of the X-ray corona. Providing an alternative perspective, \citet{Scepi2021MNRAS} and subsequently \citet{Laha2022ApJ} suggested that an advection event, carrying a magnetic flux that was opposite to that of the existing corona, could also interrupt the energy supply of the X-ray source if the disk was in a magnetically arrested disk (MAD) state.

The flux inversion scenario proposed by \citet{Scepi2021MNRAS} did not specifically address the implications for the soft X-ray light curve. There are two potential sources for the soft X-ray emission. Firstly, if the accretion flow is highly magnetized, the nonthermal emission from a MAD in the hard X-ray range can often extend into the soft X-ray band \citep{Kimura2021ApJ}. However, during stage~II the luminosity of the blackbody component was consistently above the pre-outburst level (Figure~\ref{fig:lcall}). Therefore, the soft X-ray emission cannot be merely an extension of the magnetized corona emission; otherwise, during this stage the soft X-ray energy would also fall significantly below the pre-outburst level. Additionally, in stage~III the source should follow a similar, not opposite, light curve (Figure~\ref{fig:lcall}). An alternative source for such a soft X-ray excess could be a warm corona at a larger radial scale \citep{Done2012MNRAS,Kubota2018MNRAS}. As inferred from our physical SED modeling in Section~\ref{sec:wcoronamodel}, a warm corona with $R_\mathrm{warm} \simeq 55 r_g$ is expected. In the flux inversion scenario, the advection event is assumed to have originated from a larger scale ($r \approx 180 r_g$; \citealp{Scepi2021MNRAS}), taking around 200\,days to reach the inner disk at approximately $10 R_g$. Given the warm corona's larger radius, one would expect a roughly 60-day advance for the soft X-ray emission ahead of the flux diminishing. However, as depicted in Figure~\ref{fig:lcall}, the coronal component evolves in tandem with the black-body component in stage~II and never tracks it afterward. As a result, the MAD scenario would face difficulties in explaining the behavior of the soft X-rays.

In the scenario in which the changing-look event was induced by a TDE \citep{Trakhtenbrot2019ApJ}, it is not strictly necessary that the inner disk be depleted by shocks. At the early epochs, the luminosity already exceeded the Eddington limit ($\lambda_\mathrm{E} \approx 4$), indicating super-Eddington accretion. Magnetohydrodynamic simulations of super-Eddington disks suggest that optically thick outflows are often launched during this phase \citep{Jiang2014ApJ}. Observationally, the existence of optically thick outflows could explain the broad feature at 1\,keV \citep{Ricci2020ApJL} as arising from reprocessed X-ray radiation from such a relativistic outflow \citep{Masterson2022ApJ}. \citet{Li2024paper3} calculated the size of the photosphere ($R_\mathrm{ph}$) that could give rise to the blackbody component, finding $R_\mathrm{ph}$ increased significantly at $\sim 200$ days. Once the outflow was launched, the interaction with the corona would enhance the bremsstrahlung cooling efficiency on account of the increase of gas density. A direct consequence of such an event would be a transient decrease in the coronal temperature. For example, in the case of I~Zwicky~1, the corona temperature experienced a 30\,keV decrease, amounting to a 60\% reduction, following the launch of an outflow \citep{Ding2022ApJ}. In this scenario, the luminosity of the corona would also decrease. As the gas density increases, more X-ray photons would be absorbed rather than scattered. This would lead to bremsstrahlung becoming the dominant process over Compton cooling, causing the corona to evolve along the path represented by the blue dashed line in Figure~\ref{fig:thetal}. \citet{Cao2023MNRAS} investigated the X-ray spectra of 1ES\,1927+654 using a magnetohydrodynamical outflow model and determined that 200 days represents a plausible viscous timescale in the TDE scenario. Shortly after this outflow phase, as the coronal region becomes optically thin again, Compton cooling would dominate again. The corona then will be more luminous than before, due to the larger number of seed photons from the disk after the outburst.

\subsection{The State Transition at $\sim 650$ Days}
\label{sec:650days}

\begin{figure}
\centering
\includegraphics[width=0.49\textwidth]{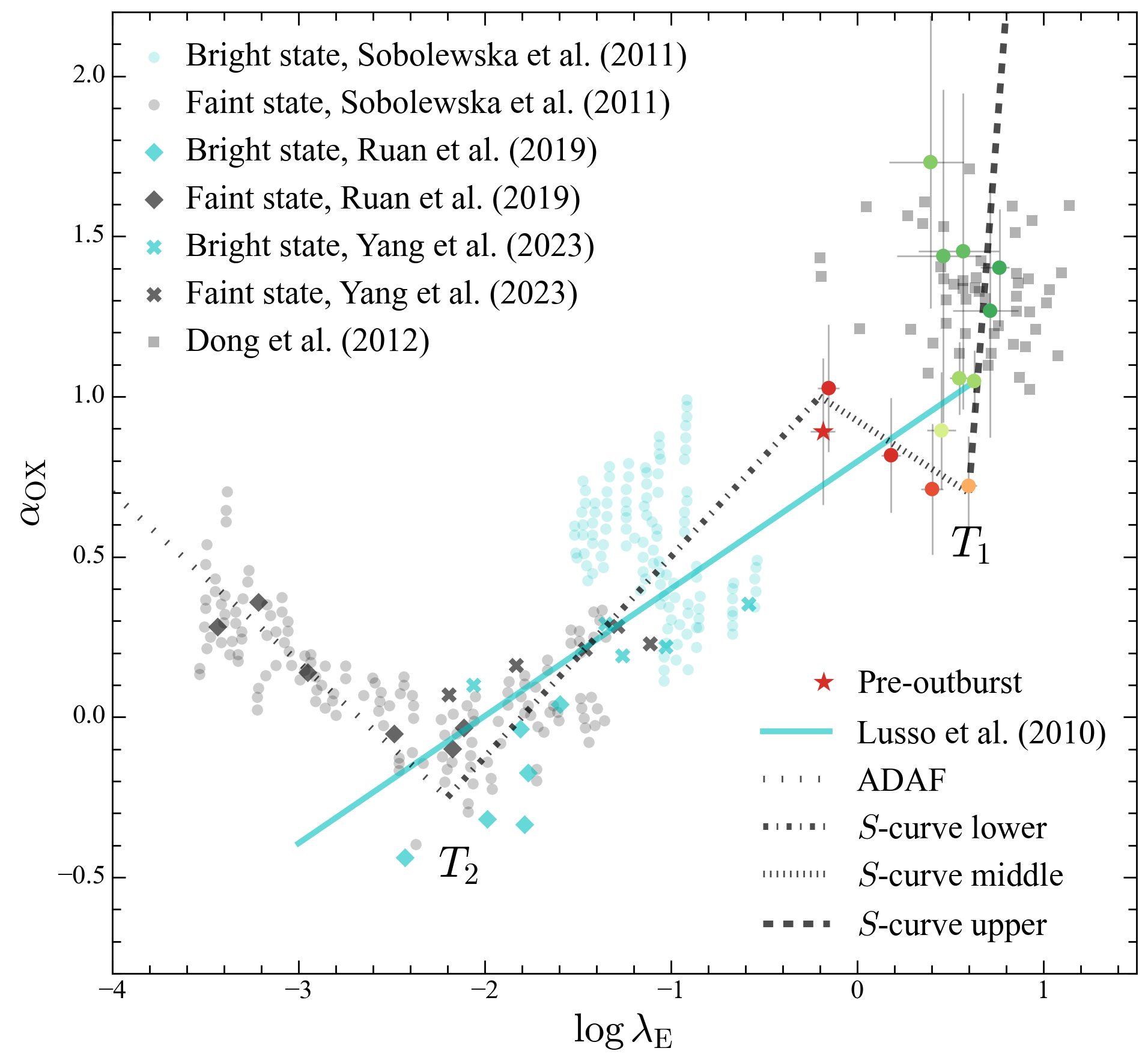}
\caption{Optical to X-ray slope ($\alpha_\mathrm{OX}$) as a function of Eddington ratio ($\lambda_\mathrm{E}$) in stages~III and IV, color-coded by the time since the outburst (23 December 2017), from the beginning (green) to the end (red); the orange circle marks 650 days after the outburst. The red star shows the value in May 2011, before the outburst. The cyan line plots the $\alpha_\mathrm{OX}-L_\mathrm{2500}$ correlation for type~1 AGNs \citep{Lusso2010AA}. The changing-look AGNs that were observed previously are represented by diamonds \citep{Ruan2019ApJ} and crosses \citep{Yang2023arXiv}, with gray denoting faint states and cyan denoting bright states. The background gray (faint states) and cyan (bright points) points represent the predicted AGN accretion state based on X-ray binaries \citep{Sobolewska2011MNRAS}. The grey squares show the intermediate-mass black holes in \citet{Dong2012ApJ}. The four black lines correspond to four different states of the accretion disk, discussed in Section~\ref{sec:650days}: sparse-dotted line for the ADAF, dash-dotted line for the gas pressure-dominated thin disk ($S$-curve lower), middle dash-dotted line for the radiation pressure-dominated thin disk ($S$-curve middle), and right dashed line for the slim disk ($S$-curve upper).  $T_1$ marks the break point of 1ES 1927+654, while $T_2$ gives the break point of the changing-look AGNs in \citet{Ruan2019ApJ}.}
\label{fig:aoxtrans}
\end{figure}

The SED and coronal properties of 1ES\,1927+654 show a significant transition point at approximately $\sim 650$\,days, which we propose can be explained by a disk-state transition from a slim disk to a standard disk. Furthermore, \citet{Li2024paper3} discovered that the mass accretion rate follows a power-law decrease ($\dot{m} \propto t^{-1.53}$) and transitioned below the Eddington limit at $t\approx 650$\,days. The evolution of the radiation efficiency reported in \citet{Li2024paper3} also provides compelling evidence of the transition from a slim to a standard disk around this time. The $\alpha-$prescription disk theory suggests that a change in the mass accretion rate would also lead to a change in the disk surface density. At a fixed radius, $\Sigma$ follows $\dot{m}$ along the so-called $S$-shape curve \citep{Abramowicz1988ApJ}: $\Sigma \propto \dot{m}$ for slim disks ($\dot{m} \gtrsim 2$, upper branch) and gas pressure-dominated standard disks ($\dot{m} \lesssim 0.5$, lower branch), distinct from $\Sigma \propto \dot{m}^{-1}$ for radiation pressure-dominated standard disks (transition branch). As elaborated in Section~\ref{sec:diskcorona}, the optical depth of the corona is closely linked to $\Sigma$, which in turn influences the fraction of X-ray energy dissipation within the inner disk according to \citet{Jiang2019aApJ}. Consequently, the bimodal nature of the $\alpha_\mathrm{OX}-L_{2500}$ correlation (Figure~\ref{fig:alphaox}) can be attributed to the transition from a slim to a radiation pressure-dominated standard disk. In Figure~\ref{fig:aoxtrans}, which shows $\alpha_\mathrm{OX}$ as a function of $\lambda_\mathrm{E}$ for stages~III and IV, the slim disk branch (dashed line) is separated by break point ($T_1$) from the thin disk branch (dense-dotted line). The slim disk states of 1ES\,1927+654 (green points) exhibit a weak trend between $\alpha_\mathrm{OX}$ and $\lambda_\mathrm{E}$, with $\lambda_\mathrm{E}$ remaining nearly constant. This can be attributed to a decrease in the radiation efficiency \citep{Li2024paper3}, which is one of the characteristics of a slim disk (e.g., \citealp{Abramowicz1988ApJ}). A similar $\Lambda$-shape break point ($T_2$) on the top left of Figure~\ref{fig:aoxtrans} has been identified both for spectroscopic studies of six changing-look AGNs \citep{Ruan2019ApJ} and simulated AGNs \citep{Sobolewska2011MNRAS}. However, it is worth noting that this break point occurs in a sub-Eddington regime ($\lambda_\mathrm{E} \approx 10^{-2}$). \citeauthor{Noda2018MNRAS} (\citeyear{Noda2018MNRAS}; see also \citealp{Ruan2019ApJ}) suggest that this may be due to a weakening of the UV radiation of the disk, which could be caused by either inner disk truncation or a lower apparent temperature. As the accretion rate drops below the critical value ($\lambda_\mathrm{E} \approx 10^{-2}$), additional UV radiation can arise from cyclo-synchrotron \citep{Narayan1995ApJ} or jet \citep{Zdziarski2003MNRAS} emission. This results in a leftward positive correlation in the $\alpha_\mathrm{OX}-\lambda_\mathrm{E}$ plot (Figure~\ref{fig:aoxtrans}).

The break point $T_2$, which was observed by \citet{Ruan2019ApJ}, can also be attributed to a transition between different disk states. Drawing an analogy with black holes of stellar mass, the decrease of $\alpha_\mathrm{OX}$ in the low-$\lambda_\mathrm{E}$ regime can be explained by a hybrid accretion disk, consisting of an inner ADAF and an outer thin disk \citep{Noda2018MNRAS,Ruan2019ApJ}. In such disks, the transition radius, at which advection is comparable to bremsstrahlung cooling, is inversely proportional to the mass accretion rate: $r_\mathrm{tr} \propto \dot{m}^{-2}$ \citep{Narayan1995ApJ,Honma1996PASJ}. Such accretion state transitions in the $\alpha_\mathrm{OX}-\lambda_\mathrm{E}$ plot are also clearly seen in the two-year multiwavelength monitoring of the TDE AT2018fyk \citep{Wevers2021ApJ}, which also shows a break point near $T_2$.

Although the $\Lambda$-shape transition observed during the changing-look event in 1ES\,1927+654 ($\log{\lambda_\mathrm{E}^{T_1}} \approx 0.5$) and six other changing-look AGNs ($\log{\lambda_\mathrm{E}^{T_2}} \approx -2$; \citealp{Ruan2019ApJ}) can be attributed to a transition in the disk state, the transition between type~1 and type~2 in changing-state AGNs does not seem to always require such a state transition. For example, although 1ES\,1927+654 crossed the $T_1$ transition at around 650 days, nonstellar continuum and broad lines were still detected \citep{Laha2022ApJ,Li2022paper1}. \citet{Yang2023arXiv} studied the broadband spectra of five changing-look AGNs with accretion rates $\lambda_\mathrm{E} \gtrsim 0.01$ and found that in their bright states the value of $\alpha_\mathrm{OX}$ increased, following the gas pressure-dominated thin disk branch shown in Figure~\ref{fig:aoxtrans} (i.e., $\log{\lambda_\mathrm{E}}\, =\, -2$ to $0.5$). These results might suggest that broad emission lines can be absent even when the disk is moderately bright and thin ($\log{\lambda_\mathrm{E}} \approx -1$; e.g., J1338$-$0124 in \citealp{Yang2023arXiv}), similar to 1ES\,1927+654 during its pre-outburst phase \citep{Trakhtenbrot2019ApJ,Li2022paper1}. Interestingly, for the five AGNs in \citet{Yang2023arXiv}, and also for 1ES\,1927+654, $\alpha_\mathrm{OX}$ decreased after the objects transitioned to type~1. Following the $\alpha_\mathrm{OX}-\Sigma$ correlation in Section~\ref{sec:diskcorona}, disk surface density would increase as the $\alpha_\mathrm{OX}$ decreases. Enhanced UV radiation and a higher disk surface density would lead to an increase in the column density of the radiation-driven wind, thereby reinforcing the BLR emission \citep{Elitzur2009ApJ,Elitzur2016MNRAS}.

\subsection{Physical Picture of the Changing-look Event in 1ES\,1927+654}
\label{sec:phypic}

\begin{figure*}
\centering
\includegraphics[width=\textwidth]{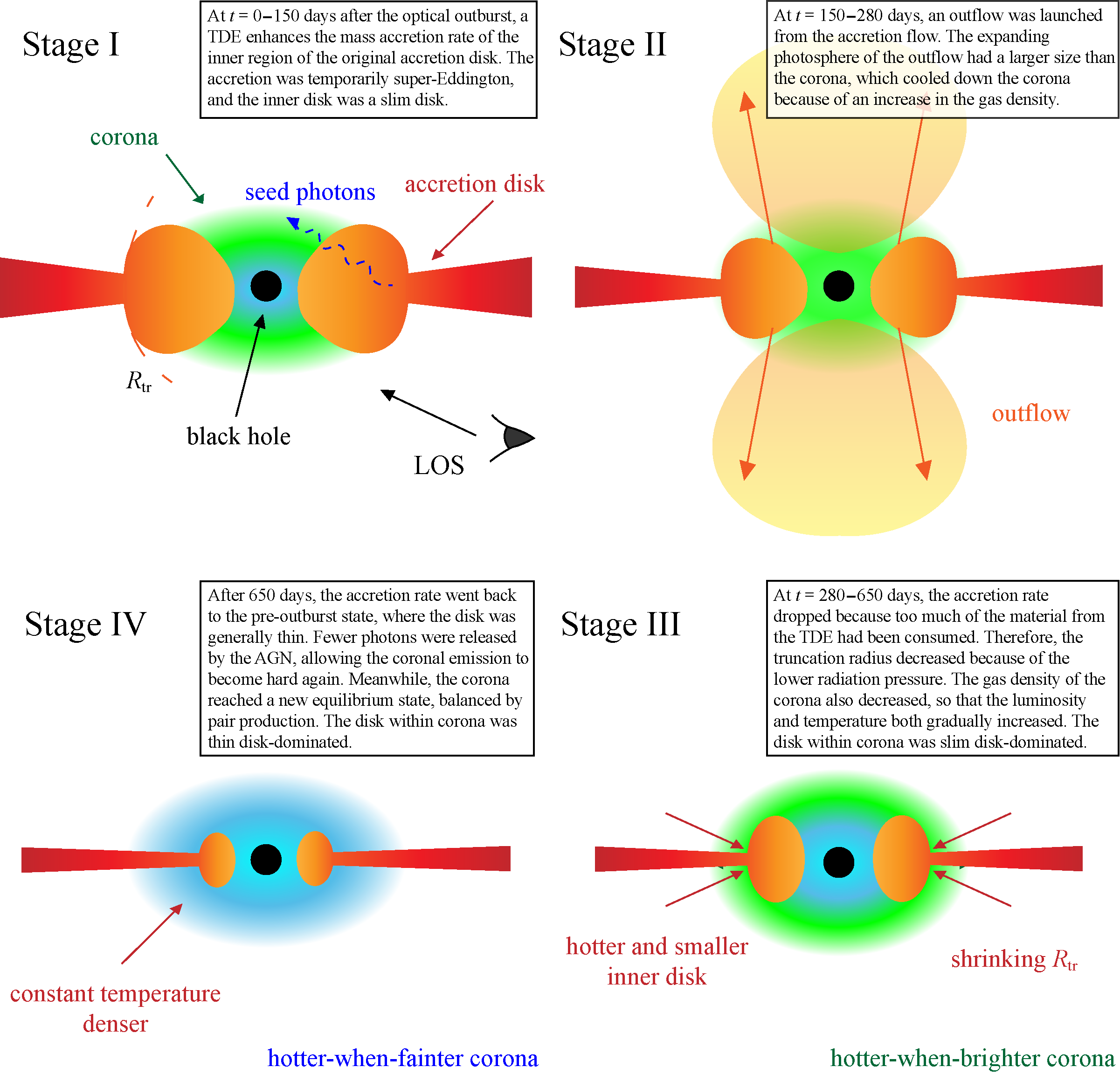}
\caption{Cartoon showing the evolution of the accretion flow during the changing-look event in 1ES\,1927+654. The four stages are defined in Section~\ref{sec:SEDevo}. The evolution of the disk-corona system in each stage is summarized in the text boxes. }
\label{fig:cartoon}
\end{figure*}

The broadband SED analysis presented in this work reveals several interesting properties associated with 1ES\,1927+654. (1) After the outburst, the SED is better described by a three-component model (Section~\ref{sec:sedmodel}), physically related to an outer thin disk (disk component), inner hot slim disk (blackbody component) and a Comptonized corona (corona component) \citep{Li2024paper3}. (2) The temperature and luminosity of the X-ray corona decreased significantly immediately after the outburst. While the luminosity recovered quickly (within $\sim 100$ days), it took about $1000$ days for the coronal temperature to return to its pre-outburst value (see also \citealp{Ricci2020ApJL,Ricci2021ApJS,Masterson2022ApJ}). (3) According to \citet{Li2024paper3}, during the event the inner truncation radius of the thin disk ($R_\mathrm{tr}$) decreased monotonically with time. The photosphere size of the inner slim disk showed a sudden increase at $\sim 200$ days and then gradually decreased to a similar value as $R_\mathrm{tr}$. (4) The optical depth of the corona ($\tau$) was tightly correlated with the inner disk properties, with the correlation being separated into two branches: a slim disk and a thin disk phase, divided by a phase transition at $\sim 650$ days. We now summarize all these results into a four-stage physical picture, as shown in Figure~\ref{fig:cartoon}:

\begin{enumerate}
\item{{\it Stage I}: The optical outburst observed in 1ES\,1927+654 was triggered by a TDE, as confirmed by the evolution of the mass accretion rate \citep{Li2024paper3}. The enhancement of the mass accretion rate to $\dot{m} \approx 10$ caused the inner regions ($r \lesssim 40 r_g$) of the accretion flow to be dominated by a slim disk. After the outburst, it is possible that because of the increase of both gas density and seed photons the cooling (bremsstrahlung) within the hard X-ray corona dominated over the heating (inverse Compton scattering). As a result, the coronal temperature decreased and the X-ray emission became ultra-soft. The three-component continuum SED model in Section~\ref{sec:finalmodel} includes contributions from the X-ray corona, the inner slim disk, and the outer thin disk. Meanwhile, the disk emission can also be reflected by the inner slim disk, producing the blackbody reflection feature at $\sim 1\,$keV \citep{Masterson2022ApJ}. The inclination angle with the respect to the disk is $\sim 60^\circ$, based on broadband SED modeling (Section~\ref{sec:wcoronamodel}).}

\item{{\it Stage II}: At around 200 days, an optically thick and relativistic outflow was launched from the inner regions of the disk. This outflow extended the slim disk photosphere size to around 200\,$r_g$, emitting thermal radiation with a temperature of $kT\approx50$ eV \citep{Li2024paper3}, which also led to a flatter spectral index in the optical ($\alpha_\nu\approx -1$; \citealp{Li2022paper1}). During the outflow phase, enhanced bremsstrahlung cooling occurs due to the increased density in the corona, leading to a substantial cooling of the corona and a consequent weakening of the hard X-ray emission \citep{Ricci2021ApJS}. After the outflow, the gas density decreased, the hot electrons reheated, and the corona quickly recovered its flux within $\sim$100\,days. At the same time, the photosphere of the inner slim disk contracted to the disk surface, resulting in a higher blackbody temperature of $kT\approx 100$ eV. }

\item{{\it Stage III}: During the period that spans between $t=280$ to $650$\,days, the mass accretion rate decreased over time but remained super-Eddington ($\dot{m} \gtrsim 2$). During this period the inner accretion flow was still dominated by a slim disk. However, as $\dot{m}$ decreased, the size of the slim disk ($R_\mathrm{tr}$) also decreased. Additionally, the surface density of the disk decreased, with more energy dissipating in the corona, leading to a harder spectrum (Section~\ref{sec:diskcorona}). During this stage, both the coronal temperature and the luminosity increased, displaying a ``hotter-when-brighter'' behavior. As the size of the slim disk shrunk, the disk temperature rose accordingly, reaching a value close to the pre-outburst level ($\sim140$\,eV). }

\item{{\it Stage IV}: After 650\,days, as $\dot{m}$ finally dropped into the sub-Eddington regime, the inner disk transitioned to the radiation pressure-dominated thin disk regime \citep{Shakura1973AA}. Consequently, in the absence of the inner puffed-up disk, there was hardly any detection of reflection during this stage \citep{Masterson2022ApJ}. This is because the blackbody emission, being the primary source, was very weak. The value of $R_\mathrm{tr}$ remained constant and was close to the pre-outburst one. In this regime, the inner temperature of the disk remained constant with time, but disk surface density, or equivalently $\tau$, increased as $\dot{m}$ dropped (Figure~\ref{fig:mdottau}). This resulted in the hardening of X-ray spectra. The corona eventually reached the pair-production line in the $\Theta_e-\ell$ plot, displaying a ``hotter-when-fainter'' behavior. }
\end{enumerate}

As discussed in Section~\ref{sec:200days}, \cite{Scepi2021MNRAS} alternatively proposed that the changing-look event in 1ES\,1927+654 was associated with an episode of magnetic flux inversion.  \citet{Laha2022ApJ} fitted the UV light curve with an exponential function and found a rather shallow slope of $t^{-0.91}$, in seeming disagreement with the TDE scenario, which predicts a much steeper slope of $t^{-5/3}$ \citep{Rees1988Nature, Lodato2011MNRAS}. Considering the fast evolution of the broadband SED, the shallow slope of the UV light curve can be explained by the fact that 1ES\,1927+654 had a pre-existing accretion disk that emitted non-negligible UV radiation even before the outburst (Figure~\ref{fig:GALEX}). As time passed, the overall spectrum became bluer, which slowed down the decrease of the UV luminosity (Figure~\ref{fig:k5100}). Furthermore, detailed modeling of the mass accretion rate indicates an evolution of the form $\dot{m} \propto t^{-1.53\pm0.1}$, fully consistent with the TDE prediction \citep{Li2024paper3}. In particular, during stage~II the X-ray flux was uncorrelated with the optical flux. However, this lack of correlation was not present throughout the entire duration of the event. For example, $\alpha_\mathrm{OX}$ showed a systematic two-branch evolution (Figure~\ref{fig:alphaox}), and the properties of the disk and corona were strongly correlated (Section~\ref{sec:diskcorona}).

%XX In the boxes, it would be best if the hyphen between two numbers (e.g., 280-650) can be longer, so that it is similar to $-$ or -- in Latex.  In Stage III, slim-disk domimated --> slim disk-dominated.  Consider making Stage I, II, III, IV just slightly smaller.
%
%  hotter-when-fainter corona
%  hotter-when-brighter corona
%reploted

\subsection{Comparison with Other AGNs}
\label{sec:compare}

1ES\,1927+654 was classified as a type~2 AGN on the \cite{Baldwin1981} optical line-intensity diagnostic diagram prior to its optical outburst, owing to the absence of both broad emission lines and substantial X-ray absorption \citep{Boller2003AA}. However, \citet{Li2022paper1} suggest that the lack of broad emission lines is due to the overall luminosity not being high enough to generate a BLR from a disk wind \citep{Elitzur2016MNRAS}. Furthermore, \citet{Li2024paper3} found that the inner disk of 1ES\,1927+654 before the outburst was located on the transition branch between the slim disk and thin disk regimes in the $\dot{m}-\Sigma$ plane, with even lower surface density than typical thin accretion disks. As a result, following the picture presented in \citet{Elitzur2016MNRAS}, it becomes even more challenging for 1ES\,1927+654 to produce broad emission lines, making it a 2-dex outlier on the correlation between X-ray and broad H$\alpha$ emission seen in type~1 AGNs \citep{Stern2012MNRAS}.

During the type~1 phase, 1ES\,1927+654 exhibited a much bluer broadband SED compared to typical AGNs reported by \citet{Elvis1994ApJS} (see Figure~A1 in \citealp{Ricci2020ApJL}), resulting in an optical bolometric correction factor that was 10 times larger than in other AGNs (Figure~\ref{fig:k5100}). However, the optical/UV SED can be well fitted by a thin disk model (Section~\ref{sec:finalmodel}), and the $\kappa_\mathrm{5100}$ value matched the prediction of a standard thin disk (Equation~\ref{equ:k5100}), given the black hole mass reported in \citet{Li2022paper1}. In the early epochs, when the optical continuum was strong, the spectral index also matched the prediction of a thin disk ($\alpha_\nu \approx 1/3$; \citealp{Li2022paper1}), which was bluer than the composite spectra of AGNs ($\alpha_\nu \approx -0.44$; \citealp{VandenBerk2001AJ}). Therefore, the higher $\kappa_\mathrm{5100}$ and $\alpha_\nu$ compared to type~1 AGNs suggest that the optical flux after the outburst originated from the outer thin disk, which is supported by the pure disk emission revealed through polarimetry in a handful of other AGNs \citep{Kishimoto2008Natur}.

The $\alpha_\mathrm{OX}$ value for 1ES\,1927+654 is generally consistent with the $\alpha_\mathrm{OX}-L_\mathrm{2500}$ correlation observed in typical type~1 AGNs \citep{Steffen2006AJ,Lusso2010AA}, with the exception of the special stage~II during which the X-ray emission was weak (Figure~\ref{fig:alphaox}). Meanwhile, similar deviations from the correlation have been found in the X-ray study of 49 AGNs with intermediate-mass black holes ($M_\mathrm{BH} \approx 10^5-10^6\,  M_\odot$ and $\lambda_\mathrm{E} \approx 2-10$; \citealp{Dong2012ApJ}). Objects in \citet{Dong2012ApJ} with similar $M_\mathrm{BH}$ as 1ES\,1927+654 occupy the same region as the stage III of 1ES\,1927+654 (Figure~\ref{fig:aoxtrans}). This suggests that highly accreting AGNs with intermediate-mass black holes \citep{Greene2020} may share a disk state similar to that of 1ES\,1927+654 before 650\,days, where the inner region is dominated by a slim disk. Building on the discussion in Section~\ref{sec:diskcorona}, the state of accretion can be inferred from the $\dot{m}-\Sigma$ curve, or $S$-curve (see review by \citealp{Yuan2014ARAA}). At a specified radius within the accretion disk, four distinct disk states are identified: ADAF, gas pressure-dominated thin disk, radiation pressure-dominated thin disk, and slim disk. These states are categorized based on the ascending order of the mass accretion rate and exhibit distinct correlations between $\dot{m}$ and $\Sigma$, as exemplified by Equations~\ref{equ:sigmaslim}--\ref{equ:sigmathinlow}. In this work, we find a correlation between $\Sigma$ and $\alpha_\mathrm{OX}$ (Equations~\ref{equ:sigmatau} and \ref{equ:tau_alphaox}), which can be used to effectively represent the $\dot{m}-\Sigma$ curve as an $\alpha_\mathrm{OX}-\lambda_\mathrm{E}$ curve in Figure~\ref{fig:aoxtrans}. Consequently, the four branches on which all AGNs align in Figure~\ref{fig:aoxtrans} correspond to the theoretical predictions of the $S$-curve: (1) $\lambda_\mathrm{E} \lesssim 0.01$, ADAF-dominated disks (left sparse-dotted branch); (2) $0.01 \lesssim \lambda_\mathrm{E} \lesssim 0.5$, gas pressure-dominated thin disks (middle dash-dotted branch); (3) $0.5 \lesssim \lambda_\mathrm{E} \lesssim 4$, radiation pressure-dominated thin disks (middle dense-dotted branch); and (4) $\lambda_\mathrm{E} \gtrsim 4$, slim disk-dominated disks (right dashed branch). Further examination of these slopes in greater detail will help shed light on these different accretion phases.

\begin{figure}
\centering
\includegraphics[width=0.49\textwidth]{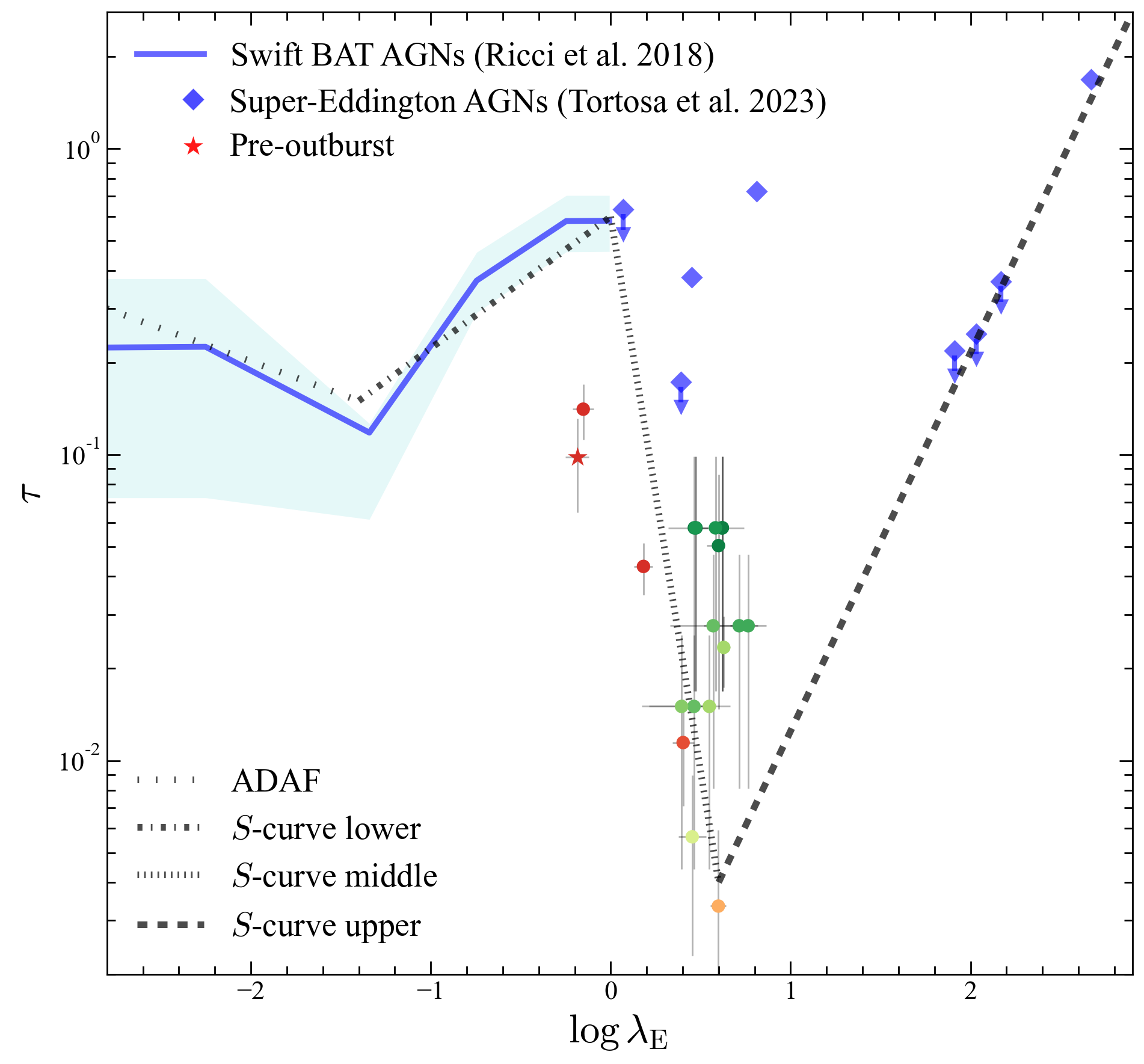}
\caption{Optical depth ($\tau$) of the X-ray corona as a function of the Eddington ratio ($\lambda_\mathrm{E}$), color-coded by the time since the outburst (23 December 2017), from the beginning (green) to the end (red); the orange circle marks 650 days after the outburst. The red star shows the value in May 2011, before the outburst. The blue curve gives the median of the Swift/BAT AGN sample from \citet{Ricci2018MNRAS}, where the cyan shaded area corresponds to the median absolute deviation. Blue diamonds show the super-Eddington AGNs analyzed by \citet{Tortosa2023MNRAS}, with arrows indicating upper limits. The four black lines correspond to the four different states of the accretion disk discussed in Section~\ref{sec:650days}: sparse-dotted line for the ADAF, dash-dotted line for the gas pressure-dominated thin disk ($S$-curve lower), middle dash-dotted line for the radiation pressure-dominated thin disk ($S$-curve middle), and right dashed line for the slim disk ($S$-curve upper).}
\label{fig:leddtau}
\end{figure}

In Section~\ref{sec:diskcorona}, we also show the $\Sigma-\tau$ correlation (Equation~\ref{equ:sigmatau} and Figure~\ref{fig:mdottau}). Therefore, the four branches in Figure~\ref{fig:aoxtrans} can be translated into regions of the $\tau-\lambda_\mathrm{E}$ space (Figure~\ref{fig:leddtau}), as an indicator of disk state. 1ES\,1927+654 shows a ``V-shape'' break (see also Figure~\ref{fig:gammatau}), with the break located in the region between the thin and slim disk regimes. We also overlay the median $\tau-\lambda_\mathrm{E}$ relation of the Swift BAT AGN sample \citep{Ricci2018MNRAS}, together with the seven super-Eddington AGNs reported by \citet{Tortosa2023MNRAS}. The four branches in Figure~\ref{fig:leddtau} share a $\lambda_\mathrm{E}$ break similar to that illustrated in Figure~\ref{fig:aoxtrans}, supporting the $\tau-\Sigma-\alpha_\mathrm{OX}$ correlation  (Section~\ref{sec:diskcorona}). The value of $\tau$ for 1ES\,1927+654 is considerably smaller than that for other AGNs, which might reflect intrinsic differences between 1ES\,1927+654 and other AGNs, such as a lower disk density \citep{Li2024paper3} or a weaker magnetic field \citep{Haardt1991ApJ}.

During stage~III 1ES\,1927+654 exhibited a ``softer-when-brighter'' corona (Section~\ref{sec:SEDevo} and \citealp{Masterson2022ApJ}), as indicated by the evolution of the hard X-ray photon index. Interestingly, the evolution of the coronal temperature shows a $\Lambda$ pattern on the $\Theta_e -\ell$ plot, hotter when brighter before $t\approx 650$\,days and cooler when brighter after $\sim 650$\,days (Figure~\ref{fig:thetal} and Section~\ref{sec:corona}). A similar trend was also found in SWIFT\,J2127.4+5654, whose cut-off energy displays a similar $\Lambda$ pattern, while the photon index shows a ``softer-when-brighter'' behavior \citep{Kang2021MNRAS}. By studying the $\Gamma-E_\mathrm{cut}$ variation patterns of seven sources, \citet{Kang2021MNRAS} conclude that pair-dominated corona (lying close to pair lines on the $\Theta_e-\ell$ plot) only display ``hotter-when-softer/brighter'' behavior, while ``cooler-when-softer/brighter'' was only found for AGNs on the left side of the pair lines. \citet{Kang2021MNRAS} suggested that the coronal cooling efficiency could be suppressed by pair production, such that the ``cooler-when-softer/brighter'' relation may reversed once it reaches the pair line, as in the case of SWIFT\,J2127.4+5654. However, 1ES\,1927+654 displays a different behavior, being ``hotter-when-softer/brighter'' when far off the pair line. Considering the plasma optical depth evolution shown in Figure~\ref{fig:gammatau}b, we suggest that the physical reason is that in stage~III, when the source was in the slim disk phase, $\tau$ decreased as $\lambda_\mathrm{E}$ decreased (Figure~\ref{fig:leddtau}), so that heating dominated over cooling, making the corona hotter. In stage~IV, when the source was in the radiation pressure-dominated thin disk phase, $\tau$ increased significantly as the $\lambda_\mathrm{E}$ decreased. Once cooling dominated heating, the corona became ``cooler-when-softer/brighter''. The above argument predicts that most type~1 AGNs, located near the middle dashed line in Figure~\ref{fig:leddtau}, would be ``hotter-when-softer/brighter'', as observed \citep{Kang2021MNRAS}.

\section{Summary}
\label{sec:summary}

Before the optical outburst on 23 December 2017 \citep{Trakhtenbrot2019ApJ}, 1ES\,1927+654 was classified as a ``true'' type~2 AGN, lacking X-ray absorption and broad-line emission in its polarized optical spectrum \citep{Boller2003AA, Tran2011ApJ}. The source also showed rapid X-ray spectral variability \citep{Boller2003AA}, as well as a FUV flux 0.87 magnitudes brighter than that of its host galaxy (Section~\ref{sec:GALEXobs}). After the outburst the source showed an increase of its blue continuum and the appearance of broad optical emission lines \citep{Trakhtenbrot2019ApJ}, while $\sim 200$ days after the outburst the X-rays first dropped sharply and then rebrightened \citep{Ricci2020ApJL,Ricci2021ApJS}, and thereafter ebbed after $\sim 800$ days \citep{Masterson2022ApJ}. Both the continuum light curve and accretion rate evolution suggest that the changing-look event was triggered by a TDE \citep{Trakhtenbrot2019ApJ,Li2024paper3}, as also evidenced by the rapidly evolving BLR, which moves on elliptical orbits \citep{Li2022paper1}. In this work, we utilized three-year X-ray and UV/optical monitoring data obtained from the Swift and XMM-Newton observatories to study the post-outburst evolution. We investigate the intrinsic SED of 1ES\,1927+654 by correcting for dust extinction and emission from the host galaxy. Modelling the broadband SED with four SED models, we found that the disk + blackbody + corona model provides the best fit to the data, offering a more detailed understanding of the complex interplay between the various components of this changing-look AGN (Table~\ref{tab:resubbf}). The dramatic SED variability can be summarized as follows:

\begin{enumerate}

\item To highlight the evolution of the different components throughout the changing-look event, we divided the event into four distinct stages, during each of  which the SED models varied differently (Figure~\ref{fig:SEDs}). During stage~I, the disk, blackbody, and corona components experienced dramatic changes in luminosity, temperature, and spectral index. In stage~II, all three components exhibited different light curve behavior, with the blackbody and corona components experiencing huge variations in amplitude. Stage~III was characterized by a dramatic decrease of the soft X-ray blackbody, while the corona remained bright and the disk varied mildly. Finally, during stage~IV all components, as well as the temperature of the X-ray corona, gradually returned to their pre-outburst state.

\item  The Eddington ratio at the beginning of the observations, after the outburst, was $\lambda_{\rm E} = 4.2$ and gradually decreased to 0.7 at $t\approx 1100$ days, with a median of 3.3 and a standard deviation of 1.1, indicating that the changing-look event following the outburst was predominantly super-Eddington (Figure~\ref{fig:lcall}).

\item The $\alpha_\mathrm{OX}-L_\mathrm{2500}$ correlation in 1ES\,1927+654 (Figure~\ref{fig:alphaox}) is generally in line with previous studies of type\,1 AGNs, except for observations during a unique period around 200\,days. A positive correlation between $\alpha_\mathrm{OX}$ and $L_\mathrm{2500}$ is observed before $t\approx650$\,days, while a negative correlation emerged after this time.

\item The optical bolometric correction at 5100\,\AA\ ($\kappa_{5100}$) is much larger than in typical luminous type\,1 AGNs and quasars (Figure~\ref{fig:k5100}). This value is consistent with the expectations of thin accretion disks of similar $M_\mathrm{BH}$ and luminosity (Equation~\ref{equ:k5100}). A negative correlation between $\kappa_{5100}$ and $\dot{m}$ is observed before $t\approx 650$\,days, and a positive correlation thereafter, the latter with a slope similar to that anticipated for a thin accretion disk.

\item Following the rapid X-ray evolution in stage~II, the power-law continuum generated by the X-ray corona exhibits a higher photon index than most AGNs, with a distinct softer-when-brighter trend observed after $t\approx300$ days. The X-ray plasma optical depth is significantly lower than those in typical AGNs (Figure~\ref{fig:gammatau}). The plasma optical depth continued to decrease until $\sim$650 days, after which it started to increase.

\item The systematic evolution of the X-ray corona temperature and the compactness parameter reveals an initially cool corona, compared to other AGNs ($\Theta_e-\ell$ plot; Figure~\ref{fig:thetal}). In stage~II, this was followed by a cooling phase dominated by bremsstrahlung, accompanied by substantial declines in both the corona luminosity and temperature.

\item  In stages~III and IV, the X-ray corona of 1ES\,1927+654 traces a $\Lambda$ pattern on the $\Theta_e-\ell$ plot. Prior to $t \approx 650$\,days (during stage~III), the source exhibited a ``hotter-when-brighter'' behavior while it remained far from the pair line. In contrast, during stage~IV, the rise in $\tau$ caused the cooling to overtake the heating, resulting in a ``cooler-when-brighter'' corona. Eventually, 1ES\,1927+654 stabilized near the pair line region, where most AGNs are typically found.

\end{enumerate}

Our SED decomposition reveals the underlying emission mechanisms at play in 1ES\,1927+654, as well as the interplay between disk and corona properties. We explore the correlation between the corona optical depth ($\tau$), disk surface density ($\Sigma$), and the energy distribution $\alpha_\mathrm{OX}$. By examining the evolution of $\tau$ about the inner part of the accretion disk, a two-branch transition is observed, with the accretion flow initially dominated by a slim disk and later evolving into a thin disk. A connection between $\Sigma$ and $\tau$ is proposed (Equation~\ref{equ:sigmatau}), as well as a relationship between $\Sigma$ and $\alpha_\mathrm{OX}$ (Equation~\ref{equ:tau_alphaox} and Figure~\ref{fig:aoxtau}). These results align with the propositions made by simulations of sub-Eddington thin accretion disks \citep{Jiang2019aApJ}, which suggest a qualitative inverse correlation between energy dissipated in the X-ray bands and disk $\Sigma$. The result indicates a density connection between the disk and the corona, supporting a ``sandwich'' corona geometry. We also found that the $\Sigma-\alpha_\mathrm{OX}$ correlation persists during the slim disk phase. While studying the $\Sigma-\alpha_\mathrm{OX}$ correlation in AGN samples may yield a large scatter, focusing on individual AGNs, such as 1ES\,1927+654, can provide more detailed information about the evolution of the inner disk surface density.

We integrate our results and previous studies of 1ES\,1927+654 into a physical picture to explain the evolution of this source (Figure~\ref{fig:cartoon}). After the TDE triggered an optical outburst, an optically thick and relativistic outflow was launched, and the interaction between the outflow and the corona explains the observed decrease in coronal temperature and luminosity at $\sim 200$ days. As the mass accretion rate drops, the inner disk state is transformed from a slim disk into a radiation pressure-dominated thin disk at $t\approx 650$\,days.

The correlation between $\Sigma$ and $\alpha_\mathrm{OX}$ found in this work can be used to effectively approximate the behavior of the theoretical $S$-curve ($\dot{m}-\Sigma$ curve; e.g., \citealt{Abramowicz1988ApJ}) as an $\alpha_\mathrm{OX}-\lambda_\mathrm{E}$ curve. The $\alpha_\mathrm{OX}-\lambda_\mathrm{E}$ plot of 1ES\,1927+654 displays a ``$\Lambda$-shaped'' break, indicating a state transition between slim and thin disks. This differs from previous studies where the break point manifests at a much lower $\lambda_\mathrm{E}$, implying a transition from a thin disk to an ADAF. By connecting these two break points, the $\alpha_\mathrm{OX}-\lambda_\mathrm{E}$ curve bifurcates into four branches, each corresponding to a different disk state (Figure~\ref{fig:aoxtrans}). Furthermore, the $\Sigma-\tau$ correlation allows these four branches to be translated into a $\tau-\lambda_\mathrm{E}$ indicator, which provides insights into the disk state. Future works focused on studying the slopes of the $\alpha_\mathrm{OX}-\lambda_\mathrm{E}$ curve in more detail will allow us to gain a better understanding of different accretion disk states.

\begin{acknowledgements}

We thank the anonymous referee for helpful suggestions. This work was supported by the National Science Foundation of China (11721303, 11991052, 12011540375, 12233001), the National Key R\&D Program of China (2022YFF0503401), and the China Manned Space Project (CMS-CSST-2021-A04, CMS-CSST-2021-A06). CR acknowledges support from the Fondecyt Regular grant 1230345 and ANID BASAL project FB210003. IA acknowledges support from the European Research Council (ERC) under the European Union's Horizon 2020 research and innovation program (grant agreement number 852097), from the Israel Science Foundation (grant number 2752/19), and from the United States - Israel Binational Science Foundation (BSF). We thank Chichuan Jin and Chris Done for many useful discussions on broadband fitting with warm corona models.

\end{acknowledgements}

\vspace{5mm}
\facilities{Swift(XRT and UVOT), XMM-Newton (EPIC and OM), NuSTAR, GALEX, Las Cumbres Observatory}

\software{GALFIT \citep{Peng2002AJ,Peng2010AJ}, SAS (v19.0.0; \citealp{Gabriel2004ASPC}), XSPEC (12.11.1; \citealp{Arnaud1996ASPC})}

%XX All your conference proceedings have the wrong format. Missing publishing city and publisher!  2023 preprints should be updated

\begin{figure}
    \figurenum{A1}
\centering
\includegraphics[width=0.49\textwidth]{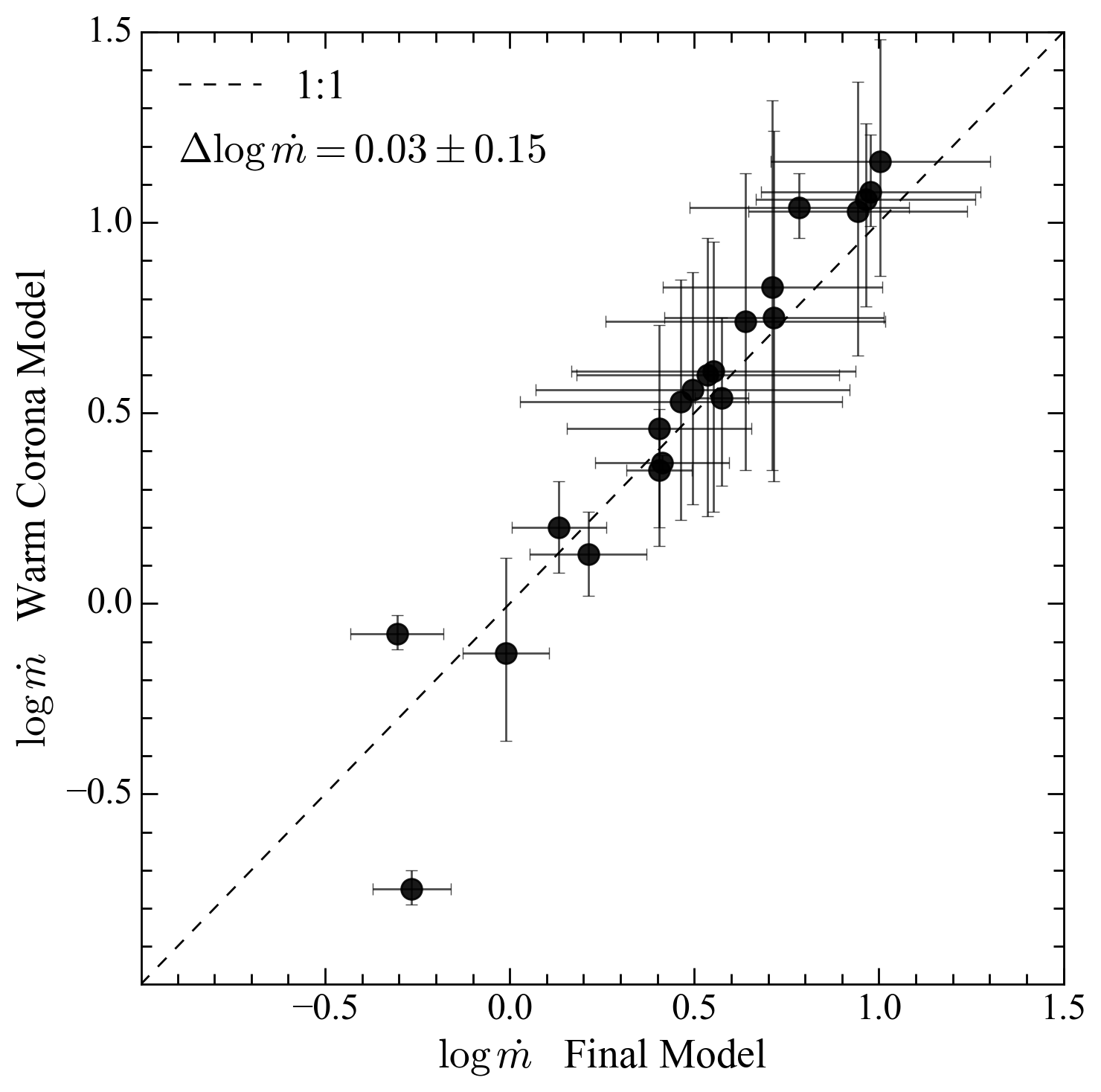}
\caption{Comparison of the dimensionless mass accretion rate ($\dot{m}$) derived from the analysis of the final SED model (x-axis, Section~\ref{sec:finalmodel}) and the warm corona model (y-axis, Section~\ref{sec:wcoronamodel}). The black dashed line shows the 1:1 relation. The median difference ($\Delta \log \dot{m}$) and standard deviation are given in the top-left corner.}
\label{fig:mdotcompare}
\end{figure}

\appendix
In Section~\ref{sec:sedmodel}, we modeled the broad-band SED for the 21 epochs using four different models. Statistics favor the model consisting of a thin disk, a single-temperature blackbody, and a Comptonization component, as the final model discussed throughout Sections~\ref{sec:sec4} and~\ref{sec:sec5}. For reference, we also present the MCMC fitting results of the other three models: the blackbody model (Table~\ref{tab:resublackbody}), the thin disk model (Table~\ref{tab:resudbb}), and the warm corona model (Table~\ref{tab:resuslim}). In Figure~\ref{fig:mdottau}, we adopt the dimensionless mass accretion rate ($\dot{m}$) from \citet{Li2024paper3}, which is derived from solving the disk equations based on the SED fitting results of our final SED model, presented in Section~\ref{sec:finalmodel}. In the warm corona model, $\dot{m}$ is a direct model parameter of {\tt agnslim}, as shown in Table~\ref{tab:resuslim}. We compared the $\dot{m}$ derived from the two different SED models in Figure~\ref{fig:mdotcompare}. The median difference is 0.03 dex, with a standard deviation of 0.15 dex. Considering that the median error bar of $\dot{m}$ is $\sim$0.25 dex, the two SED models yields consistent values of $\dot{m}$.

%XX get rid of the -1.0 on the X-axis.
% Reploted

\input{tab_bbfit}

\input{tab_dbbfit}

\input{tab_slimfit}

\end{document}

%% file: tab_opspec.tex
%LCH 2023.12.25
%LCH 2024.08.12

\startlongtable
\begin{deluxetable*}{c c c l c c c c c c }
\tabletypesize{\scriptsize}
\tablecaption{Broadband Spectroscopy  \label{tab:obses}}
\tablehead{
\colhead{No.} & 
\multicolumn{3}{c}{Soft X-ray Observations} &
\multicolumn{3}{c}{Hard X-ray Observations} &
\multicolumn{3}{c}{Optical Observations}
\\
&
\colhead{Date} & 
\colhead{Telescope} & 
\colhead{OBSID Soft} &
\colhead{Energy (keV)} &
\colhead{Telescope} & 
\colhead{OBSID Hard} &
\colhead{Date} &
\colhead{Telescope} &
\colhead{Aperture ($\arcsec$)} 
\\	  \colhead{(1)}
        & \colhead{(2)} 
	& \colhead{(3)}
        & \colhead{(4)}
	& \colhead{(5)}
        & \colhead{(6)}
 	& \colhead{(7)}
	& \colhead{(8)}
        & \colhead{(9)}
	& \colhead{(10)}
        }
\startdata
        0       & 20-05-2011  &    XMM   &   0671860201 & 0.3-6.0 &  ---           &   ---                                    &      ---      &   ---  &  ---   \\
	1       & 17-05-2018  &  Swift   &  00010682001 & 0.2-3.0 &  ---           &   ---                                    &  05-14-2018   &   LCO  &  2.0   \\ 
	2       & 31-05-2018  &  Swift   &  00010682002 & 0.2-1.4 &  ---           &   ---                                    &  05-28-2018   &   LCO  &  2.0   \\ 
	3       & 05-06-2018  &    XMM   &   0830191101\tablenotemark{a}  & 0.3-2.0 &  ---           &   ---        &  06-05-2018   &   LCO  &  2.0   \\ 
	4       & 14-06-2018  &  Swift   &  00010682003 & 0.2-1.3 &  ---           &   ---                                    &  06-14-2018   &   APO  &  1.5   \\ 
	5       & 10-07-2018  &  Swift   &  00010682004 & 0.2-1.0 &  ---           &   ---                                    &  07-06-2018   &   LCO  &  2.0   \\ 
	6       & 24-07-2018  &  Swift   &  00010682005 & 0.2-0.6 &  ---           &   ---                                    &  07-27-2018   &   LCO  &  2.0   \\ 
	7       & 07-08-2018  &  Swift   &  00010682006 & 0.3-1.0 &  ---           &   ---                                    &  08-12-2018   &   LCO  &  2.0   \\ 
	8       & 23-08-2018  &  Swift   &  00010682007 & 0.2-1.5 &  ---           &   ---                                    &  08-12-2018   &   LCO  &  2.0   \\ 
        9       & 03-10-2018  &  Swift   &  00010682008 & 0.2-3.5 &  ---           &   ---                                    &  09-08-2018   &   LCO  &  2.0   \\ 
       10       & 19-10-2018  &  Swift   &  00010682009 & 0.2-3.5 &  ---           &   ---                                    &  11-13-2018   &   LCO  &  2.0   \\ 
       11       & 23-10-2018  &  Swift   &  00010682010 & 0.2-3.5 &  ---           &   ---                                    &  11-13-2018   &   LCO  &  2.0   \\
       12       & 21-11-2018  &  Swift   &  00010682011 & 0.3-3.5 &  ---           &   ---                                    &  11-13-2018   &   LCO  &  2.0   \\ 
       13       & 06-12-2018  &  Swift   &  00010682012 & 0.3-2.0 &  ---           &   ---                                    &  11-13-2018   &   LCO  &  2.0   \\ 
       14       & 12-12-2018  &    XMM   &   0831790301\tablenotemark{a}  & 0.3-3.0 &       NuSTAR   &  90401641002   &  11-13-2018   &   LCO  &  2.0   \\
       15       & 28-03-2019  &  Swift   &  00010682014 & 0.3-2.5 &  ---           &   ---                                    &  03-19-2019   &   LCO  &  2.0   \\
       16       & 07-05-2019  &    XMM   &   0843270101\tablenotemark{a}  & 0.3-3.0 &       NuSTAR   &  90501618002   &  05-19-2019   &   LCO  &  2.0   \\
       17       & 02-11-2019  &    XMM   &   0843270201\tablenotemark{a}  & 0.3-3.0 &       NuSTAR   &  60502034002   &  09-10-2019   &   LCO  &  2.0   \\ 
       18       & 03-05-2020  &    XMM   &   0863230101\tablenotemark{a}  & 0.3-3.0 &       NuSTAR   &  60502034002   &  09-10-2019   &   LCO  &  2.0   \\ 
       19       & 16-09-2020  &    XMM   &   0863230201\tablenotemark{a}  & 0.3-3.0 &       NuSTAR   &  60602003002   &  09-10-2019   &   LCO  &  2.0   \\ 
       20       & 12-01-2021  &    XMM   &   0863230301 & 0.3-3.0 &       NuSTAR   &  60602003004 &                              09-10-2019   &   LCO  &   2.0     
\enddata
\tablecomments{ Summary of optical and X-ray observations (Section~\ref{sec:sec2}). Col. (1): Number of the broadband dataset used for the broadband SED analysis in Section~\ref{sec:sec3}. Col. (2): Date of the X-ray observations. Col. (3): Telescope used to take the soft X-ray spectra and simultaneous photometric data: XMM = XMM-Newton with EPIC and OM instruments; Swift with XRT and UVOT instruments. Col. (4) Observation ID of the soft X-ray observation. Col. (5): Energy range used in spectral fitting. Col. (6): Telescope used to take the hard X-ray spectra. Col. (7): Observation ID of the hard X-ray observation. Col. (8): Date of the optical spectroscopic observation. Col. (9): Telescope used to take the optical spectra: APO = 3.5~m telescope at Apache Point Observatory; LCO = Las Cumbres Observatory. Col. (10): Aperture size to extract the optical spectra.
\tablenotetext{a}{Pile-up was detected in the PN camera of XMM-Newton; we adopted an annular region to extract the spectra (see details in Section~\ref{sec:xmmobs}).}}
\end{deluxetable*}

%% file: photometry.tex
%LCH 2023.12.25
%
%XX Apparent (i.e. observed) magnitudes: symbol is "m", not "M". The latter is only for absolute magnitudes. However, there is no need for "m" at all. Just use the filter name, as is customary.

\startlongtable
\begin{deluxetable*}{lcccccccc}
\tablecaption{Summary of Photometric Analysis of OM/UVOT Images of 1ES~1927+654 \label{tab:photometry}}
\tabletypesize{\scriptsize}
\tablehead{
      \colhead{No.}   &
      \colhead{Instrument} &
      \colhead{Date}   &
      \colhead{${\rm UVW2}$} &
      \colhead{${\rm UVM2}$ }   &
      \colhead{${\rm UVW1}$ } & 
      \colhead{${U}$ } &
      \colhead{${B}$ } &
      \colhead{${V}$ } \\
      \colhead{} &
      \colhead{} &
      \colhead{} &
      \colhead{(mag)} &
      \colhead{(mag)} &
      \colhead{(mag)} &
      \colhead{(mag)} &
      \colhead{(mag)} &
      \colhead{(mag)} \\
\colhead{(1)} &
\colhead{(2)} &
\colhead{(3)} &
\colhead{(4)} &
\colhead{(5)} &
\colhead{(6)} &
\colhead{(7)} &
\colhead{(8)} &
\colhead{(9)}  
}
\startdata
  0 &   OM & 20-05-2011 & ---            &  $16.93\pm0.29$ &  $15.85\pm0.46$ &  ---            &  ---            &  $16.16\pm0.14$ \\
  1 & UVOT & 17-05-2018 & $14.63\pm0.05$ &  $14.42\pm0.06$ &  $14.43\pm0.05$ &  $14.48\pm0.12$ &  $15.45\pm0.17$ &  $15.54\pm0.21$ \\
  2 & UVOT & 31-05-2018 & $14.59\pm0.06$ &  $14.77\pm0.05$ &  $14.48\pm0.06$ &  $14.57\pm0.08$ &  $15.45\pm0.12$ &  $14.98\pm0.23$ \\
  3 &   OM & 05-06-2018 & $14.72\pm0.21$ &  $14.70\pm0.11$ &  $14.64\pm0.16$ &  $15.06\pm0.44$ &  $15.86\pm0.27$ &  $15.40\pm0.21$ \\
  4 & UVOT & 14-06-2018 & $14.80\pm0.04$ &  $14.72\pm0.05$ &  $14.55\pm0.08$ &  $14.62\pm0.08$ &  $15.46\pm0.13$ &  $14.93\pm0.16$ \\
  5 & UVOT & 10-07-2018 & $15.07\pm0.04$ &  $14.99\pm0.06$ &  $14.85\pm0.06$ &  $15.02\pm0.10$ &  $15.94\pm0.15$ &  $15.38\pm0.16$ \\
  6 & UVOT & 24-07-2018 & $14.89\pm0.05$ &  $14.98\pm0.04$ &  $14.73\pm0.06$ &  $14.65\pm0.09$ &  $15.79\pm0.10$ &  $15.44\pm0.15$ \\
  7 & UVOT & 07-08-2018 & $15.04\pm0.05$ &  $14.65\pm0.07$ &  $14.89\pm0.07$ &  $15.36\pm0.09$ &  $16.02\pm0.13$ &  $15.45\pm0.17$ \\
  8 & UVOT & 23-08-2018 & $14.94\pm0.06$ &  $15.07\pm0.05$ &  $15.19\pm0.05$ &  $15.19\pm0.08$ &  $15.97\pm0.15$ &  $15.64\pm0.19$ \\
  9 & UVOT & 03-10-2018 & $15.54\pm0.09$ &  $15.19\pm0.13$ &  $15.49\pm0.13$ &  $15.50\pm0.30$ &  $16.54\pm0.31$ &  $15.85\pm0.45$ \\
 10 & UVOT & 19-10-2018 & $15.11\pm0.07$ &  $15.34\pm0.08$ &  $15.27\pm0.09$ &  $15.45\pm0.17$ &  $16.19\pm0.20$ &  $15.48\pm0.19$ \\
 11 & UVOT & 23-10-2018 & $15.22\pm0.07$ &  $15.30\pm0.06$ &  $15.40\pm0.09$ &  $15.42\pm0.15$ &  $16.40\pm0.16$ &  $15.88\pm0.18$ \\
 12 & UVOT & 21-11-2018 & $15.51\pm0.05$ &  $15.20\pm0.09$ &  $15.46\pm0.08$ &  $15.62\pm0.15$ &  $16.69\pm0.30$ &  $15.96\pm0.28$ \\
 13 & UVOT & 06-12-2018 & $15.22\pm0.07$ &  $15.27\pm0.07$ &  $15.07\pm0.11$ &  $15.52\pm0.15$ &  $16.41\pm0.22$ &  $15.79\pm0.23$ \\
 14 & UVOT & 12-12-2018 & $15.32\pm0.07$ &  $15.23\pm0.07$ &  $15.45\pm0.10$ &  $15.68\pm0.17$ &  $16.49\pm0.27$ &  $15.68\pm0.21$ \\
 15 & UVOT & 28-03-2019 & $15.80\pm0.05$ &  $15.40\pm0.09$ &  $15.32\pm0.11$ &  $15.31\pm0.17$ &  $16.54\pm0.13$ &  $15.80\pm0.23$ \\
 16 &   OM & 07-05-2019 & $15.84\pm0.24$ &  $15.77\pm0.15$ &  $15.56\pm0.13$ &  $15.87\pm0.20$ &  $16.69\pm0.25$ &  ---            \\
 17 &   OM & 02-11-2019 & ---            &  ---            &  ---            &  $16.19\pm0.34$ &  $16.80\pm0.22$ &  ---            \\
 18 &   OM & 03-05-2020 & ---            &  ---            &  ---            &  $16.48\pm0.35$ &  $16.81\pm0.23$ &  $16.29\pm0.25$ \\
 19 &   OM & 16-09-2020 & ---            &  ---            &  ---            &  $16.55\pm0.33$ &  $16.83\pm0.22$ &  $16.34\pm0.24$ \\
 20 &   OM & 12-01-2021 & ---            &  ---            &  ---            &  $16.72\pm0.34$ &  $16.96\pm0.25$ &  $16.38\pm0.27$ \\
     \enddata
      \tablecomments{Summary of integrated magnitudes for the Swift UVOT and XMM-Newton OM images.  Col. (1): Number of the broadband dataset. Col. (2): Instrument. Col. (3): Date of the observations.
 Cols. (4)--(9): Total integrated magnitude for the UVW2, UVM2, UVW1, $U$, $B$, and $V$ band, respectively. For observations with multiple exposures, we use the median magnitude. 
       }
       \end{deluxetable*}

%% file: host_phot.tex
%LCH 2023.12.28
%LCH 2024.08.12
%
%\startlongtable
\begin{deluxetable}{cc}
\tablecaption{Optical/UV Host Galaxy Flux of 1ES\,1927+654} \label{tab:hostflux}
\tabletypesize{\scriptsize}
\setlength{\tabcolsep}{20pt}
\tablehead{
      \colhead{Band}             &
      \colhead{Flux}    \\
       &
 ($10^{-15}\rm \, erg\,s^{-1}\,cm^{-2}$)
}
\startdata
      FUV  & $0.39\pm 0.13$  \\
      NUV  & $0.59\pm 0.16$  \\
      UVW2 & $0.62\pm 0.12$  \\
      UVM2 & $0.55\pm 0.15$  \\
      UVW1 & $0.59\pm 0.25$  \\
      $U$  & $0.63\pm 0.26$  \\
      $B$  & $1.03\pm 0.26$  \\
      $V$  & $1.01\pm 0.13$
     \enddata
      \tablecomments{The optical/UV flux of the host galaxy of 1ES\,1927+654. The GALEX FUV and NUV fluxes are derived from extrapolating the best-fit host galaxy model in Section~\ref{sec:intSED}. The XMM-Newton six-band fluxes are derived from image decomposition, as illustrated in Section~\ref{sec:xmmobs}.
       }
      \end{deluxetable}

%% file: tab_result.tex
%LCH 2023.12.28
%LCH 2024.08.12
%
\startlongtable
\begin{longrotatetable}
\begin{deluxetable*}{cccccccccccccccc}
%\tabletypesize{\tiny}
%\movetabledown=600mm
%\movetabledown=3mm
\setlength{\tabcolsep}{2.5pt}
\tablecaption{The Results of Broadband SED Fitting \label{tab:resubbf}}
\tablehead{
\colhead{No.} &
\colhead{$kT_\mathrm{in}$} &
\colhead{$R_\mathrm{in}$} &
\colhead{$\log{L_\mathrm{disk}}$} &
\colhead{$kT_\mathrm{bb}$} &
\colhead{$\log{L_\mathrm{bb}}$} &
\colhead{$kT_\mathrm{cor}$} &
\colhead{$\Gamma$} &
\colhead{$\log{L_\mathrm{cor}}$} &
\colhead{$E_\mathrm{Gau}$} &
\colhead{$\sigma_\mathrm{Gau}$} &
\colhead{$\log{L_\mathrm{Gau}}$} &
\colhead{$N_\mathrm{H}^{i}$} &
\colhead{$N_\mathrm{Nu}$} &
\colhead{$\log{L_\mathrm{bol}}$} &
\colhead{Stat./dof}  \\
 &
\colhead{(eV)} &
\colhead{($10^{12}\rm cm$)} &
\colhead{($\rm erg\,s^{-1}$)} &
\colhead{(eV)} &
\colhead{($\rm erg\,s^{-1}$)} &
\colhead{(keV)} &
 &
\colhead{($\rm erg\,s^{-1}$)} &
\colhead{(keV)} &
\colhead{(keV)} &
\colhead{($ \rm erg\,s^{-1}$)} &
\colhead{($10^{21} \; \rm cm^{-2}$)} &
 &
\colhead{($ \rm erg\,s^{-1}$)} &
  \\
\colhead{(1)}  &
\colhead{(2)} &
\colhead{(3)}   &
\colhead{(4)} &
\colhead{(5)} &
\colhead{(6)} &
\colhead{(7)} &
\colhead{(8)} &
\colhead{(9)}  &
\colhead{(10)} &
\colhead{(11)}  &
\colhead{(12)} &
\colhead{(13)}   &
\colhead{(14)} &
\colhead{(15)} &
\colhead{(16)}
}
\startdata
0&$52.8^{+1.7}_{-1.8}$&$0.97^{+0.08}_{-0.08}$&$43.98^{+0.07}_{-0.08}$&$135.09^{+3.17}_{-3.00}$&$42.83^{+0.05}_{-0.05}$&$71.90^{+86.24}_{-66.96}$&$2.64^{+0.03}_{-0.07}$&$43.09^{+0.01}_{-0.02}$&---&---&---&$0.70^{+0.10}_{-0.11}$&---&$44.06^{+0.06}_{-0.07}$&1.10\\
1&$30.0\pm5.0$\tablenotemark{$\dagger$}&$7.80^{+0.05}_{-0.05}$&$44.83^{+0.01}_{-0.01}$&$106.31^{+7.15}_{-9.06}$&$43.64^{+0.20}_{-0.10}$&$2.00\pm1.0$&$3.40\pm0.30$&$42.20^{+0.23}_{-0.45}$&$1.00\pm0.10$&$0.10\pm0.10$&$41.70^{+0.23}_{-0.37}$&$0.32^{+0.40}_{-0.22}$&---&$44.86^{+0.02}_{-0.01}$&2.65\\
2&$30.0\pm5.0$&$7.72^{+0.05}_{-0.05}$&$44.82^{+0.01}_{-0.01}$&$80.89^{+1.51}_{-0.67}$&$43.75^{+0.14}_{-0.15}$&$2.00\pm1.0$&$3.40\pm0.30$&$41.69^{+0.37}_{-0.54}$&$1.00\pm0.10$&$0.10\pm0.10$&$40.80^{+0.34}_{-0.54}$&$2.18^{+0.52}_{-0.50}$&---&$44.86^{+0.02}_{-0.02}$&2.89\\
3&$30.0\pm5.0$&$7.52^{+0.18}_{-0.39}$&$44.81^{+0.02}_{-0.05}$&$98.89^{+1.07}_{-1.04}$&$43.60^{+0.02}_{-0.03}$&$5.83^{+2.87}_{-3.20}$&$3.40\pm0.30$&$41.41^{+0.16}_{-0.30}$&$1.00^{+0.01}_{-0.01}$&$0.11^{+0.01}_{-0.01}$&$41.74^{+0.04}_{-0.04}$&$0.24^{+0.05}_{-0.05}$&---&$44.84^{+0.02}_{-0.04}$&2.53\\
4&$30.0\pm5.0$&$7.94^{+0.03}_{-0.03}$&$44.85^{+0.00}_{-0.00}$&$84.70^{+5.05}_{-5.17}$&$43.30^{+0.10}_{-0.10}$&$2.00\pm1.0$&$3.40\pm0.30$&$39.75^{+0.03}_{-0.03}$&---&---&---&$1.49^{+0.01}_{-0.01}$&---&$44.86^{+0.01}_{-0.01}$&58.0\\
5&$30.0\pm5.0$&$6.75^{+0.03}_{-0.03}$&$44.71^{+0.00}_{-0.00}$&$46.61^{+22.41}_{-9.25}$&$42.92^{+0.55}_{-0.63}$&$2.00\pm1.0$&$3.40\pm0.30$&$40.07^{+0.11}_{-0.21}$&---&---&---&$1.49^{+0.01}_{-0.01}$&---&$44.71^{+0.02}_{-0.01}$&77.1\\
6&$30.0\pm5.0$&$6.40^{+0.05}_{-0.05}$&$44.66^{+0.01}_{-0.01}$&$33.73^{+7.81}_{-2.75}$&$43.69^{+0.49}_{-0.82}$&$2.00\pm1.0$&$3.40\pm0.30$&$40.22^{+0.03}_{-0.08}$&---&---&---&$1.41^{+0.07}_{-0.15}$&---&$44.70^{+0.08}_{-0.04}$&5.58\\
7&$30.0\pm5.0$&$6.38^{+0.05}_{-0.04}$&$44.66^{+0.01}_{-0.01}$&$46.07^{+11.06}_{-6.40}$&$43.76^{+0.42}_{-0.53}$&$2.00\pm1.0$&$3.40\pm0.30$&$40.08^{+0.07}_{-0.09}$&---&---&---&$1.42^{+0.06}_{-0.12}$&---&$44.71^{+0.08}_{-0.04}$&3.82\\
8&$34.1^{+8.0}_{-3.1}$&$5.11^{+0.70}_{-1.26}$&$44.66^{+0.11}_{-0.25}$&$46.11^{+4.72}_{-3.59}$&$44.32^{+0.25}_{-0.29}$&$2.00\pm1.0$&$3.40\pm0.30$&$41.79^{+0.23}_{-0.48}$&$0.83^{+0.04}_{-0.02}$&$0.09^{+0.02}_{-0.02}$&$41.54^{+0.15}_{-0.22}$&$1.39^{+0.08}_{-0.16}$&---&$44.82^{+0.16}_{-0.26}$&1.65\\
9&$32.7^{+4.8}_{-2.1}$&$4.49^{+0.42}_{-0.76}$&$44.48^{+0.08}_{-0.16}$&$84.82^{+2.60}_{-2.49}$&$44.84^{+0.04}_{-0.08}$&$2.00\pm1.0$&$3.70\pm0.30$&$43.02^{+0.20}_{-0.37}$&$1.00\pm0.10$&$0.24^{+0.35}_{-0.13}$&$42.04^{+0.26}_{-0.49}$&$3.63^{+0.21}_{-0.22}$&---&$45.00^{+0.05}_{-0.10}$&2.12\\
10&$35.6^{+7.0}_{-4.4}$&$4.47^{+0.87}_{-0.96}$&$44.62^{+0.15}_{-0.21}$&$83.11^{+3.68}_{-2.18}$&$44.64^{+0.15}_{-0.18}$&$2.00\pm1.0$&$3.70\pm0.30$&$43.57^{+0.24}_{-0.44}$&$1.00\pm0.10$&$0.22^{+0.03}_{-0.03}$&$43.14^{+0.09}_{-0.13}$&$1.82^{+0.34}_{-0.39}$&---&$44.95^{+0.15}_{-0.20}$&1.03\\
11&$34.7^{+6.8}_{-3.6}$&$4.58^{+0.72}_{-0.98}$&$44.60^{+0.13}_{-0.21}$&$91.44^{+11.72}_{-7.42}$&$44.38^{+0.29}_{-0.29}$&$2.00\pm1.0$&$3.70\pm0.30$&$42.98^{+0.20}_{-0.46}$&$1.00\pm0.10$&$0.15^{+0.17}_{-0.04}$&$42.53^{+0.15}_{-0.24}$&$1.77^{+0.65}_{-0.65}$&---&$44.81^{+0.19}_{-0.24}$&1.43\\
12&$36.0^{+8.8}_{-4.6}$&$4.10^{+0.82}_{-1.03}$&$44.56^{+0.16}_{-0.25}$&$103.09^{+9.77}_{-9.25}$&$44.12^{+0.31}_{-0.24}$&$2.00\pm1.0$&$4.00\pm0.30$&$43.01^{+0.16}_{-0.30}$&$0.90\pm0.10$&$0.10\pm0.10$&$42.56^{+0.16}_{-0.24}$&$1.54^{+0.74}_{-0.56}$&---&$44.70^{+0.20}_{-0.25}$&1.18\\
13&$36.8^{+8.0}_{-5.1}$&$4.06^{+0.90}_{-0.93}$&$44.59^{+0.17}_{-0.23}$&$106.29^{+8.44}_{-8.30}$&$43.59^{+0.21}_{-0.17}$&$2.00\pm1.0$&$4.00\pm0.30$&$42.29^{+0.25}_{-0.44}$&$0.80\pm0.10$&$0.20\pm0.10$&$42.64^{+0.10}_{-0.12}$&$0.97^{+0.49}_{-0.42}$&---&$44.63^{+0.18}_{-0.22}$&1.08\\
14&$49.9^{+1.3}_{-1.6}$&$2.67^{+0.12}_{-0.09}$&$44.76^{+0.04}_{-0.03}$&$118.95^{+1.38}_{-1.57}$&$43.97^{+0.02}_{-0.03}$&$1.93^{+0.63}_{-0.39}$&$3.80^{+0.10}_{-0.11}$&$43.82^{+0.03}_{-0.03}$&$1.04^{+0.01}_{-0.01}$&$0.17^{+0.01}_{-0.01}$&$42.69^{+0.04}_{-0.04}$&$0.40^{+0.03}_{-0.04}$&$0.77^{+0.05}_{-0.04}$&$44.87^{+0.04}_{-0.03}$&1.19\\
15&$50.0\pm5.0$&$2.34^{+0.02}_{-0.03}$&$44.64^{+0.01}_{-0.01}$&$117.77^{+14.64}_{-10.72}$&$44.05^{+0.10}_{-0.12}$&$2.00\pm1.0$&$4.00\pm0.30$&$43.75^{+0.16}_{-0.33}$&$1.00\pm0.10$&$0.19^{+0.07}_{-0.04}$&$42.82^{+0.14}_{-0.23}$&$0.95^{+0.18}_{-0.28}$&---&$44.79^{+0.04}_{-0.05}$&1.47\\
16&$54.8^{+3.4}_{-3.7}$&$1.67^{+0.16}_{-0.13}$&$44.51^{+0.08}_{-0.07}$&$114.04^{+5.80}_{-5.65}$&$43.48^{+0.18}_{-0.38}$&$6.89^{+19.87}_{-3.44}$&$4.28^{+0.11}_{-0.15}$&$44.12^{+0.05}_{-0.05}$&$0.99^{+0.02}_{-0.02}$&$0.19^{+0.01}_{-0.01}$&$42.80^{+0.08}_{-0.08}$&$0.36^{+0.04}_{-0.05}$&$0.98^{+0.05}_{-0.04}$&$44.69^{+0.08}_{-0.08}$&1.16\\
17&$74.7^{+2.0}_{-1.9}$&$1.06^{+0.06}_{-0.06}$&$44.66^{+0.05}_{-0.05}$&$140.74^{+2.95}_{-2.93}$&$42.87^{+0.40}_{-0.55}$&$14.53^{+60.65}_{-9.58}$&$4.42^{+0.09}_{-0.15}$&$44.36^{+0.01}_{-0.02}$&$1.08^{+0.02}_{-0.02}$&$0.08^{+0.02}_{-0.02}$&$42.01^{+0.11}_{-0.11}$&$0.47^{+0.04}_{-0.04}$&$0.70^{+0.04}_{-0.04}$&$44.84^{+0.04}_{-0.05}$&1.04\\
18&$63.4^{+3.1}_{-3.1}$&$1.14^{+0.09}_{-0.08}$&$44.44^{+0.06}_{-0.07}$&$137.88^{+4.88}_{-4.54}$&$42.94^{+0.40}_{-0.56}$&$4.23^{+5.24}_{-1.21}$&$4.02^{+0.15}_{-0.12}$&$44.20^{+0.03}_{-0.03}$&$0.97^{+0.04}_{-0.04}$&$0.12^{+0.05}_{-0.04}$&$41.96^{+0.18}_{-0.20}$&$0.26^{+0.08}_{-0.09}$&$0.77^{+0.05}_{-0.04}$&$44.64^{+0.06}_{-0.06}$&0.99\\
19&$57.5^{+2.7}_{-2.4}$&$1.20^{+0.08}_{-0.08}$&$44.31^{+0.06}_{-0.06}$&$153.30^{+3.65}_{-3.65}$&$43.45^{+0.05}_{-0.04}$&$111.23^{+60.28}_{-62.17}$&$2.96^{+0.04}_{-0.04}$&$43.53^{+0.02}_{-0.02}$&---&---&---&$0.22^{+0.14}_{-0.13}$&$0.85^{+0.04}_{-0.03}$&$44.42^{+0.05}_{-0.05}$&0.86\\
20&$51.5^{+1.5}_{-1.4}$&$1.11^{+0.08}_{-0.08}$&$44.05^{+0.06}_{-0.06}$&$143.11^{+2.99}_{-3.16}$&$42.64^{+0.04}_{-0.04}$&$107.49^{+63.37}_{-62.46}$&$2.42^{+0.05}_{-0.05}$&$42.81^{+0.01}_{-0.01}$&---&---&---&$0.61^{+0.12}_{-0.12}$&$1.08^{+0.05}_{-0.05}$&$44.09^{+0.06}_{-0.06}$&1.13
\enddata
\tablecomments{MCMC results of broadband SED fitting by the final model (Section~\ref{sec:finalmodel}). Col. (1): Number of the broadband dataset. Col. (2): Effective temperature of the thin disk at the inner radius. Col. (3): Radius of the inner boundary of the thin disk. Col. (4): Integrated luminosity of the thin disk. Col. (5): Temperature of the blackbody. Col. (6): Integrated luminosity of the blackbody. Col. (7): Electron temperature of the hard X-ray corona. Col. (8): The hard X-ray photon index. Col. (9): Integrated luminosity of the corona. Col. (10): Central energy of the broad Gaussian line at $\sim 1\,$keV. Col. (11): Width of the broad Gaussian line. Col. (12): Integrated luminosity of the broad Gaussian line. Col. (13): Column density of the intrinsic neutral absorber. Col. (14): Cross-calibration constant applied to the NuSTAR data. Col. (15): Bolometric luminosity of the AGN. Col. (16): Statistics divided by the degree of freedom from the best-fit model.
\tablenotetext{ \dagger }{Uncertainties shown after $\pm$ were not derived from the MCMC sampling, but based on the selected parameter grid (see details in Section~\ref{sec:finalmodel}).}
}
\end{deluxetable*}
\end{longrotatetable}

%% file: tab_bbfit.tex
%LCH 2024.08.12

\startlongtable
\begin{longrotatetable}
\begin{deluxetable*}{ccccccccccc}
%\tabletypesize{\tiny}
%\movetabledown=600mm
%\movetabledown=3mm
\setlength{\tabcolsep}{2.5pt}
\tablecaption{The Results of Broadband SED Fitting by the Blackbody Model \label{tab:resublackbody}}
\tablehead{
\colhead{No.} &
\colhead{$kT_\mathrm{bb}$} &
\colhead{$\log{N_\mathrm{bb}}$} &
\colhead{$\Gamma$} &
\colhead{$\log{N_\mathrm{pow}}$} &
\colhead{$E_\mathrm{Gau}$} &
\colhead{$\sigma_\mathrm{Gau}$} &
\colhead{$\log{N_\mathrm{Gau}}$} &
\colhead{$N_\mathrm{H}^{i}$} &
\colhead{$N_\mathrm{Nu}$} &
\colhead{Stat./dof}  \\
 &
\colhead{(eV)} &
 &
\colhead{(keV)} &
\colhead{($\rm photons \, keV^{-1} \, cm^{-2}$)} &
\colhead{(keV)} &
\colhead{(keV)} &
\colhead{($\rm photons \, cm^{-2}$)} &
\colhead{($10^{20} \; \rm cm^{-2}$)} &
 &
  \\
\colhead{(1)}  &
\colhead{(2)} &
\colhead{(3)}   &
\colhead{(4)} &
\colhead{(5)} &
\colhead{(6)} &
\colhead{(7)} &
\colhead{(8)} &
\colhead{(9)}  &
\colhead{(10)} &
\colhead{(11)}   
}
\startdata
0  & $185.07^{+3.96}_{-4.03}$      & $-4.46^{+0.04}_{-0.04}$ & $2.53^{+0.04}_{-0.05}$ & $-2.54^{+0.01}_{-0.01}$ & ---                    & ---                    & ---                     & $3.38^{+0.62}_{-0.59}$ & ---                    &  1.22\\
1  & $131.74^{+2.93}_{-3.52}$      & $-3.60^{+0.05}_{-0.05}$ & $4.00\pm0.30         $ & $-3.10^{+0.01}_{-0.02}$ & $1.00\pm0.10$          & $0.30\pm0.10$          & $-4.56^{+0.01}_{-0.02}$ & $0.85^{+0.13}_{-0.13}$ & ---                    &  53.3\\
2  & $80.20^{+5.46}_{-7.08}$       & $-3.85^{+0.21}_{-0.09}$ & $4.00\pm0.30         $ & $-3.81^{+0.23}_{-0.45}$ & $1.00\pm0.10$          & $0.20\pm0.10$          & $-4.12^{+0.24}_{-0.39}$ & $1.46^{+3.73}_{-1.65}$ & ---                    &  84.3\\
3  & $99.29^{+1.59}_{-0.95}$       & $-3.32^{+0.15}_{-0.14}$ & $2.80\pm0.30         $ & $-4.25^{+0.35}_{-0.57}$ & $1.00^{+0.01}_{-0.01}$ & $0.11^{+0.01}_{-0.01}$ & $-3.42^{+0.39}_{-0.54}$ & $1.81^{+0.36}_{-0.38}$ & ---                    &  45.4\\
4  & $101.19^{+1.12}_{-1.01}$      & $-4.23^{+0.02}_{-0.03}$ & $4.00\pm0.30         $ & $-8.48^{+0.13}_{-0.23}$ & ---                    & ---                    & ---                     & $0.93^{+0.21}_{-0.19}$ & ---                    &  327 \\
5  & $201.13^{+12.70}_{-14.01}$    & $-5.70^{+0.09}_{-0.09}$ & $3.10\pm0.30         $ & $-4.78^{+0.03}_{-0.02}$ & ---                    & ---                    & ---                     & $0.29^{+0.00}_{-0.00}$ & ---                    &  732 \\
6  & $349.56^{+167.33}_{-83.02}$   & $-7.77^{+0.54}_{-0.67}$ & $4.00\pm0.30         $ & $-5.45^{+0.12}_{-0.19}$ & ---                    & ---                    & ---                     & $0.02^{+0.00}_{-0.00}$ & ---                    &  236 \\
7  & $101.82^{+23.43}_{-8.07}$     & $-8.12^{+0.49}_{-0.86}$ & $4.00\pm0.30         $ & $-4.46^{+0.03}_{-0.07}$ & ---                    & ---                    & ---                     & $2.87^{+0.13}_{-0.26}$ & ---                    &  144 \\
8  & $103.00^{+23.03}_{-15.53}$    & $-5.58^{+0.41}_{-0.50}$ & $4.00\pm0.30         $ & $-3.82^{+0.07}_{-0.09}$ & $0.94^{+0.05}_{-0.06}$ & $0.10^{+0.04}_{-0.04}$ & $-5.82^{+0.07}_{-0.08}$ & $2.14^{+0.09}_{-0.18}$ & ---                    &  51.0\\
9  & $122.32^{+11.81}_{-9.60}$     & $-3.36^{+0.27}_{-0.26}$ & $2.80\pm0.30         $ & $-3.02^{+0.20}_{-0.48}$ & $1.00\pm0.10$          & $0.07^{+0.05}_{-0.03}$ & $-3.55^{+0.24}_{-0.45}$ & $2.79^{+0.17}_{-0.35}$ & ---                    &  16.4\\
10 & $105.00^{+3.21}_{-2.68}$      & $-3.03^{+0.04}_{-0.09}$ & $2.80\pm0.30         $ & $-2.31^{+0.19}_{-0.38}$ & $1.00\pm0.10$          & $0.23^{+0.06}_{-0.06}$ & $-2.21^{+0.17}_{-0.37}$ & $3.21^{+0.17}_{-0.18}$ & ---                    &  15.0\\
11 & $139.92^{+5.62}_{-3.11}$      & $-3.43^{+0.14}_{-0.17}$ & $4.00\pm0.30         $ & $-2.65^{+0.25}_{-0.48}$ & $1.00\pm0.10$          & $0.10^{+0.14}_{-0.05}$ & $-3.22^{+0.23}_{-0.39}$ & $2.75^{+0.49}_{-0.65}$ & ---                    &  22.5\\
12 & $134.39^{+18.09}_{-10.85}$    & $-3.39^{+0.29}_{-0.29}$ & $3.40\pm0.30         $ & $-3.15^{+0.21}_{-0.48}$ & $0.90\pm0.10$          & $0.10\pm0.10$          & $-3.70^{+0.20}_{-0.42}$ & $0.10^{+0.03}_{-0.04}$ & ---                    &  21.0\\
13 & $116.72^{+9.50}_{-9.29}$      & $-3.58^{+0.33}_{-0.25}$ & $2.80\pm0.30         $ & $-4.11^{+0.15}_{-0.32}$ & $0.90\pm0.10$          & $0.20\pm0.10$          & $-3.76^{+0.17}_{-0.32}$ & $1.66^{+0.81}_{-0.63}$ & ---                    &  23.1\\
14 & $143.92^{+10.83}_{-11.76}$    & $-3.02^{+0.24}_{-0.20}$ & $3.46^{+0.28}_{-0.21}$ & $-1.90^{+0.25}_{-0.47}$ & $1.02^{+0.02}_{-0.02}$ & $0.21^{+0.02}_{-0.02}$ & $-2.48^{+0.22}_{-0.47}$ & $3.45^{+1.67}_{-1.26}$ & $0.84^{+0.06}_{-0.06}$ &  6.58\\
15 & $155.66^{+2.18}_{-1.79}$      & $-3.10^{+0.02}_{-0.03}$ & $3.40\pm0.30         $ & $-2.04^{+0.03}_{-0.03}$ & $1.00\pm0.10$          & $0.14^{+0.06}_{-0.05}$ & $-2.63^{+0.03}_{-0.03}$ & $2.93^{+0.22}_{-0.31}$ & ---                    &  13.0\\
16 & $173.78^{+23.04}_{-15.67}$    & $-3.14^{+0.11}_{-0.11}$ & $3.54^{+0.12}_{-0.31}$ & $-1.72^{+0.16}_{-0.33}$ & $0.99^{+0.02}_{-0.02}$ & $0.16^{+0.03}_{-0.03}$ & $-2.65^{+0.18}_{-0.36}$ & $2.59^{+0.65}_{-0.90}$ & $0.91^{+0.06}_{-0.05}$ &  63.8\\
17 & $210.33^{+10.90}_{-10.23}$    & $-2.85^{+0.18}_{-0.40}$ & $3.45^{+0.09}_{-0.20}$ & $-1.56^{+0.06}_{-0.05}$ & $1.05^{+0.02}_{-0.02}$ & $0.09^{+0.05}_{-0.03}$ & $-5.34^{+0.05}_{-0.06}$ & $3.52^{+0.28}_{-0.43}$ & $0.66^{+0.04}_{-0.04}$ &  3.05\\
18 & $205.88^{+4.74}_{-3.67}$      & $-3.01^{+0.42}_{-0.62}$ & $3.31^{+0.20}_{-0.32}$ & $-1.64^{+0.01}_{-0.02}$ & $0.98^{+0.04}_{-0.04}$ & $0.04^{+0.03}_{-0.02}$ & $-3.77^{+0.01}_{-0.02}$ & $3.25^{+0.27}_{-0.28}$ & $0.68^{+0.05}_{-0.05}$ &  2.89\\
19 & $177.66^{+6.47}_{-6.18}$      & $-3.57^{+0.48}_{-0.50}$ & $2.85^{+0.07}_{-0.13}$ & $-1.97^{+0.03}_{-0.03}$ & ---                    & ---                    & ---                     & $1.31^{+0.80}_{-1.08}$ & $0.80^{+0.04}_{-0.05}$ &  1.56\\
20 & $156.04^{+3.60}_{-3.49}$      & $-4.78^{+0.05}_{-0.04}$ & $2.53^{+0.05}_{-0.05}$ & $-2.78^{+0.02}_{-0.02}$ & ---                    & ---                    & ---                     & $1.50^{+1.07}_{-0.90}$ & $1.20^{+0.06}_{-0.06}$ &  1.36
\enddata
\tablecomments{MCMC results of broadband SED fitting by the blackbody model (Section~\ref{sec:2cbbody}). Col. (1): Number of the broadband dataset. Col. (2): Temperature of the blackbody. Col. (3): Normalized luminosity of the blackbody, in units of $L_{39}/D_{10}^2$, where $L_{39}$ is the source luminosity in units of $10^{39} \rm erg\,s^{-1}$ and $D_{10}^2$ is the distance to the source in units of 10 kpc. (4): The hard X-ray photon index. Col. (5): Normalized flux of the corona at 1 keV. Col. (6): Central energy of the broad Gaussian line at $\sim 1\,$keV. Col. (7): Width of the broad Gaussian line. Col. (8): Integrated flux of the broad Gaussian line. Col. (9): Column density of the intrinsic neutral absorber. Col. (10): Cross-calibration constant applied to the NuSTAR data. Col. (11): Statistics divided by the degree of freedom from the best-fit model.
}
\end{deluxetable*}
\end{longrotatetable}

%% file: tab_dbbfit.tex
%LCH 2024.08.12

\startlongtable
\begin{longrotatetable}
\begin{deluxetable*}{ccccccccccc}
%\tabletypesize{\tiny}
%\movetabledown=600mm
%\movetabledown=3mm
\setlength{\tabcolsep}{2.5pt}
\tablecaption{The Results of Broadband SED Fitting by the Thin Disk Model \label{tab:resudbb}}
\tablehead{
\colhead{No.} &
\colhead{$kT_\mathrm{in}$} &
\colhead{$\log{N_\mathrm{dbb}}$} &
\colhead{$\Gamma$} &
\colhead{$\log{N_\mathrm{pow}}$} &
\colhead{$E_\mathrm{Gau}$} &
\colhead{$\sigma_\mathrm{Gau}$} &
\colhead{$\log{N_\mathrm{Gau}}$} &
\colhead{$N_\mathrm{H}^{i}$} &
\colhead{$N_\mathrm{Nu}$} &
\colhead{Stat./dof}  \\
 &
\colhead{(eV)} &
 &
\colhead{(keV)} &
\colhead{($\rm photons \, keV^{-1} \, cm^{-2}$)} &
\colhead{(keV)} &
\colhead{(keV)} &
\colhead{($\rm photons \, cm^{-2}$)} &
\colhead{($10^{20} \; \rm cm^{-2}$)} &
 &
  \\
\colhead{(1)}  &
\colhead{(2)} &
\colhead{(3)}   &
\colhead{(4)} &
\colhead{(5)} &
\colhead{(6)} &
\colhead{(7)} &
\colhead{(8)} &
\colhead{(9)}  &
\colhead{(10)} &
\colhead{(11)}   
}
\startdata
0  & $46.21^{+1.11}_{-1.10}$      & $5.95^{+0.04}_{-0.01}$  & $2.65^{+0.05}_{-0.05}$ & $-2.47^{+0.01}_{-0.01}$ & ---                    & ---                   &  ---                   & $1.94^{+0.42}_{-0.39}$    & ---                    &  1.11\\
1  & $84.55^{+2.04}_{-2.07}$      & $6.43^{+0.05}_{-0.02}$  & $4.30\pm0.30         $ & $-2.51^{+0.01}_{-0.02}$ &          $0.80\pm0.10$ &         $0.40\pm0.10$ &$-4.14^{+0.01}_{-0.02}$ & $35.97^{+4.58}_{-5.09}$   & ---                    &  2.78\\
2  & $64.13^{+4.51}_{-5.06}$      & $6.74^{+0.20}_{-0.48}$  & $4.30\pm0.30         $ & $-3.24^{+0.23}_{-0.40}$ &          $1.00\pm0.12$ &         $0.30\pm0.10$ &$-4.25^{+0.22}_{-0.34}$ & $33.94^{+39.11}_{-24.47}$ & ---                    &  2.69\\
3  & $84.74^{+1.52}_{-0.65}$      & $6.34^{+0.15}_{-0.43}$  & $2.50\pm0.30         $ & $-3.81^{+0.38}_{-0.53}$ & $1.00^{+0.01}_{-0.01}$ &$0.11^{+0.01}_{-0.01}$ &$-3.98^{+0.37}_{-0.51}$ & $33.79^{+8.87}_{-5.80}$   & ---                    &  38.8\\
4  & $1.71^{+0.02}_{-0.02}$       & $11.45^{+0.02}_{-0.28}$ & $4.30\pm0.30         $ & $-3.69^{+0.14}_{-0.29}$ &                    --- &                   --- &                    --- & $0.01^{+0.01}_{-0.01}$    & ---                    &  5.23\\
5  & $2.05^{+0.11}_{-0.14}$       & $10.98^{+0.08}_{-0.04}$ & $4.30\pm0.30         $ & $-4.58^{+0.03}_{-0.03}$ &                    --- &                   --- &                    --- & $0.01^{+0.01}_{-0.01}$    & ---                    &  11.8\\
6  & $79.02^{+39.22}_{-15.50}$    & $6.35^{+0.51}_{-0.19}$  & $4.30\pm0.30         $ & $-8.26^{+0.11}_{-0.24}$ &                    --- &                   --- &                    --- & $85.93^{+0.64}_{-0.54}$   & ---                    &  5.61\\
7  & $3.60^{+0.75}_{-0.25}$       & $10.03^{+0.49}_{-0.08}$ & $4.30\pm0.30         $ & $-4.18^{+0.03}_{-0.08}$ &                    --- &                   --- &                    --- & $0.01^{+0.01}_{-0.01}$    & ---                    &  1.31\\
8  & $36.56^{+9.35}_{-4.59}$      & $7.15^{+0.49}_{-0.09}$  & $4.30\pm0.30         $ & $-3.76^{+0.07}_{-0.08}$ & $0.96^{+0.04}_{-0.08}$ &$0.10^{+0.05}_{-0.04}$ &$-4.67^{+0.06}_{-0.08}$ & $0.01^{+0.01}_{-0.01}$    & ---                    &  1.96\\
9  & $101.71^{+12.23}_{-7.59}$    & $5.81^{+0.26}_{-0.53}$  & $3.70\pm0.30         $ & $-2.58^{+0.23}_{-0.49}$ &          $0.90\pm0.10$ &$0.06^{+0.06}_{-0.03}$ &$-4.41^{+0.19}_{-0.33}$ & $36.49^{+2.12}_{-4.37}$   & ---                    &  2.01\\
10 & $5.55^{+0.20}_{-0.17}$       & $9.33^{+0.04}_{-0.45}$  & $3.70\pm0.30         $ & $-1.71^{+0.20}_{-0.32}$ &          $0.90\pm0.10$ &$0.21^{+0.07}_{-0.08}$ &$-3.13^{+0.18}_{-0.37}$ & $9.76^{+0.49}_{-0.53}$    & ---                    &  1.03\\
11 & $99.47^{+4.63}_{-2.53}$      & $5.93^{+0.15}_{-0.44}$  & $4.30\pm0.30         $ & $-2.22^{+0.23}_{-0.34}$ &          $1.10\pm0.10$ &$0.15^{+0.10}_{-0.05}$ &$-3.24^{+0.20}_{-0.37}$ & $32.17^{+6.14}_{-6.66}$   & ---                    &  1.39\\
12 & $99.51^{+13.25}_{-8.15}$     & $5.88^{+0.26}_{-0.54}$  & $4.30\pm0.30         $ & $-2.42^{+0.20}_{-0.43}$ &          $0.90\pm0.10$ &         $0.20\pm0.10$ &$-3.56^{+0.21}_{-0.36}$ & $32.34^{+14.10}_{-12.22}$ & ---                    &  1.16\\
13 & $87.97^{+8.71}_{-7.76}$      & $6.04^{+0.30}_{-0.27}$  & $4.30\pm0.30         $ & $-2.71^{+0.15}_{-0.33}$ &          $0.70\pm0.10$ &         $0.10\pm0.10$ &$-3.31^{+0.12}_{-0.27}$ & $36.41^{+18.79}_{-11.66}$ & ---                    &  1.17\\
14 & $217.41^{+15.20}_{-19.07}$   & $3.53^{+0.23}_{-0.45}$  & $2.04^{+0.29}_{-0.46}$ & $-2.57^{+0.25}_{-0.38}$ & $1.01^{+0.02}_{-0.02}$ &$0.20^{+0.02}_{-0.02}$ &$-2.92^{+0.26}_{-0.41}$ & $0.01^{+0.01}_{-0.01}$    & $0.87^{+0.05}_{-0.06}$ &  8.60\\
15 & $126.18^{+1.23}_{-2.06}$     & $5.50^{+0.02}_{-0.03}$  & $3.10\pm0.30         $ & $-1.92^{+0.03}_{-0.03}$ &          $1.00\pm0.08$ &$0.17^{+0.07}_{-0.05}$ &$-2.77^{+0.03}_{-0.04}$ & $30.73^{+2.55}_{-2.89}$   & ---                    &  1.54\\
16 & $237.14^{+30.87}_{-19.12}$   & $3.40^{+0.10}_{-0.32}$  & $1.97^{+0.27}_{-0.25}$ & $-2.39^{+0.20}_{-0.31}$ & $1.03^{+0.02}_{-0.02}$ &$0.15^{+0.03}_{-0.03}$ &$-2.75^{+0.19}_{-0.35}$ & $0.01^{+0.01}_{-0.01}$    & $0.92^{+0.05}_{-0.05}$ &  7.84\\
17 & $292.00^{+16.74}_{-14.86}$   & $3.25^{+0.17}_{-0.05}$  & $2.98^{+0.15}_{-0.14}$ & $-5.57^{+0.05}_{-0.04}$ & $1.01^{+0.02}_{-0.02}$ &$0.09^{+0.04}_{-0.02}$ &$-4.84^{+0.05}_{-0.06}$ & $0.01^{+0.01}_{-0.01}$    & $0.83^{+0.07}_{-0.06}$ &  9.53\\
18 & $298.08^{+7.32}_{-4.62}$     & $3.07^{+0.39}_{-0.02}$  & $2.98^{+0.11}_{-0.12}$ & $-5.51^{+0.01}_{-0.02}$ & $0.99^{+0.03}_{-0.04}$ &$0.06^{+0.02}_{-0.02}$ &$-3.97^{+0.01}_{-0.02}$ & $0.01^{+0.01}_{-0.01}$    & $0.77^{+0.06}_{-0.06}$ &  8.98\\
19 & $310.79^{+9.73}_{-10.41}$    & $2.52^{+0.45}_{-0.03}$  & $2.98^{+0.13}_{-0.13}$ & $-5.51^{+0.03}_{-0.03}$ &                    --- &                   --- &                    --- & $0.01^{+0.01}_{-0.01}$    & $0.82^{+0.03}_{-0.05}$ &  10.6\\
20 & $236.76^{+5.30}_{-5.72}$     & $1.88^{+0.05}_{-0.02}$  & $1.95^{+0.06}_{-0.05}$ & $-3.01^{+0.02}_{-0.02}$ &                    --- &                   --- &                    --- & $0.01^{+0.01}_{-0.01}$    & $1.18^{+0.05}_{-0.05}$ &  5.08
\enddata
\tablecomments{MCMC results of broadband SED fitting by the thin disk model (Section~\ref{sec:2cdisk}). Col. (1): Number of the broadband dataset. Col. (2): Effective temperature of the thin disk at the inner radius. Col. (3): Normalized luminosity of the thin disk,  $N_\mathrm{disk} \equiv (R_\mathrm{in}/D_{10})^2 \cos i$, detailed in Section~\ref{sec:2cdisk}. (4): The hard X-ray photon index. Col. (5): Normalized flux of the corona at 1 keV. Col. (6): Central energy of the broad Gaussian line at $\sim 1\,$keV. Col. (7): Width of the broad Gaussian line. Col. (8): Integrated flux of the broad Gaussian line. Col. (9): Column density of the intrinsic neutral absorber. Col. (10): Cross-calibration constant applied to the NuSTAR data. Col. (11): Statistics divided by the degree of freedom from the best-fit model.
}
\end{deluxetable*}
\end{longrotatetable}

%% file: tab_slimfit.tex
%LCH 2024.08.12

\startlongtable
\begin{longrotatetable}
\begin{deluxetable*}{cccccccccccccc}
%\tabletypesize{\tiny}
%\movetabledown=600mm
%\movetabledown=3mm
\setlength{\tabcolsep}{2.5pt}
\tablecaption{The Results of Broadband SED Fitting by the Warm Corona Model \label{tab:resuslim}}
\tablehead{
\colhead{No.} &
\colhead{$\log{\dot{m}}$} &
\colhead{$kT_\mathrm{hot}$} &
\colhead{$R_\mathrm{out}^\mathrm{hot}$} &
\colhead{$\Gamma_\mathrm{hot}$} &
\colhead{$kT_\mathrm{warm}$} &
\colhead{$R_\mathrm{out}^\mathrm{warm}$} &
\colhead{$\Gamma_\mathrm{warm}$} &
\colhead{$E_\mathrm{Gau}$} &
\colhead{$\sigma_\mathrm{Gau}$} &
\colhead{$\log{N_\mathrm{Gau}}$} &
\colhead{$N_\mathrm{H}^{i}$} &
\colhead{$N_\mathrm{Nu}$} &
\colhead{Stat./dof}  \\
 &
 &
\colhead{(keV)} &
\colhead{($R_g$)} &
&
\colhead{(eV)} &
\colhead{($R_g$)} &
 &
\colhead{(keV)} &
\colhead{(keV)} &
 &
\colhead{($10^{21} \; \rm cm^{-2}$)} &
 &
  \\
\colhead{(1)}  &
\colhead{(2)} &
\colhead{(3)}   &
\colhead{(4)} &
\colhead{(5)} &
\colhead{(6)} &
\colhead{(7)} &
\colhead{(8)} &
\colhead{(9)}  &
\colhead{(10)} &
\colhead{(11)}  &
\colhead{(12)}  &
\colhead{(13)}  &
\colhead{(14)}    
}
\startdata
0  & $-0.08^{+0.05}_{-0.04}$   & $92.10^{+77.12}_{-45.23} $ & $5.27^{+0.11}_{-0.11}$  & $2.10^{+0.06}_{-0.05}$ & $143.09^{+2.55}_{-2.74}$   & 10.54\tablenotemark{$\dagger$} & $2.00^{+0.04}_{-0.05}$ & ---                    & ---                   &  ---                   & $15.95^{+2.84}_{-2.77}$   & ---                    &  2.43\\
1  &  $1.08^{+0.15}_{-0.09}$   & $2.00\pm1.0              $ & $2.53^{+0.05}_{-0.05}$  & $3.70\pm0.30         $ & $238.14^{+4.65}_{-5.12}$   & 5.06                           & $4.70^{+0.12}_{-0.11}$ &          $1.10\pm0.10$ &         $0.10\pm0.10$ &$-4.02^{+0.05}_{-0.02}$ & $13.61^{+2.14}_{-1.94}$   & ---                    &  1.66\\
2  &  $1.06^{+0.20}_{-0.28}$   & $2.00\pm1.0              $ & $2.68^{+0.16}_{-0.26}$  & $4.30\pm0.30         $ & $60.24^{+3.24}_{-4.27}$    & 5.37                           & $1.01^{+0.06}_{-0.09}$ &          $1.00\pm0.10$ &         $0.10\pm0.10$ &$-3.97^{+0.18}_{-0.17}$ & $24.20^{+27.01}_{-14.68}$ & ---                    &  1.28\\
3  &  $1.03^{+0.34}_{-0.38}$   & $2.08^{+1.75}_{-1.20}    $ & $2.53^{+0.05}_{-0.02}$  & $2.20\pm0.30         $ & $138.59^{+2.67}_{-0.91}$   & 5.06                           & $4.68^{+0.08}_{-0.05}$ & $1.00^{+0.01}_{-0.01}$ &$0.10^{+0.01}_{-0.01}$ &$-3.41^{+0.14}_{-0.19}$ & $13.95^{+4.08}_{-3.50}$   & ---                    &  4.52\\
4  &  $1.16^{+0.32}_{-0.30}$   & $2.00\pm1.0              $ & $2.53^{+0.03}_{-0.03}$  & $4.30\pm0.30         $ & $60.17^{+0.68}_{-0.62}$    & 5.06                           & $4.70^{+0.05}_{-0.05}$ &                    --- &                   --- &                    --- & $45.00^{+10.08}_{-10.07}$ & ---                    &  49.0\\
5  &  $1.04^{+0.09}_{-0.08}$   & $2.00\pm1.0              $ & $24.31^{+6.44}_{-5.49}$ & $4.30\pm0.30         $ & $76.15^{+3.94}_{-5.17}$    & 48.62                          & $1.00^{+0.06}_{-0.07}$ &                    --- &                   --- &                    --- & $65.00^{+0.43}_{-0.48}$   & ---                    &  69.9\\
6  &  $0.75^{+0.49}_{-0.43}$   & $2.00\pm1.0              $ & $2.89^{+1.25}_{-0.62}$  & $4.00\pm0.30         $ & $53.93^{+27.77}_{-10.77}$  & 5.78                           & $3.00^{+1.36}_{-0.67}$ &                    --- &                   --- &                    --- & $45.99^{+0.27}_{-0.40}$   & ---                    &  7.47\\
7  &  $0.83^{+0.49}_{-0.48}$   & $2.00\pm1.0              $ & $2.84^{+0.73}_{-0.24}$  & $2.20\pm0.30         $ & $55.82^{+15.14}_{-4.85}$   & 5.68                           & $1.01^{+0.22}_{-0.08}$ &                    --- &                   --- &                    --- & $35.25^{+1.61}_{-3.35}$   & ---                    &  3.57\\
8  &  $0.74^{+0.39}_{-0.39}$   & $2.00\pm1.0              $ & $3.02^{+0.73}_{-0.48}$  & $4.30\pm0.30         $ & $50.00^{+13.58}_{-6.55}$   & 6.03                           & $4.42^{+1.17}_{-0.63}$ & $1.01^{+0.05}_{-0.06}$ &$0.12^{+0.04}_{-0.04}$ &$-3.61^{+0.15}_{-0.09}$ & $29.88^{+1.28}_{-2.41}$   & ---                    &  1.48\\
9  &  $0.46^{+0.27}_{-0.31}$   & $2.00\pm1.0              $ & $3.89^{+0.41}_{-0.35}$  & $3.70\pm0.30         $ & $125.64^{+12.39}_{-9.24}$  & 7.78                           & $2.00^{+0.20}_{-0.18}$ &          $1.00\pm0.10$ &$0.09^{+0.05}_{-0.05}$ &$-3.21^{+0.26}_{-0.30}$ & $16.64^{+0.92}_{-1.61}$   & ---                    &  2.14\\
10 &  $0.61^{+0.34}_{-0.37}$   & $2.00\pm1.0              $ & $3.97^{+0.11}_{-0.14}$  & $2.80\pm0.30         $ & $190.10^{+5.85}_{-5.30}$   & 7.94                           & $2.38^{+0.07}_{-0.06}$ &          $1.10\pm0.10$ &$0.14^{+0.06}_{-0.06}$ &$-3.78^{+0.14}_{-0.19}$ & $7.49^{+0.45}_{-0.51}$    & ---                    &  1.03\\
11 &  $0.60^{+0.36}_{-0.37}$   & $2.00\pm1.0              $ & $2.89^{+0.13}_{-0.07}$  & $4.30\pm0.30         $ & $143.09^{+6.13}_{-3.85}$   & 5.77                           & $1.87^{+0.09}_{-0.05}$ &          $1.10\pm0.10$ &$0.13^{+0.05}_{-0.05}$ &$-2.96^{+0.16}_{-0.16}$ & $9.95^{+1.95}_{-2.46}$    & ---                    &  1.24\\
12 &  $0.53^{+0.32}_{-0.31}$   & $2.00\pm1.0              $ & $3.13^{+0.44}_{-0.28}$  & $2.20\pm0.30         $ & $135.88^{+16.71}_{-10.94}$ & 6.27                           & $1.94^{+0.28}_{-0.13}$ &          $1.00\pm0.10$ &         $0.10\pm0.10$ &$-3.12^{+0.30}_{-0.34}$ & $11.96^{+4.88}_{-3.95}$   & ---                    &  1.02\\
13 &  $0.56^{+0.31}_{-0.30}$   & $2.00\pm1.0              $ & $2.53^{+0.23}_{-0.23}$  & $4.30\pm0.30         $ & $130.91^{+12.75}_{-11.16}$ & 5.06                           & $2.29^{+0.16}_{-0.19}$ &          $0.90\pm0.10$ &         $0.10\pm0.10$ &$-2.90^{+0.29}_{-0.27}$ & $16.30^{+8.53}_{-5.07}$   & ---                    &  0.96\\
14 &  $0.54^{+0.21}_{-0.23}$   & $2.40^{+2.66}_{-1.28}    $ & $5.39^{+0.43}_{-0.39}$  & $3.11^{+0.31}_{-0.25}$ & $163.44^{+12.44}_{-11.70}$ & 10.78                          & $1.70^{+0.13}_{-0.14}$ & $1.06^{+0.03}_{-0.03}$ &$0.12^{+0.03}_{-0.03}$ &$-2.85^{+0.21}_{-0.27}$ & $6.58^{+3.98}_{-2.77}$    & $0.67^{+0.12}_{-0.08}$ &  1.29\\
15 &  $0.37^{+0.02}_{-0.02}$   & $2.00\pm1.0              $ & $5.39^{+0.06}_{-0.06}$  & $3.10\pm0.30         $ & $167.19^{+1.66}_{-2.46}$   & 10.78                          & $1.57^{+0.02}_{-0.02}$ &          $1.10\pm0.08$ &$0.13^{+0.04}_{-0.04}$ &$-2.73^{+0.02}_{-0.03}$ & $7.53^{+0.58}_{-0.82}$    & ---                    &  1.43\\
16 &  $0.13^{+0.11}_{-0.11}$   & $2.03^{+3.01}_{-0.77}    $ & $6.77^{+0.79}_{-0.58}$  & $2.46^{+0.29}_{-0.27}$ & $190.51^{+27.34}_{-16.68}$ & $48.86^{+7.03}_{-4.85}  $      & $1.71^{+0.22}_{-0.17}$ & $1.03^{+0.03}_{-0.04}$ &$0.11^{+0.04}_{-0.03}$ &$-2.86^{+0.11}_{-0.14}$ & $6.58^{+1.13}_{-2.12}$    & $0.55^{+0.07}_{-0.07}$ &  1.42\\
17 &  $0.35^{+0.16}_{-0.15}$   & $2.11^{+2.22}_{-1.01}    $ & $6.38^{+0.28}_{-0.31}$  & $2.40^{+0.16}_{-0.16}$ & $249.37^{+12.57}_{-13.30}$ & $167.35^{+10.68}_{-9.67}$      & $2.00^{+0.09}_{-0.10}$ & $1.00^{+0.03}_{-0.02}$ &$0.18^{+0.04}_{-0.03}$ &$-2.56^{+0.07}_{-0.04}$ & $6.58^{+0.80}_{-1.09}$    & $0.51^{+0.08}_{-0.08}$ &  1.29\\
18 &  $0.20^{+0.12}_{-0.12}$   & $2.53^{+5.24}_{-1.21}    $ & $6.24^{+0.12}_{-0.12}$  & $2.22^{+0.13}_{-0.13}$ & $253.63^{+5.14}_{-5.56}$   & $156.21^{+4.39}_{-3.70} $      & $2.00^{+0.04}_{-0.04}$ & $0.98^{+0.03}_{-0.04}$ &$0.15^{+0.03}_{-0.03}$ &$-2.52^{+0.07}_{-0.06}$ & $6.58^{+0.55}_{-0.55}$    & $0.56^{+0.07}_{-0.07}$ &  2.07\\
19 & $-0.13^{+0.25}_{-0.23}$   & $2.04^{+3.21}_{-1.01}    $ & $7.54^{+0.29}_{-0.22}$  & $2.22^{+0.12}_{-0.13}$ & $223.09^{+6.63}_{-6.71}$   & $160.44^{+8.27}_{-6.06} $      & $2.00^{+0.07}_{-0.06}$ &                    --- &                   --- &                    --- & $6.58^{+2.04}_{-2.25}$    & $0.70^{+0.04}_{-0.04}$ &  1.98\\
20 & $-0.75^{+0.05}_{-0.04}$   & $198.00^{+63.37}_{-62.46}$ & $9.61^{+0.25}_{-0.23}$  & $2.35^{+0.07}_{-0.07}$ & $198.21^{+5.22}_{-3.90}$   & $18.82^{+0.44}_{-0.45}  $      & $2.03^{+0.04}_{-0.05}$ &                    --- &                   --- &                    --- & $6.58^{+4.36}_{-4.13}$    & $1.07^{+0.06}_{-0.05}$ &  1.25
\enddata
\tablecomments{MCMC results of broadband SED fitting by the warm corona model (Section~\ref{sec:wcoronamodel}). Col. (1): Number of the broadband dataset. Col. (2): Dimensionless mass accretion rate, $\dot{m} \equiv \dot{M}/\dot{M}_\mathrm{Edd}$. Cols. (3)--(5): Electron temperature ($kT_\mathrm{hot}$), outer radius ($R_\mathrm{out}^\mathrm{hot}$), and photon index ($\Gamma_\mathrm{hot}$) for the hot Comptonizing component. Cols. (6)--(8): Electron temperature ($kT_\mathrm{warm}$), outer radius ($R_\mathrm{out}^\mathrm{warm}$), and photon index ($\Gamma_\mathrm{warm}$) for the warm Comptonizing component. Col. (9): Central energy of the broad Gaussian line at $\sim 1\,$keV. Col. (10): Width of the broad Gaussian line. Col. (11): Integrated flux of the broad Gaussian line. Col. (12): Column density of the intrinsic neutral absorber. Col. (13): Cross-calibration constant applied to the NuSTAR data. Col. (14): Statistics divided by the degree of freedom from the best-fit model.
\tablenotetext{ \dagger }{We linked $R_\mathrm{out}^\mathrm{warm} = 2 \, R_\mathrm{out}^\mathrm{hot}$ to mitigate parameter degeneracy, as suggested by \citet{Kubota2018MNRAS}.}
}
\end{deluxetable*}
\end{longrotatetable}